\documentclass[aps,prb,twocolumn,superscriptaddress,floatfix,showpacs,10pt,letterpaper]{revtex4}
\pdfoutput=1

\usepackage{amsthm}

\newtheoremstyle{wenthm}% name of the style to be used
  {3pt}% measure of space to leave above the theorem. E.g.: 3pt
  {3pt}% measure of space to leave below the theorem. E.g.: 3pt
  {\slshape}% name of font to use in the body of the theorem
  {}% measure of space to indent
  {\bfseries}% name of head font
  {:}% punctuation between head and body
  {.5em}% space after theorem head; " " = normal interword space
  {}% Manually specify head

\theoremstyle{wenthm}

\newtheorem{topinvE}{SPT invariant}

\newcommand{\topinv}[1]{\begin{topinvE}\dbl{#1}\end{topinvE}}

\newcommand{\RZ}{\R/\Z}

\begin{document}

\begin{titlepage}

\title{
Symmetry-protected topological invariants of\\
symmetry-protected topological phases of interacting bosons and fermions
}

\author{Xiao-Gang Wen}
\affiliation{Perimeter Institute for Theoretical Physics, Waterloo, Ontario, N2L 2Y5 Canada}
\affiliation{Department of Physics, Massachusetts Institute of
Technology, Cambridge, Massachusetts 02139, USA}

\begin{abstract}
Recently, it was realized that quantum states of matter can be classified as
long-range entangled (LRE) states (\ie with non-trivial topological order) and
short-range entangled (SRE) states (\ie with trivial topological order).  We
can use group cohomology class $\cH^d(SG,\R/\Z)$ to systematically describe the
SRE states with a symmetry $SG$ [referred as symmetry-protected trivial (SPT)
or  symmetry-protected topological (SPT) states] in $d$-dimensional space-time.
In this paper, we study the physical properties of those SPT states, such as
the fractionalization of the quantum numbers of the global symmetry on some
designed point defects, and the appearance of fractionalized SPT states on some
designed defect lines/membranes.  Those physical properties are SPT
invariants of the SPT states which allow us to experimentally or numerically
detect those SPT states, \ie to measure the elements in $\cH^d(G, \R/\Z)$ that
label different SPT states.  For example, 2+1D bosonic SPT states with $Z_n$
symmetry are classified by a $\Z_n$ integer $m \in \cH^3(Z_n, \R/\Z)=\Z_n$.  We
find that $n$ \emph{identical} monodromy defects, in a $Z_n$ SPT state labeled
by $m$, carry a total $Z_n$-charge $2m$ (which is not a multiple of $n$ in
general).

\end{abstract}


\maketitle

\end{titlepage}

{\small \setcounter{tocdepth}{1} \tableofcontents }

\section{Introduction}

\subsection{Beyond symmetry breaking and quantum entanglement}

Landau symmetry breaking theory\cite{L3726,GL5064,LanL58} was regarded as the
standard theory to describe all phases and phase transitions. However, in 1989,
through a theoretical study of chiral spin liquid\cite{KL8795,WWZ8913} in connection with high $T_c$
superconductivity, we realized that there exists a new kind of orders --
topological order.\cite{Wtop,WNtop,Wrig} Topological order cannot be
characterized by the local order parameters associated with the symmetry
breaking.  Instead, it is characterized/defined by (a) the robust ground state
degeneracy that depend on the spatial topologies\cite{Wtop,WNtop} and (b) the
modular representation of the degenerate ground states,\cite{Wrig,KW9327} just
like superfluid order is characterized/defined by zero-viscosity and quantized
vorticity.  In some sense, the robust ground state degeneracy and the modular
representation of the degenerate ground states  can be viewed as a type of
``topological order parameters'' for topologically ordered states.  Those
``topological order parameters'' are also referred as topological invariants of
topological order.

\begin{figure}[b]
\begin{center}
\includegraphics[scale=0.49]{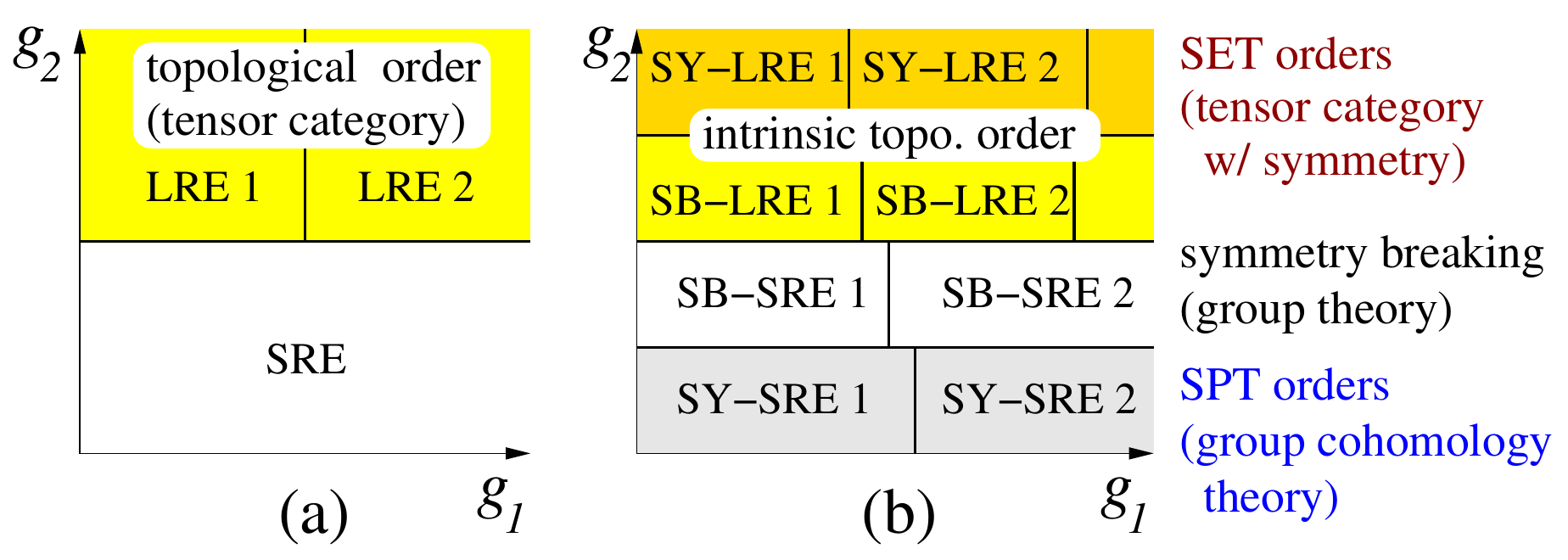}
%Fig. 1
\end{center}
\caption{
(Color online)
(a) The possible gapped phases for a class of Hamiltonians $H(g_1,g_2)$ without
any symmetry.  (b) The possible gapped phases for the class of
Hamiltonians $H_\text{symm}(g_1,g_2)$ with a symmetry.  The yellow regions in
(a) and (b) represent the phases with long range entanglement.  Each phase is
labeled by its entanglement properties and symmetry breaking properties.  SRE
stands for short range entanglement, LRE for long range entanglement, SB for
symmetry breaking, SY for no symmetry breaking.  SB-SRE phases are the Landau
symmetry breaking phases.  The SY-SRE phases are the SPT phases.  The SY-LRE
phases are the SET phases.
}
\label{topsymm}
\end{figure}

We know that, microscopically, superfluid order is originated from boson or
fermion-pair condensation.  Then, what is the  microscopic origin of
topological order?  Recently, it was found that, microscopically, topological
order is related to long range entanglement.\cite{LW0605,KP0604} In fact, we
can regard topological order as pattern of long range
entanglement\cite{CGW1038} defined through local unitary (LU)
transformations.\cite{LWstrnet,VCL0501,V0705} The notion of topological orders
and quantum entanglement leads to a point of view of quantum phases and quantum
phase transitions (see Fig.  \ref{topsymm}):\cite{CGW1038} for gapped quantum
systems without any symmetry, their quantum phases can be divided into two
classes: short-range entangled (SRE) states and long-range entangled (LRE)
states.

SRE states are states that can be transformed into direct product states via LU
transformations. All SRE states can be transformed into each other via  LU
transformations, and thus all SRE states belong to the same phase (see Fig.
\ref{topsymm}a).  LRE states are states that cannot be transformed into  direct
product states via LU transformations. There are LRE states that cannot be
connected to each other through LU transformations. Those LRE states represent
different quantum phases, which are nothing but the topologically ordered
phases.  Chiral spin liquids,\cite{KL8795,WWZ8913} fractional quantum Hall
states\cite{TSG8259,L8395}, $Z_2$ spin liquids,\cite{RS9173,W9164,MS0181}
non-Abelian fractional quantum Hall states,\cite{MR9162,W9102,WES8776,RMM0899}
\etc are examples of topologically ordered phases.

Topological order and long-range entanglement, as truly new phenomena, even
require new mathematical language to describe them.  It appears that tensor
category theory\cite{FNS0428,LWstrnet,CGW1038,GWW1017} and simple current
algebra\cite{MR9162,LWW1024} may be part of the new  mathematical language.
Using the new language, we have developed a systematic and quantitative theory
for non-chiral topological orders in 2D interacting boson and fermion
systems.\cite{LWstrnet,CGW1038,GWW1017} Also for chiral 2D topological orders
with only Abelian statistics, we find that we can use integer $K$-matrices to
classify them.\cite{BW9045,R9002,FK9169,WZ9290,BM0535,KS1193}

\subsection{Short-range entangled  states with symmetry}

For gapped quantum systems with symmetry, the structure of phase diagram is
much richer (see Fig. \ref{topsymm}b).  Even SRE states now can belong to
different phases, which include the well known Landau symmetry breaking states.
But even SRE states that do not break any symmetry can belong to different
phases, despite they all have trivial topological order and vanishing symmetry
breaking order parameters.  The 1D Haldane phase for spin-1
chain\cite{H8364,AKL8877,GW0931,PBT0959} and topological
insulators\cite{KM0501,BZ0602,KM0502,MB0706,FKM0703,QHZ0824} are non-trivial
examples of SRE phases that do not break any symmetry.  We will refer this kind
of phases as symmetry-protected trivial (SPT) phases or symmetry-protected
topological (SPT) phases.\cite{GW0931,PBT0959} Note that the SPT phases have no
long range entanglement and have trivial topological orders.

It turns out that there is no gapped bosonic LRE state in 1+1D (\ie topological
order does not exist in 1+1D).\cite{VCL0501} So all 1D gapped bosonic states
are either symmetry breaking states or SPT states.  This realization led to a
complete classification of all 1+1D gapped bosonic quantum
phases.\cite{CGW1107,SPC1139,CGW1128}

In \Ref{CLW1141,CGL1172,CGL1204}, the classification of 1+1D SPT phases is
generalized to any dimensions: \frm{For gapped bosonic systems in $d$
space-time dimensions with an on-site symmetry group $SG$, the SPT phases that
do not break the symmetry are described by the elements in $\cH^d[SG, \R/\Z]$
-- the group cohomology class of $SG$.} Such a systematic understanding of SPT
states was obtained by thinking those states as ``trivial'' short range
entangled states rather then topologically ordered states. The group cohomology
theory predicted several new bosonic topological insulators and bosonic
topological superconductors, as well as many other new quantum phases with
different symmetries and in different dimensions.  This led to an intense
research activity on SPT
states.\cite{LS0903,LV1219,LW1224,CW1217,HW1267,HW1232,VS1258,X1299,LL1209,HW1227,LL1263,YW1221,OCX1226,XS1372,WS1334,BCF1372,CG1303,CWL1321,CLV1301,YW1372,MKF1335,GL1369,LV1334}

What are the ``topological order parameters'' or more precisely SPT invariants that
can be used to characterize SPT states?  One way to characterize SPT states is
to gauge the on-site symmetry and use the introduced gauge field as an
effective probe for the SPT order.\cite{LG1220} This will be the main theme of
this paper.  After we integrate out the matter fields, a non-trivial SPT phase
will leads to a non-trivial quantized gauge topological term.\cite{HW1267} So
one can use the induced gauge topological terms, as the ``topological order
parameters'' or SPT invariants, to characterize the  SPT phases.  It
turns out that the quantized  gauge topological terms for gauge group $SG$ is
also classified by the same group cohomology class $\cH^d[SG, \R/\Z]$.  Thus
the gauge-probe will allow us to fully characterize the SPT phases.  We will use
the structure of $\cH^d[SG, \R/\Z]$ as a guide to help us to construct the
SPT invariants for the SPT states.  Another general way to obtain
SPT invariants is to study boundary states, which is effective for both
topological order\cite{W9211,WW1263,L1355} and SPT
order.\cite{CGW1107,VS1258}

We like to point out that the gauge approach can also be applied to fermion
systems.  \frm{We can use the elements in $\cH^d[G_f, \R/\Z]$ to characterize
fermionic SPT states\cite{GW1248} in $d$ space-time dimensions with a full
symmetry group $G_f$ (see section \ref{Gf}).}
  However, it is not clear if every
element in $\cH^d[G_f, \R/\Z]$ can be realized by fermion systems or not. It is
also possible that two different elements in $\cH^d[G_f, \R/\Z]$ may correspond
to the same fermionic SPT state. Despite the incomplete result, we can still
use $\cH^d[G_f, \R/\Z]$ to guide us to construct the SPT invariants for
fermionic SPT states.

\subsection{Long-range entangled  states with symmetry}

For gapped LRE states with symmetry, the possible quantum phases should be even
richer than SRE states. We may call those phases Symmetry Enriched Topological
(SET) phases. At moment, we do not have a classification or a systematic
description of SET phases. But we have some partial results.

Projective symmetry group (PSG) was introduced in 2002 to study the SET
phases.\cite{W0213,W0303a,WV0623} The PSG describes how the quantum numbers of
the symmetry group $SG$ get fractionalized on the gauge
excitations.\cite{W0303a} When the gauge group $GG$ is Abelian, the PSG
description of the SET phases can be be expressed in terms of group cohomology:
The different SET states with symmetry $SG$ and gauge group $GG$ can be
(partially) described by a subset of $\cH^2(SG,GG)$.\cite{EH1293}

One class of SET states in $d$ space-time dimensions with global symmetry $SG$
are described by weak-coupling gauge theories with gauge group $GG$ and
quantized topological terms (assuming the weak-coupling gauge theories are
gapped, that can happen when the space-time dimension $d=3$ or when $d>3$ and
the gauge group $GG$ is finite).  Those SET states (\ie the quantized
topological terms) are described by the elements in $\cH^d(PSG,
\R/\Z)$,\cite{MR1235,HW1227} where the group $PSG$ is an extension of $SG$ by
$GG$: $SG=PSG/GG$.  Or in other words, we have a short exact sequence
\begin{align}
 1\to GG\to PSG \to SG \to 1.
\end{align}
We will denote $PSG$ as $PSG = GG \gext SG$.  Many examples of the
SET states can be found in \Ref{W0213,KLW0834,KW0906,LS0903,YFQ1070}.

Although we have a systematic understanding of SPT phases and some of the SET
phases in term of $\cH^d(SG, \R/\Z)$ and $\cH^d(PSG, \R/\Z)$, however, those
constructions do not tell us to how to experimentally or numerically measure the
elements in  $\cH^d(SG, \R/\Z)$ or $\cH^d(PSG, \R/\Z)$ that label the different
SPT or SET phases.  We do not know, even given an exact ground state wave
function, how to determine which SPT or SET phase the ground state belongs to.

In this paper, we will address this important question.  We will find physical
ways to the detect different SPT/SET phases and to measure the elements in
$\cH^d(SG, \R/\Z)$ or $\cH^d(PSG, \R/\Z)$.  This is achieved by gauging the
symmetry group $SG$ (\ie coupling the $SG$ quantum numbers to a $SG$ gauge
potential $A^\text{SG}$). Note that $A^\text{SG}$ is treated as a
non-fluctuating probe field.  By study the topological response of the system
to various $SG$ gauge configurations, we can measure the elements in $\cH^d(SG,
\R/\Z)$ or $\cH^d(PSG, \R/\Z)$. Those  topological response are the measurable
SPT invariants (or ``topological order parameters'') that characterize
the SPT/SET phases.  Tables \ref{TopInv} and \ref{TopInvF}, and the
\textbf{SPT invariant} statements in the paper, describe the many
constructed SPT invariants, and represent the main results of the
paper.

\section{SPT invariants of SPT states: a general discussion}

Because of the duality relation between the SPT states and the SET states
described by weak-coupling gauge theories\cite{LG1220,MR1235,HW1227} (see
appendix \ref{dual}), in this paper, we will mainly discuss the physical
properties and the SPT invariants of the SPT state.  The physical
properties and the topological invariants of the SET states can be obtained
from the physical properties and the SPT invariants of corresponding
SPT states via the duality relation.

\subsection{Gauge the symmetry and twist the space}

Let us consider a system with symmetry group $G$ in $d$ space-time dimensions.
The ground state of the system is a SPT state described by an element $\nu_d$
in $\cH^d(G,\R/Z)$.  But how to physically measure $\nu_d$?  Following the idea
of \Ref{LG1220}, we will propose to measure $\nu_d$ by ``gauging'' the symmetry
$G$, \ie by introducing a $G$ gauge potential $A_\mu(x^i)$ to couple to the
quantum numbers of $G$.  The $G$ gauge potential $A_\mu$ is a fixed probe
field, not a dynamical field.  We like to consider how the system responds to
various $G$ gauge configurations described by $A_\mu$.  We will show that the
topological responses allow us to fully measure the cocycle $\nu_d$ that
characterizes the SPT phase, at least for the many cases considered.  Those
topological responses are the SPT invariants that we are looking for.

There are several topological responses that we can use to construct
SPT invariants: \enu{
\item We set up a time independent $G$ gauge configuration $A_\mu(x^i)$.
If the gauge configuration is invariant under a subgroup $GG$
of $G$: $A_\mu(x^i) = h^{-1} A_\mu(x^i) h$, $h\in GG$ , then we can study
the conserved $GG$ quantum number of the ground state under such gauge
configuration.  Some times, the ground states may be degenerate which form a
higher dimensional representation of $GG$.

In particular, the time independent $G$ gauge configuration may be chosen to be
a monopole-like or other soliton-like gauge configuration.  The quantum number
of the unbroken symmetry carried by those defects can be SPT invariants
of the SPT states.

We can also remove $n$ identical regions $D(i)$, $i=1,\cdots,n$, from the space
$M_{d-1}$ to get a $(d-1)$-dimensional manifold $M'_{d-1}$ with $n$ ``holes''.
Then we consider a flat $G$ gauge configuration $A_\mu(x^i)$ on $M'_{d-1}$
such that the gauge fields near the boundary  of those ``holes'', $\prt D(i)$,
are identical.  We then measure the  conserved $GG$ quantum number on the
ground state for such $G$ gauge configuration.  We will see that the $GG$
quantum number may not be multiples of $n$, indicating a non-trivial SPT
phases.

%\item  We start with a $G$ gauge configuration $A_\mu(x^i)$ in space, and
%then use an element $h \in GG \subset G$ to transform $A_\mu(x^i)$ to
%$A^h_\mu(x^i) = h^{-1} A_\mu(x^i) h$.  Let $|h\>$ be the ground state of
%the system with the  gauge configuration described by $A^h_\mu(x^i)$.  Now,
%we allow $h$ to be time dependent and derive the effective theory for $h$.  The
%effective theory is obtained from the coherent state $|h\>$ using the coherent
%state path integral approach, where the phase-space Lagrangian is given by
%\begin{align}
%L(h,\dot h)= \ii \<h| \frac{\dd}{\dd t}|h\>-\<h| H(A^h)|h\>
%\end{align}
%where $H(A^h)$ is the Hamiltonian with $A^h_\mu(x^i)$ gauge
%configuration.  Note that $\<h| H(A^h)|h\>$ is independent of $h$.  This will
%allow us to determine the $GG$ quantum number of the ground state.  Again, we
%consider space with $n$ identical holes and consider only flat $G$ gauge
%configurations.

\item We may choose the space to have a form $M_k \times M_{d-k-1}$ where $M_k$
is a closed $k$-dimensional manifold or a closed $k$-dimensional manifold with
$n$ identical holes.  $M_{d-k-1}$ is a closed $(d-k-1)$-dimensional manifold.
We then put a  $G$ gauge configuration $A_\mu(x^i)$ on $M_k$, or a flat $G$
gauge configuration on $M_k$ if $M_k$ has $n$ holes.  In the large $M_{d-k-1}$
limit, our system can be viewed as a system in $(d-k-1)$-dimensional space with
a symmetry $GG$, where $GG \subset G$ is formed by the symmetry transformations
that leave the $G$ gauge configuration invariant.  The ground state of the
system is a SPT state characterized by cocycles in $\cH^{d-k}(GG,\RZ)$.  The
mapping from the  gauge configurations on $M_k$ to $\cH^{d-k}(GG,\RZ)$ is our
SPT invariant.

\item
We can have a family of $G$ gauge configurations $A_\mu(x^i)$ that have the
same energy. As we go around a loop in such a family of $G$ gauge
configurations, the corresponding ground states will generate a geometric phase
(or non-Abelian geometric phases if the ground states are degenerate).
Sometimes, the (non-Abelian) geometric phases are also SPT invariants
which allow us to probe and measure the cocycles.  One such type of the
SPT invariants is the statistics of the $G$  gauge vortices in 2+1D
or monopoles in 3+1D.

\item
The above topological responses can be measured in a Hamiltonian formulation of
the system.  In the imaginary-time path-integral formulation of the system
where the space-time manifold $M_d$ can have an arbitrary topology, we can have
a most general construction of SPT invariants.  We simply put a
nearly-flat $G$ gauge configuration on a closed  space-time manifold $M_d$ and
evaluate the path integral. We will obtain a partition function $Z(M_d,A_\mu)$
which is a function of the space-time topology $M_d$ and the nearly-flat gauge
configuration $A_\mu$.  In the limit of the large volume $V=\la^d V_0$ of the
space-time (\ie $\la \to \infty$), $Z(M_d,A_\mu)$ has a form
(assuming we only scale the space-time volume without any change
in shape)
\begin{align}
 Z(M_d,A_\mu)=\ee^{- \sum_{n=1}^d f_n \la^n} Z_\text{top}(M_d,A_\mu),
\end{align}
where $ Z_\text{top}(M_d,A_\mu)$ is independent of the scaling factor $\la$.
$Z_\text{top}(M_d,A_\mu)$ is a SPT invariant that allows us to fully
measure the elements in $\cH^d(G,\R/Z)$ that describe the SPT
phases.\cite{DW9093,HW1267,HWW1295}  In fact, $ Z_\text{top}(M_d,A_\mu)$ is the
partition function for the pure topological term
$W^\text{gauge}_\text{top}(g,A)$ in \eqn{Sset}.

We like to point out that if $Z_\text{top}(M_d,A_\mu)$ contain a Chern-Simons
term (\ie $Z_\text{top}(M_d,A_\mu)=\ee^{\ii \int \cL_{CS}}$), then it describes
an SPT phase that is labeled by an element in the free part of $\cH^d(G,\R/Z)$.
If $Z_\text{top}(M_d,A_\mu)$ is a topological term whose value is independent of
any small perturbations of $A_\mu$, then it describes an SPT phase that is
labeled by an element in the torsion part of $\cH^d(G,\R/Z)$.\cite{HW1267}

}

\subsection{Cup-product,  K\"unneth formula, and SPT invariants}

The cohomology class $\cH^d(G,\RZ)\cong \cH^{d+1}(G,\Z)$ is not only an Abelian
group.  The direct sum of $\cH^{d+1}(G,\Z)$, $\cH^*(G,\Z)=\oplus_d
\cH^d(G,\Z)$, also has a cup-product that makes $\cH^*(G,\Z)$ into a ring:
\begin{align}
 \nu_{d_1} \cup \nu_{d_1} &=\nu_{d_1+d_2}, &
 \nu_{d_1+d_2} &\in \cH^{d_1+d_2}(G,\Z),
\nonumber\\
  \nu_{d_1} &\in \cH^{d_1}(G,\Z), &
 \nu_{d_1} &\in \cH^{d_1}(G,\Z).
\end{align}
We also have the K\"unneth formula
\begin{align}
\label{kunn1}
\cH^d&(GG  \times SG,\RZ) = \oplus_{k=0}^{d} \cH^{k}[SG,\cH^{d-k}(GG,\RZ)]
\nonumber\\
 &= \oplus_{k=0}^{d} \cH^{k}[GG,\cH^{d-k}(SG,\RZ)]
\nonumber\\
&= \left( \oplus_{k=1}^{d} \cH^{k-1}(GG,\RZ) \otimes_\Z \cH^{d-k}(SG,\RZ) \right) \oplus
\nonumber\\
&\ \ \ \,
\left( \oplus_{k=0}^{d} \cH^{k}(GG,\RZ)\boxtimes_\Z \cH^{d-k}(SG,\RZ) \right)
.
\end{align}
(see \eqn{kunnU} and \eqn{ucfGG}). Both of the above two results relate
cocycles at higher dimensions to cocycles at lower dimensions.  The structures
of the cup-product and  K\"unneth formula give us some quite direct hints on
how to construct SPT invariants that probe the cocycles.  For example,
consider a SPT state with symmetry $GG\times SG$, which is label by an element
in $\cH^d(GG \times SG,\RZ)$. If such an element belong to $\cH^{k-1}(GG,\RZ)
\otimes_\Z \cH^{d-k}(SG,\RZ)$ or $\cH^{k}(GG,\RZ) \boxtimes_\Z
\cH^{d-k}(SG,\RZ)$, then we can choose the space-time to have a topology
$M_k\times M_{d-k}$.  Next we try to design a defect that couple to
$\cH^{k-1}(GG,\RZ)=\cH^{k}(GG,\Z)$ or $\cH^{k}(GG,\RZ)$ on $M_k$ and try to
measure the response described by $\cH^{d-k}(SG,\RZ)$ on $M_{d-k}$.  (See also
SPT invariant \ref{U1U13dTop} and \ref{Zn1Zn22dTop}.) For
simple examples, see sections \ref{U1U13d} and \ref{Zn1Zn22d}.

In this paper, we will review some known SPT invariants, such as Hall
conductance and defect statistics, for some simple SPT states, such as 2+1D
$U(1)\times U(1)$, $U(1)\times Z_n$, and $Z_m\times Z_n$ SPT
states,\cite{{LS0903,LG1220,LV1219,HW1227,GL1369,LV1334}} and 3+1D $U(1)\rtimes
Z_2^T$ SPT state.\cite{MKF1335} We will also introduce some additional
SPT invariants, such as total quantum number of the identical monodromy
defects that can be created on a closed space and the dimension reduction of
SPT states, for those  simple SPT states.  We compare those SPT
invariants to the structures of the cup-product and  K\"unneth formula. This
comparison helps us to understand the relation between the
cup-product/K\"unneth-formula and the SPT invariants.  

We will discuss SPT invariants in many examples of SPT states, starting
from simple ones. Each example offers a little bit of new features than
previous example.  We hope that, through those examples, we will build some
intuitions of constructing SPT invariants for general SPT states.  Such
an intuition and understanding, in turn, help us to construct new SPT
invariants for more complicated SPT states (see SPT invariant
\ref{U1U13dTop} and \ref{Zn1Zn22dTop}).  The new understanding allows us to
construct SPT invariants for more general  SPT states in 1+1D, 2+1D
and 3+1D. The main results are summarize in Table \ref{TopInv}.

%In the following, we will illustrate the above constructions
%of SPT invariants using some simple examples. We will show that the
%constructed SPT invariants can fully characterize those SPT phases.

\section{SPT invariants of SPT states with simple symmetry groups}

\subsection{Bosonic $Z_n$ SPT phases}
\label{bZ2}

\subsubsection{0+1D}

In $1$-dimensional space-time, the bosonic  SPT states with symmetry
$Z_n=\{g^{(k)}=\ee^{2\pi k \ii/n}| k=0,\cdots,n-1\}$ are described by the
cocycles in $\cH^1(Z_n,\RZ)=\Z_n$.  How to measure the cocycles in
$\cH^1(Z_n,\RZ)$? What is the measurable SPT invariants that allow us
to characterize the $Z_n$ SPT states?

One way to construct a SPT invariant is to gauge the $Z_n$ global
symmetry in the action that describes that SPT state, and obtain a $Z_n$-gauge
theory $\cL(g_i,h_{ij})$, where $h_{ij}\in Z_n$ is the $Z_n$-gauge
``connection'' on the link connecting vertices $i$ and $j$, and $g_i\in Z_n$ is
the ``matter'' field that describes the SPT state (if we set $h_{ij}=1$).
Due to the gauge invariance, $\cL(g_i,h_{ij})$ has a form
$\cL(g_i,h_{ij})=\cL(g_i^{-1}h_{ij}g_j)$ (see \eqn{Vnud}).

After integral out the ``matter'' fields $g_i$, we obtain a SPT
invariant which appears as a topological term in the $Z_n$-gauge theory $
Z_\text{top}(M_d,A_\mu) =Z_\text{top}(M_d,h_{ij}) $.  (Note that, in a $Z_n$
gauge theory, $h_{ij}$ is the gauge ``connection'' $A_\mu$.) The $Z_n$-gauge
topological term can be expressed in term of cocycles $\om_1(h_{ij})$:
\begin{align}
\label{Z2top1d}
 Z_\text{top}(S_1,A_\mu)=\ee^{\ii 2\pi \sum_i  \om_1(h_{i,i+1})} ,
\end{align}
where we have assumed that the space-time is a circle $S_1$ formed by a ring of
vertices labeled by $i$.

In fact, before we integrate out that ``matter''
field $g_i$, the partition function for an ideal fixed-point SPT Lagrangian is
given by (see \eqn{Vnud})
\begin{align}
 Z(S_1,A_\mu)=\sum_{\{g_i\}}
\ee^{\ii 2\pi \sum_i  \om_1(g_i^{-1}h_{i,i+1}g_{i+1})} ,
\end{align}
where $\sum_{\{g_i\}}$ sums over all the $g_i$ configurations on $S_1$.  Since
$\ee^{\ii 2\pi \sum_i  \om_1(g_i^{-1}h_{i,i+1}g_{i+1})}$ is \emph{independent}
of $\{g_i\}$, we can integrate out $g_i$ easily and obtain \eqn{Z2top1d}.

A $Z_n$-gauge configuration on  $S_1$ is given by
$Z_n$ group elements $h_{i,i+1}$ on each link $(i,i+1)$.  We may view the
cocycle $\om_1$ as a ``discrete differential form'' and use the differential
form notion to express the above topological action amplitude
(which is also a $Z_n$-gauge topological term)
\begin{align}
\label{Z2CS1d}
 Z_\text{top}(S_1,A_\mu)=\ee^{\ii 2\pi \int_{S_1} \om_1(h_{i,i+1})} .
\end{align}
For more details on such a notation, see appendix \ref{dform}.
The cocycle condition (see appendix \ref{gcoh})
ensures that
\begin{align}
 Z_\text{top}(S_1,A_\mu)=\ee^{\ii 2\pi \int_{S_1} \om_1(h_{i,i+1})}=1
\end{align}
if $h_{i,i+1}=g^{i+1}g^{-1}_i$ is a pure $Z_n$-gauge.

The cocycles in $\cH^1(Z_n,\RZ)=\Z_n$ are labeled by $m=0,\cdots,n-1$ with
$m=0$ corresponding to the trivial cocycle.  The $m^{th}$ cocycle is given by
\begin{align}
 \om_1(g^{(k)})= \text{mod}(m k/n,1)
\end{align}

We note that the above cocycle $\om_1(h_{i,i+1})$
is a torsion element in  $\cH^1(Z_n,\RZ)$. It gives rise to
a quantized topological term $Z_\text{top}(S_1,A_\mu)$:
\begin{align}
 \ee^{\ii 2\pi \int_{S_1} \om_1(h_{i,i+1})} &=\ee^{2\pi m k \ii/n},\ & &\text{ if } \prod_i h_{i,i+1} =g^{(k)}.
\end{align}
Such a partition function is a SPT invariant.  Its non-trivial
dependence on the total $Z_n$ flux through the circle, $g^{(k)}=\prod_i
h_{i,i+1}$, implies that the SPT state is non-trivial.

The above partition function also implies that the ground state of the system
carries a $Z_n$ quantum number $m$.  Thus the non-trivial $Z_n$ quantum number
of the  ground state $m\neq 0$ also measure the non-trivial cocycle in
$\cH^1(Z_n,\RZ)$.

\subsubsection{Monodromy defect}

In $3$-dimensional space-time, the bosonic $Z_n$ SPT states are described by
the cocycles in $\cH^3(Z_n,\RZ)=\Z_n$.  To find the SPT invariants for
such a case, let us introduce the notion of monodromy defect.\cite{LG1220}

\begin{figure}[tb]
\begin{center}
\includegraphics[scale=0.4]{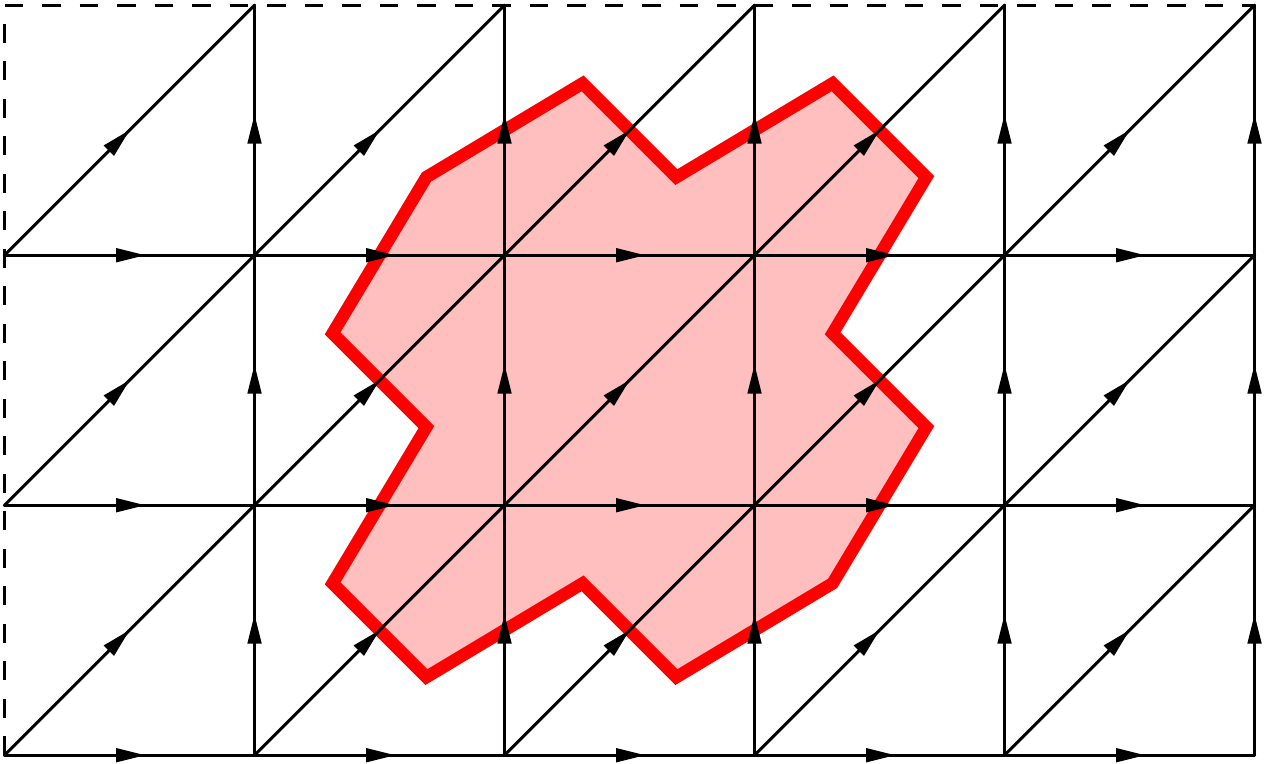} \end{center}
%Fig. 2
\caption{ (Color online)
A 2D lattice on a torus.  A $Z_n$ transformation is performed on the sites in
the shaded region.  The $Z_n$ transformation changes the Hamiltonian term on
the triangle $(ijk)$ across the boundary from $H_{ijk}$ to $H'_{ijk}$.
}
\label{ltrans}
\end{figure}

Let us assume that the 2D lattice Hamiltonian for a SPT state
with symmetry $G$ has a
form (see Fig. \ref{ltrans})
\begin{align}
 H=\sum_{(ijk)} H_{ijk},
\end{align}
where $\sum_{(ijk)}$ sums over all the triangles in Fig. \ref{ltrans} and
$H_{ijk}$ acts on the states on site-$i$, site-$j$, and site-$k$:
$|g_ig_jg_k\>$.  (Note that the states on site-$i$ are labeled by $g_i \in
G$.)  $H$ and $H_{ijk}$ are invariant under the global $G$ transformations.

Let us perform a  $G$ transformation only in the shaded region in Fig.
\ref{ltrans}. Such a transformation will change $H$ to $H'$.  However, only the
Hamiltonian terms on the triangles $(ijk)$ across the boundary are changed from
$H_{ijk}$ to $H'_{ijk}$.  Since the  $G$ transformation is an unitary
transformation, $H$ and $H'$ have the same energy spectrum.  In other words the
boundary in Fig.  \ref{ltrans} (described by $H'_{ijk}$'s) do not cost any
energy.

\begin{figure}[tb]
\begin{center}
\includegraphics[scale=0.4]{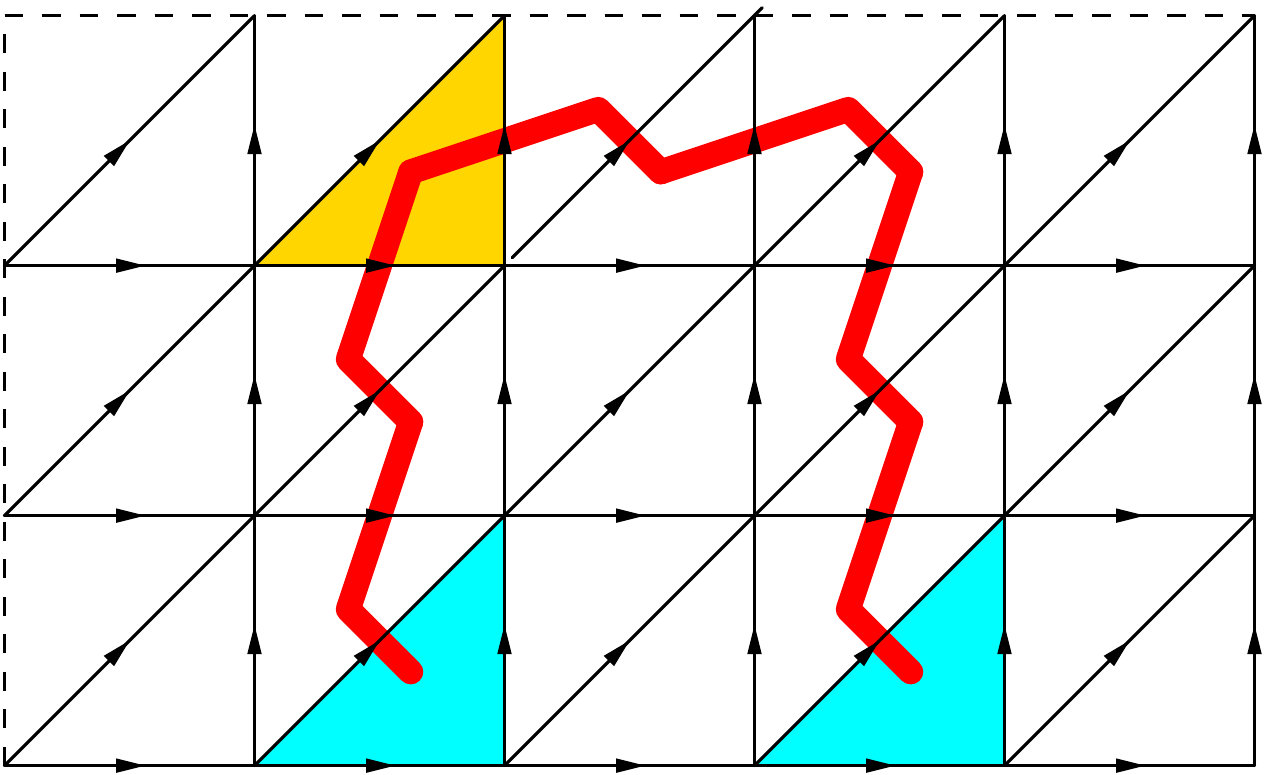} \end{center}
%Fig. 3
\caption{ (Color online)
A $Z_2$-gauge configuration with two \emph{identical} $Z_2$ vertices (or two
monodromy defects) on a torus.  Such a $Z_2$-gauge configuration has
$U^\text{bulk}_{-1}=-1$ (each yellow triangle contributes a factor $-1$).  Thus
$U^\text{bulk}_g$ forms a 1D representation of $Z_2$ with a $Z_2$-charge 1.
}
\label{z2gauge}
\end{figure}

Now let us consider a Hamiltonian on a lattice with a ``cut'' (see Fig.
\ref{z2gauge})
\begin{align}
\t H= { \sum_{(ijk)}}' H_{ijk} +{\sum_{(ijk)}}^\text{cut} H'_{ijk}
\end{align}
where $ \sum'_{(ijk)}$ sums over the triangles not on the cut and
$\sum^\text{cut}_{(ijk)}$ sums over the triangles that are divided into
disconnected pieces by the cut.  The triangles at the ends of the cut have no
Hamiltonian terms.  We note that the cut carries no energy. Only the ends of
cut cost energies.  Thus we say that the cut corresponds to two monodromy
defects.  The Hamiltonian $\t H$ defines the two  monodromy defects.

We also like to point out that the above procedure to obtain $\t H$ is actually
the ``gauging'' of the $G$ symmetry.  $\t H$ is a gauged Hamiltonian that
contain a $G$ vortex-antivortex pair  at the ends of the cut.

To summarize, a system with on-site symmetry $G$ can have many monodromy
defects, labeled by the group elements that generate the twist along the cut.
When $G$ is singly generated, we will call the monodromy defect generated by
the natural generator of $G$ as elementary monodromy defect.  In this case,
other  monodromy defects can be viewed a bound states of several elementary
monodromy defects.  In the rest of this paper, we will only consider the
elementary monodromy defects.

\subsubsection{2+1D: total $Z_n$-charge of $n$
\emph{identical} monodromy defects}
\label{Zn3D}

The SPT invariant to detect the cocycle in $\cH^3(Z_n,\RZ)$ is the
$Z_n$ quantum number of $n$ \emph{identical} monodromy defects created by the
twist $g^{(1)}\in Z_n$ (see Fig.  \ref{z2gauge}).  Note that the  monodromy
defects created by $g^{(1)}$ are the elementary  monodromy defects.  Other
elementary  monodromy defects can be viewed as bound states of the elementary
monodromy defects.  Also note that the monodromy defects or the $Z_n$-vortices
are identical which correspond to the same kind of $\begin{matrix}
\includegraphics[height=0.25in]{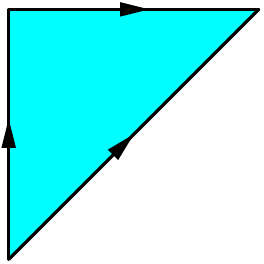} \end{matrix} $ triangles.

Since $\cH^3(Z_n,\RZ)=\Z_n$, the 2+1D $Z_n$ SPT states are labeled by
$m=0,\cdots,n-1$, with the corresponding 3-cocycle given by
\begin{align}
\label{om3Zn1}
 \om_3(g^{(k_1)},g^{(k_2)},g^{(k_3)})
&=\ee^{m\frac{2\pi \ii}{n^2} k_1(k_2+k_3-[k_2+k_3]_n)},
\nonumber\\
g^{(k)} &=\ee^{\frac{2\pi k \ii}{n}},
\end{align}
where
%we have used the integers $g_1,g_2,g_3$ to represent the $Z_n$ group elements
%and
%$[k_1+k_2]_n$ to represent group multiplication.  Here
$[k]_n$ is a short-hand notation for
\begin{align}
 [k]_n \equiv \text{mod}(k,n).
\end{align}
In appendix \ref{Zncharge}, we show that \topinv{$n$
\emph{identical} monodromy defects generated by $g^{(1)}$ twist in 2+1D
$Z_n$ SPT states on a torus always carry a total $Z_n$-charge $2m$, if the
$Z_n$ SPT states are described by the $m^{th}$ cocycle in $\cH^3(Z_n,\RZ)$.}
When $n=$odd, we find that the total $Z_n$-charge of $n$ \emph{identical}
monodromy defects allows us to completely characterize the 2+1D $Z_n$ SPT
states.  However, when $n=$even, the total $Z_n$-charge of $n$ \emph{identical}
monodromy defects only allows us to distinguish $n/2$ different $Z_n$ SPT
states.  The $m$ and $m+\frac{n}{2}$ $Z_n$ SPT states give rise to the same
total $Z_n$ charge, and cannot be distinguished this way.

We like to point out that when constructing the above SPT invariant,
we have assumed that the system has an additional translation symmetry although
the existence of the $Z_n$ SPT states do not require the translation symmetry.
We use the translation symmetry to make identical  monodromy defects,
which allow us to construct the above SPT invariant.

\subsubsection{2+1D: the statistics of the monodromy defects}
\label{ZnSta}

To construct new SPT invariant that can  distinguish $m$ and
$m+\frac{n}{2}$ $Z_n$ SPT states, we will consider the statistics of the
(elementary) monodromy defects.\cite{LG1220} To compute the statistics of the
monodromy defects we will use the duality relation between the $Z_n$ SPT states
and the twisted $Z_n$ gauge theory discovered by Levin and Gu.\cite{LG1220} The
(twisted) $Z_n$ gauge theory can be studied using $U(1)\times U(1)$
Chern-Simons theory.\cite{KLW0834,LS0903,LV1219,HW1227}

The $Z_n$ SPT states are described by $\cH^3(Z_n,\R/\Z)=\{m|m=0,\cdots,n-1\}$.
Thus, the $Z_n$ integer $m$ labels different 2+1D $Z_n$ SPT states.  The dual
gauge theory description of the $Z_n$ SPT state (labeled by $m$) is given by
\begin{align}
 \cL+W_\text{top}=\frac{1}{4\pi} K_{IJ} a_{I\mu}\prt_\nu a_{J\la} + ...
\end{align}
with
\begin{align}
\label{cszn}
K=
\bpm
  -2m & n\\
  n  & 0\\
\epm .
\end{align}
The $K$-matrix with $K_{11}=-2m$ correspond to the 3-cocycle in
\eqn{om3Zn1}.\cite{HW1227}
Note that, here, $a_{I\mu}$ are dynamical gauge fields whose charges are
quantized as integers. They are not the fixed probe gauge fields which are
denoted by capital letter $A_\mu$.  Two $K$-matrices $K_1$ and $K_2$ are
equivalent $K_1\sim K_2$ (\ie give rise to the same theory) if $K_1=U^TK_2U$
for an integer matrix with det$(U)=\pm 1$. We find that $K(m) \sim K(m+n)$.
Thus only $m=0,\cdots,n-1$ give rise to nonequivalent $K$-matrices.

A particle carrying $l_I$ $a^I_\mu$-charge will have a statistics
\begin{align}
 \th_l=\pi l_I (K^{-1})^{IJ}l_J .
\end{align}
A particle carrying $l_I$ $a^I_\mu$-charge will have a mutual statistics with a
particle carrying $\t l_I$ $a^I_\mu$-charge:
\begin{align}
\label{mutual}
 \th_{l,\t l}=2\pi l_I (K^{-1})^{IJ}\t l_J .
\end{align}

A particle with a unit of $Z_n$-charge is described by a particle with a
unit $a^1_\mu$-charge.  Using
\begin{align}
 K^{-1}=\frac{1}{n^2}\bpm
0&n\\
n&2m \\
\epm ,
\end{align}
we find that the $Z_n$-charge (the unit $a^1_\mu$-charge)
are always bosonic.

The $Z_n$ monodromy defect in the original theory corresponds to
$2\pi/n$-flux in $a^1_\mu$, since the unit $a^1_\mu$-charge corresponds to the
$Z_n$-charge in the original theory. We note that a particle carry $l_I$
$a^I_\mu$-charge created a $l_2\pi$ flux in $a^1_\mu$.  So a unit
$a^2_\mu$-charge always represent a $Z_n$ monodromy defect.  But such a $Z_n$
monodromy defect may not be a \emph{pure} $Z_n$ monodromy defect. It may carry
some additional $Z_n$-charges.

Since the $Z_n$ monodromy defect correspond to $2\pi/n$-flux in $a^1_\mu$, by
itself, a single monodromy defect is not an allowed excitation.  However, $n$
identical $Z_n$ monodromy defects (\ie $n$ particles that each carries a  unit
$a^2_\mu$-charge) correspond to $2\pi$-flux in $a^1_\mu$ which is an allowed
excitation.  Then, what is the total $Z_n$ charge of $n$ identical $Z_n$
monodromy defects (\ie $n$ units of $a^2_\mu$-charges)?  We note that $n$ units
of $a^2_\mu$-charges can be viewed as a bound state of a particle with
$(l_1,l_2)=(-2m,n)$ $a^I_\mu$-charges and a particle with $(l_1,l_2)=(2m,0)$
$a^I_\mu$-charges.  The  particle with $(l_1,l_2)=(2m,n)$ $a^I_\mu$-charges is
a trivial excitation that carry zero $Z_n$ charge, since
$(l_1,l_2)=(-2m,n)$ is a row of the $K$-matrix. The particle with
$(l_1,l_2)=(2m,0)$ $a^I_\mu$-charges carries $2m$ $Z_n$ charges.
Thus,  $n$ identical $Z_n$
monodromy defects (described by $n$ particles that each carries a  unit
$a^2_\mu$-charge) have $2m$ total $Z_n$ charges, which
agrees with the result obtained the in last section.

A particle that carries a  unit $a^2_\mu$-charge is only one way to realize the
$Z_n$ monodromy defect.  A generic $Z_n$ monodromy defect that may carry a
different $Z_n$-charge corresponds to $\v l^M=(l^M_1,1)$ $a^I_\mu$-charge.  The
statistics of such generic $Z_n$ monodromy defect is
\begin{align}
\label{thMZn}
 \th_M=
\pi (\v l^M)^T K^{-1} \v l^M =2\pi(\frac{l^M_1}{n} +\frac{m}{n^2})
.
\end{align}
We find that
\topinv{the statistical angle $\th_M$ of an elementary monodromy defect is a
SPT invariant that allows us to fully characterize the 2+1D bosonic
$Z_n$ SPT states.\cite{LG1220} In particular mod$(\frac{\th_M}{2\pi},\frac 1
n)=\frac{m}{n^2}$ where $m\in \cH^3(Z_n,\R/\Z)=\Z_n$ labels the different
$Z_n$ SPT states.}
We note that such a SPT invariant can full detect the 3-cocycles in
$\cH^3(Z_n,\RZ)$.

\subsubsection{$Z_n$-gauge topological term in 2+1D}

Just like the 0+1D case, we can also construct a SPT invariant and
probe the 3-cocycles in $\cH^3(Z_n,\RZ)$ by gauging the global $Z_n$ symmetry.
After integrating out the matter fields, we obtain a $Z_n$-gauge topological
term. Such a  $Z_n$-gauge topological term correspond to a 3-cocycle $\om_3$ in
$\cH^3(Z_n,\RZ)$ which describes the $Z_n$ SPT states.  In fact, the
$Z_n$-gauge topological term can be directly expressed in terms of the
3-cocycle $\om_3(h_{ij})$ (using the differential form notation in appendix
\ref{dform}):
\begin{align}
\label{ZnCS3d}
 \ee^{\ii 2\pi \int_{M_3} \om_3(h_{ij})},
\end{align}
where $M_3$ is the 3-dimensional space-time and $h_{ij}$ the $Z_n$-gauge
``connection'' in the link $ij$.  Such a $Z_n$-gauge topological term is a
generalization of the Chern-Simons term to a discrete group $Z_n$.

%The differential form notation can help us to perform the following
%calculation: Let us consider a non-trivial $Z_2$ SPT state on space-time
%manifold $M_2\times S_1$.  We add a $Z_2$-monodromy defect through the circle
%$S_1$ (\ie add $Z_2$ $\pi$-flux through the circle $S_1$).  In the small $S_1$
%limit, we may rewrite the $Z_2$-gauge topological term as
%\begin{align}
% \ee^{\ii 2\pi \int_{M_2\times S_1} \om_3(g_{i,i+1})}=
%\ee^{\ii 2\pi \int_{M_3} \om_2(g_{i,i+1})},
%\end{align}
%where $\om_2(g_{ij})$ is given by $U(g,g_{01},g_{12},g_{02})$ calculated
%before
%\begin{align}
% \ee^{\ii 2\pi  \om_2(g_{01},g_{12})}=
%U(-1,g_{01},g_{12},g_{02}).
%\end{align}
%From \eqn{Ugggg}, we see that
%the above $\om_2$ is a non-trivial 2-cocycle.
%

\subsubsection{4+1D}
\label{Zn5dsec}

We can also generalize the above construction to 5-dimensional space-time where
$Z_n$ SPT states are described by $\cH^5(Z_n,\RZ)=\Z_n$.  We choose the 4+1D
space-time to have a topology $M_2\times M_3$ where $M_3$ and $M_2$ are two
closed 2+1D and 2D manifolds.  We then create  $n$  \emph{identical} $Z_n$
monodromy defects on $M_2$.  In the large $M_3$ limit, we may view our 4+1D
$Z_n$ SPT state on space-time $M_3\times M_2$ as a 2+1D $Z_n$ SPT state on
$M_3$ which is described by $\cH^3(Z_n,\RZ)$. We have \topinv{in a 4+1D $Z_n$ SPT
state labeled by $m \in \cH^5(Z_n,\RZ)=\Z_n$ on space-time $M_3\times M_2$, $n$
\emph{identical} $Z_n$-vortices (\ie $Z_n$-monodromy defects) on $M_2$, induce
a 2+1D $Z_n$ SPT state labeled by $3m \in \cH^3(Z_n,\RZ)$ on $M_3$ in the small
$M_2$ limit.} We will show the above result when we discuss the $U(1)$ SPT
states in 4+1D (see section \ref{U15dsec}).

In the section \ref{Zn3D}, we have discussed how to detect the
cocycles in $\cH^3(Z_n,\RZ)$, by creating  $n$  \emph{identical} $Z_n$
monodromy defects on $M_2$, and then measure the $Z_n$-charge of the ground
state.  So the cocycles in $\cH^5(Z_n,\RZ)$ can be measured by creating  $n$
\emph{identical} $Z_n$-monodromy defects on $M_2$ and  $n$  \emph{identical}
$Z_n$-monodromy defects on $M_2'$. Then we  measure the $Z_n$-charge of the
corresponding  ground state.

The above construction of $Z_n$ SPT invariant is motivated by the
following mathematical result.  First $\cH^{2k+1}(Z_n,\RZ)\simeq
\cH^{2k+2}(Z_n,\Z)$.  The generating cocycle $c_{2k+2}$ in $\cH^{2k+2}(Z_n,\Z)$
can be expressed as a wedge product $c_{2k+2}=c_2 \wedge c_2\wedge \cdots
\wedge c_2$ where $c_2$ is the  generating cocycle in $\cH^{2}(Z_n,\Z)$.  Since
$\cH^{2}(Z_n,\Z)\simeq \cH^{1}(Z_n,\RZ)$, we can replace one of $c_2$ in
$c_{2k+2}=c_2 \wedge c_2\wedge \cdots \wedge c_2$ by $\th_1$ in
$\cH^{1}(Z_n,\RZ)$, and write $c_{2k+2}=\th_1 \wedge c_2\wedge \cdots \wedge
c_2$.  Note that $c_2\wedge \cdots \wedge c_2$ describes the topological gauge
configuration on $2k$ dimensional space, while $\th_1$ describes the 1D
representation of $Z_n$.  This motivates us to use a $Z_n$  gauge configuration
on $2k$ dimensional space to generate a non-trivial $Z_n$-charge in the ground
state.  In the next section, we use the similar idea to construct the
SPT invariant for bosonic $U(1)$ SPT states.

\subsection{Bosonic $U(1)$ SPT phases}

\subsubsection{0+1D}

In $1$-dimensional space-time, the bosonic SPT states with symmetry $U(1)
=\{\ee^{\ii\th}\}$ are described by the cocycles in $\cH^1[U(1),\RZ]=\Z$.  Let
us first study the SPT invariant from the topological partition
function.

A non-trivial cocycle in
$\cH^1[U(1),\RZ]=\Z$ labeled integer $m$ is given by
\begin{align}
 \om_1(\ee^{\ii\th}) = \ee^{\ii m \th}.
\end{align}
Let us assume the space-time is a circle $S_1$ formed by a ring of vertices
labeled by $i$.  A flat $U(1)$-gauge configuration on  $S_1$ is given the
$U(1)$ group
elements $\ee^{\ii\th_{i,i+1}}$ on each link $(i,i+1)$.  The topological part of the
partition function for such a flat $U(1)$-gauge configuration is determined by the above
cocycle $\om_1$
\begin{align}
 Z_\text{top}(S_1,A_\mu)=\ee^{\ii 2\pi \sum_i  \om_1(g_{i,i+1})} .
\end{align}
We note that the above $\om_1(g_{i,i+1})$
is a free element in  $\cH^1[U(1),\RZ]$. So it gives rise to
a Chern-Simons-type topological term $Z_\text{top}(S_1,A_\mu)$:
\begin{align}
 Z_\text{top}(S_1,A_\mu) &=
\ee^{\ii m \sum_i\th_{i,i+1}}
=
\ee^{\ii m \int_{S_1} A}
\end{align}
where $A$ is the $U(1)$-gauge potential one-form.  (Note that $\int_{S_1} A$ is
the $U(1)$ Chern-Simons term in 1D, and \eqn{Z2CS1d} can be viewed as a
discrete 1D Chern-Simons term for $Z_n$-gauge theory.) Such a partition
function is a SPT invariant.  When $m\neq 0$, its non-trivial
dependence on the total $U(1)$ flux through the circle, $\sum_i\th_{i,i+1}
=\oint \dd t A_0=\int_{S_1} A$, implies that the SPT state is non-trivial.

The above partition function also implies that the ground state of the system
carries a $U(1)$ quantum number $m$.  Thus the non-trivial $U(1)$ quantum
number $m$ of the  ground state also measure the non-trivial cocycle in
$\cH^1[U(1),\RZ]$.

\subsubsection{2+1D}
\label{bU1SPT}

In $3$-dimensional space-time, the bosonic $U(1)$ SPT states are described by
the cocycles in $\cH^3(U(1),\RZ)=\Z$.  How to measure the cocycles in
$\cH^3(U(1),\RZ)$?  One way is to ``gauge'' the $U(1)$ symmetry and put the
``gauged'' system on a 2D closed space $M_2$.  We choose a $U(1)$-gauge
configuration on $M_2$ such that there is a unit of $U(1)$-flux. We then
measure the $U(1)$-charge $q$ of the ground state on $M_2$.  We will show that
$q$ is an even integer and $q/2 =m \in \Z$ is the SPT invariant that
characterize the $U(1)$ SPT states.  In fact, such a   SPT invariant is
actually the quantized Hall conductance:
\topinv{The SPT invariant for 2+1D bosonic $U(1)$ SPT phases is given by
quantized Hall conductance which is quantized as even integers
$\si_{xy}=\frac{2m}{2\pi},   \ \ m\in \Z$.\cite{LV1219,LW1224,CW1217,SL1204}}

To show the above result, let us use the result that all 2+1D Abelian bosonic
topological orders can be described by $U^\ka (1)$ Chern-Simons theory
characterized by an even $K$-matrix:\cite{WZ9290}
\begin{align}
\label{CS}
\cL=
\frac{1}{4\pi} K_{IJ} a_{I\mu}\prt_\nu a_{J\la}\eps^{\mu\nu\la}
+
\frac{1}{2\pi} q_{I} A_{\mu}\prt_\nu a_{I\la}\eps^{\mu\nu\la}
+\cdots
\end{align}
The SPT states have a trivial topological order and are special cases of 2+1D
Abelian topological order.  Thus the  SPT states can be described by even
$K$-matrices with det$(K)=1$ and a zero signature.  In particular, we can use a
$U(1)\times U(1)$ Chern-Simons theory to describe the $U(1)$ SPT
state,\cite{LV1219,SL1204} with the $K$-matrix and the charge vector $\v q$
given by:\cite{BW9045,R9002,WZ9290}
\begin{align}
\label{Kq}
K=\bpm
0 & 1 \\
1 & 0 \\
\epm,\ \ \ \
\v q=\bpm
1  \\
m  \\
\epm, \ \ \ m\in \Z.
\end{align}
Note that, here, $a_{I\mu}$ are dynamical gauge fields. They are not
fixed probe gauge fields which are denoted by capital letter $A_\mu$.
The Hall conductance is given by
\begin{align}
 \si_{xy}= (2\pi)^{-1} \v q^T K^{-1} \v q = \frac{2 m}{2\pi}.
\end{align}

If we write the topological partition function as
$Z_\text{top}(M_d,A_\mu)=\ee^{\ii \int \dd^d x \cL_\text{top}}$, the above Hall
conductance implies that topological partition function is given by
a 3D Chern-Simons term (obtained from \eq{CS} by integrating out
$a_{I\mu}$'s)
\begin{align}
 \cL_\text{top} =
\frac{2m}{4\pi} A_\mu\prt_\mu A_\la \eps^{\mu\nu\la}
=\frac{2m}{4\pi} AF
\end{align}
where $F$ is the $U(1)$ field strength two-form.  Note that, in comparison,
\eqn{ZnCS3d} can be viewed as a discrete 3D Chern-Simons term for $Z_n$-gauge
theory.

The above result can be generalized to
other continuous symmetry group. For example:
\topinv{The SPT invariant for 2+1D bosonic $SU(2)$ SPT phases is given by
quantized spin Hall conductance which is quantized as half-integers
$\si_{xy}=\frac{m/2}{2\pi}, \ \ m\in \Z$.\cite{LW1224}}
\topinv{The SPT invariant for 2+1D bosonic $SO(3)$ SPT phases is given by
quantized spin Hall conductance which is quantized as even-integers
$\si_{xy}=\frac{2m}{2\pi}, \ \ m\in \Z$.\cite{LW1224}}

\subsubsection{4+1D}
\label{U15dsec}

In 5-dimensional space-time, the bosonic $U(1)$ SPT states are labeled by an
integer $m \in \cH^5(U(1),\RZ)=\Z$.  Again, one can construct a SPT
invariant to measure $m$ by ``gauging'' the $U(1)$ symmetry and putting the
``gauged'' system on a 4D closed space $M_4$.  We choose a $U(1)$-gauge
configuration on $M_4$ such that
\begin{align}
 \int_{M_4} \frac{F^2}{8\pi^2}  =1,
\end{align}
where $F$ is the two-form $U(1)$-gauge field strength and $F^2\equiv F\wedge F$
is the wedge product of differential forms.  We then measure the $U(1)$-charge
$q$ of the ground state induced by the $U(1)$-gauge configuration. Here the
potential SPT invariant $q$ must be an integer.

However, not all the integer  SPT invariants are realizable.  We find
that the bosonic $U(1)$ SPT states can only realized the SPT
invariants $q=6m$.  This is because, after integrating out that matter fields,
the  bosonic $U(1)$ SPT states are described by the following $U(1)$-gauge
topological term (see discussions in section \ref{U1U15d})
\begin{align}
\label{U15d}
 \cL_\text{top}= \frac{m}{(2\pi)^2} AF^2 .
\end{align}
Such a topological term implies that \topinv{$\int_{M_4} \frac{F^2}{8\pi^2}  =1$
gauge configuration on space $M_4$ will induce $6m$ $U(1)$-charges, for a
bosonic 4+1D $U(1)$ SPT state labeled by $m\in \cH^5(U(1),\RZ)=\Z$.} Thus $m/6$
measures the cocycles in $\cH^5(U(1),\RZ)$.

Again, one can also construct another SPT invariant by putting the
``gauged'' system on a 4+1D space-time with topology $M_2\times M_3$.  We
choose a $U(1)$-gauge configuration on $M_2$ such that
\begin{align}
 \int_{M_2} \frac{F}{2\pi}  =1.
\end{align}
In the large $M_3$ limit, we may view the 4+1D
system on $M_2\times M_3$ as a 2+1D system on $M_3$.
The 4+1D Chern-Simons topological term \eqn{U15d}  on $M_2\times M_3$ reduces to
a 2+1D Chern-Simons topological term on $M_3$:
\begin{align}
\label{U13d}
 \cL_\text{top}= \frac{3m}{2\pi} AF .
\end{align}
Such a  2+1D Chern-Simons topological term implies that the 4+1D $U(1)$ SPT on
on $M_2\times M_3$ reduces to a 2+1D $U(1)$ SPT labeled by $3m$ on $M_3$ in the
large $M_3$ limit.
To summarize,
\topinv{in a 4+1D $U(1)$ SPT state labeled by $m \in \cH^5[U(1),\RZ]=\Z$ on
space-time $M_3\times M_2$, $2\pi$ $U(1)$ flux on $M_2$ induces a 2+1D $Z_n$ SPT
state on $M_3$ labeled by $3m \in \cH^3[U(1),\RZ]$ in the large $M_3$ limit.}

We may embed the $Z_n$ group into the $U(1)$ group and view the $U(1)$ SPT
states as an $Z_n$ SPT state.  By comparing the $Z_n$ SPT invariants
and the $U(1)$ SPT invariants, we find that a $U(1)$ SPT state labeled
by $m \in \cH^d[U(1),\RZ]$ correspond to a $Z_n$ SPT state labeled by mod$(m,n)
\in \cH^d(Z_n,\RZ)$.

\subsection{Bosonic $Z_2^T$ SPT phases}

We have been constructing SPT invariants by gauging
the on-site symmetry. However, since we do not know how to gauge the time
reversal symmetry $Z_2^T$, to construct the SPT invariants for  $Z_2^T$
SPT phases, we have to use a different approach.

\subsubsection{1+1D}

We first consider bosonic $Z_2^T$ SPT states in 1+1 dimensions, where $Z_2^T$
is the anti-unitary time reversal symmetry.  The $Z_2^T$ SPT states are
described by $\cH^2[Z_2^T,(\R/\Z)_T]$, which is given by
\begin{align}
& \cH^2[Z_2^T,(\R/\Z)_T] = \Z_2 =\{m\}
\end{align}
Here $(\R/\Z)_T$ is the module $\R/\Z$. The subscript $T$ just stresses that
the time reversal symmetry $T$ has a non-trivial action on the module $\R/\Z$:
$T\cdot x =-x, \  x\in \RZ$.

We see that $m=0,1$ labels different 1+1D $Z_2^T$ SPT
states.  To measure $m$, we put the system on a finite line $I_1$.  At an end
of the line, we get degenerate states that form a projective representation of
$Z_2^T$, which is classified by $\cH^2[
Z_2^T,(\R/\Z)_T]$.\cite{CGW1107,SPC1139,CGW1128} We find that \topinv{a 1+1D
bosonic $Z_2^T$ SPT state labeled by $m$ has a degenerate Kramer doublet at an
open boundary if $m=1$.}

\subsubsection{3+1D}

The 3+1D $Z_2^T$ SPT states are
described by $\cH^4[Z_2^T,(\R/\Z)_T]$, which is given by
\begin{align}
& \cH^4[Z_2^T,(\R/\Z)_T] = \Z_2 =\{m\}
\end{align}
\Ref{VS1258,WS1334} have constructed several potential symmetry protected
SPT invariants for the $Z_2^T$ SPT states. Here we will give a brief
review of those potential SPT invariants.

The first way to construct the potential SPT invariants is to consider
a 3+1D $Z_2^T$ SPT state with a boundary.  We choose the boundary interaction
in such a way that the boundary state is gapped and does not break the
symmetry.  In this case, the 2+1D boundary state must  be a topologically
ordered state.  It was shown in \Ref{VS1258,WS1334} that if the boundary state
is a 2+1D $Z_2$ topologically ordered state\cite{RS9173,W9164}  and if the
$Z_2$-charge and the $Z_2$-vortex excitations in the  $Z_2$ topologically
ordered state are both Kramer doublets under the time-reversal symmetry, then
the 3+1D bulk $Z_2^T$ SPT state must be non-trivial.  Also if the boundary
state is a 2+1D ``all fermion $Z_2$ liquid''\cite{VS1258,WS1334,BCF1372},  then
the 3+1D bulk $Z_2^T$ SPT state must be non-trivial as well.
Both the above two SPT invariants can be realized by 3+1D states that
contain no topologically non-trivial particles.\cite{WS1334}

The second way to construct the potential SPT invariants is to break
the time reversal symmetry explicitly at the boundary only. We break the
symmetry in such a way that the ground state at the boundary is gapped without
any degeneracy.  Since there is no ground state degeneracy, there is no
excitations with fractional statistics at the boundary.  We may also break the
time reversal symmetry in the opposite way to obtain the time-reversal
partner of the above gapped non-degenerate ground state.  Now, let us
consider a domain wall between the above two ground states with opposite
time-reversal symmetry breaking.  Since there is no excitations with
fractional statistics at the boundary, the low energy edge state on the domain
wall must be a chiral boson theory described by an integer $K$-matrix which is
even and det$(K)=1$:
\begin{align}
\label{L1D}
 \cL_{1+1D} & = \frac{1}{4\pi} [
K_{IJ} \prt_t \phi_I \prt_x \phi_J-
V_{IJ} \prt_x \phi_I \prt_x \phi_J ]
\\
&\ \ \
+ \sum_l \sum_{J=1,2} [ c_{J,l} \ee^{\ii l \t K_{JI}\phi_I} + h.c.
]
,
\nonumber
\end{align}
where the field $\phi_I (x,t)$ is a map from the 1+1D space-time to a circle
$2\pi \RZ$, and $V$ is a positive definite real matrix.

If we modify the domain wall, while keeping the surface state unchanged, we
may obtain a different low energy effective chiral boson theory on the domain
wall described by a different even $K$-matrix, $K'$, with det$(K')=1$.  We say
the $K'$ matrix is equivalent to $K$.  According to \Ref{PMN1372}, the
equivalent classes of even $K$-matrices with det$(K)=1$ are given by
\begin{align}
 K= K_{E_8}\oplus \cdots \oplus K_{E_8},
\end{align}
where $K_{E_8}$ is the $K$-matrix that describes the $E_8$ root lattice.

When $K$ is a direct sum of even number $n$ of $K_{E_8}$'s, such a domain wall
can be produced by a pure 2D bosonic system, where the boundary ground state is
the bosonic quantum Hall state described by a
$K$-matrix\cite{BW9045,R9002,FK9169,WZ9290,BM0535,KS1193} that is a direct sum
of $n/2$  $K_{E_8}$'s.  The time-reversal partner is the bosonic quantum Hall
state described by a $K$-matrix that is a direct sum of $n/2$ $-K_{E_8}$'s.  In
this case, the edge state on the domain wall does not reflect any
non-trivialness of 3+1D bulk.  So if $K$ is a direct sum of even number $n$ of
$K_{E_8}$'s, it will represent a trivial potential SPT invariant.

When $K$ is a direct sum of an odd number of $K_{E_8}$'s, then, there is no
way to use a pure 2D bosonic system to produce such an edge state on the
domain wall.  Thus if the  domain wall between the time-reversal partners of
boundary ground states is described by a 1+1D chiral boson theory with a
$K$-matrix $K_{E_8}$ (or a direct sum of an odd number of $K_{E_8}$), then the
3+1D bosonic $Z_2^T$ SPT state is non-trivial.  It was suggested that such a
$K_{E_8}$ SPT invariant is the same as the all-fermion-$Z_2$-liquid
SPT invariant.\cite{VS1258,WS1334}

\subsection{Fermionic $U^f(1)$ SPT phases}
\label{Uf1}

Although the SPT invariant described above is motivated by the group
cohomology theory that describes the bosonic SPT states, however, the obtained
SPT invariant can be used to characterize/define fermionic SPT
phases.

The general theory of interacting fermionic SPT phases is not as well
developed as the bosonic SPT states.  (A general theory of \emph{free} fermion
SPT phases were developed in \Ref{K0886,SCR1101,AK1154}, which include the
non-interacting topological
insulators\cite{KM0501,BZ0602,KM0502,MB0706,R0922,FKM0703,QHZ0824} and the
non-interacting topological
superconductors.\cite{SMF9945,RG0067,R0664,QHR0901,SF0904}).  The first attempt
was made in \Ref{GW1248} where a group super-cohomology theory was developed.
However, the  group super-cohomology theory can only describe a subset of
fermionic SPT phases.  A more general theory is needed to describe all
fermionic SPT phases.

Even though the general theory of interacting fermionic SPT phases is not as
well developed, this does not prevent us to use the same SPT invariants
constructed by bosonic SPT states to study fermionic SPT states.  We hope the
study of the SPT invariants may help us to develop the more general
theory for interacting fermionic SPT phases.

\subsubsection{Symmetry in fermionic systems}
\label{Gf}

A fermionic system always has a $Z_2^f$ symmetry generated by $P_f\equiv
(-)^{N_F}$ where $N_F$ is the total fermion number.  Let us use $G_f$ to denote
the full symmetry group of the fermion system. $G_f$ always contain $Z_2^f$ as
a normal subgroup.  Let $G_b\equiv G_f/Z_2^f$ which represents the ``bosonic''
symmetry. We see that $G_f$ is an extension of $G_b$ by $Z_2^f$, described by the short exact sequence:
\begin{align}
\label{es}
 1\to Z_2^f \to G_f \to G_b \to 1 .
\end{align}
People some times use $G_b$ to describe the symmetry in fermionic systems and
some times  use $G_f$ to describe the symmetry.  Both $G_b$ and $G_f$ do not
contain the full information about the symmetry properties of a fermion system.
To completely describe the symmetry of a fermion system, we need to use  the
short exact sequence \eq{es}.  However, for simplicity, we will still use $G_f$
to refer the symmetry in fermion systems. When we say that a fermion system has
a $G_f$ symmetry, we imply that we also know how $Z_2^f$ is embedded in $G_f$
as a normal subgroup. (Note that $P_f$ always commute with any elements in
$G_f$: $ [P_f, g]=0,\  g \in G_f$.)

\subsubsection{SPT invariant for fermionic $U^f(1)$ SPT phases}

In this section, we are going to discuss the SPT invariant for the
simplest fermionic SPT states, which is a system with a full symmetry group
$G_f=U^f(1)$. The full symmetry group  contains $Z_2^f$ as a subgroup such that
odd $U^f(1)$-charges are always fermions.  We will use the SPT
invariant developed in the last section to study fermionic SPT states with a
$U^f(1)$ symmetry in $3$-dimensional space-time.  To construct the SPT
invariance, we first ``gauge'' the $U^f(1)$ symmetry, and then put the fermion
system on a 2D closed space $M_2$ with a $U^f(1)$ gauge configuration that
carries a unit of the gauge flux $\int_{M_2} \frac{F}{2\pi}=1$.  We then
measure the $U^f(1)$-charge $q$ of the ground state on $M_2$ induced by the
$U^f(1)$ gauge configuration.  Such a $U^f(1)$-charge is a SPT
invariant that can be used to characterize the fermionic $U^f(1)$ SPT phases.

Do we have other SPT invariant?  We may choose $M_2=S_1\times S_1$
(where $S_d$ is a $d$-dimensional sphere).  However, on $S_1\times S_1$ we do
not have additional \emph{discrete} topological $U^f(1)$ gauge configurations
except those described by the $U^f(1)$-flux $\int_{M_2} \frac{F}{2\pi}$
discussed above.  (We need \emph{discrete} topological gauge configurations to
induce discrete $U^f(1)$-charges.) This suggests that we do not have other
SPT invariant and the fermionic $U^f(1)$ SPT states are described by
integers $\Z$.  In fact, the integer $q$ is nothing but the integral quantized
Hall conductance $\si_{xy}=\frac{q}{2\pi}$.

The above just show that every fermionic $U^f(1)$ SPT state can be
characterized by an integer $q$. But we do not know if every integer $q$ can be
realized by a fermionic $U^f(1)$ SPT state or not.  To answer this question, we
note that a fermionic $U^f(1)$ SPT state is an Abelian state. So it can
described by a $U(1)\times \cdots\times U(1)$ Chern-Simons theory with an odd
$K$-matrix and a charge vector $\v q$.\cite{WZ9290} Let us first assume that
the $K$-matrix is two dimensional.  In this case, the fermionic $U^f(1)$ SPT
state must be described by a $U(1)\times U(1)$ Chern-Simons theory in \eqn{CS}
with the $K$-matrix and the charge vector $\v q$ of the form\cite{WZ9290}
\begin{align}
\label{Kq}
K=\bpm
1 & 0 \\
0 & -1 \\
\epm,\ \ \ \
\v q=\bpm
2m_1+1  \\
2m_2+1  \\
\epm, \ \ \ m_{1,2}=\text{ integers}.
\end{align}
We require the elements of $\v q$ to be odd integers since odd $U^f(1)$-charges
are always fermions.  The Hall conductance is given by
\begin{align}
 \si_{xy}= (2\pi)^{-1} \v q^T K^{-1} \v q =
\frac{4[m_1(m_1+1)-m_2(m_2+1)]}{2\pi}.
\end{align}
We find that
\topinv{the SPT invariant for 2+1D fermionic  $U^f(1)$ SPT phases is given
by quantized Hall conductance which is quantized as 8 times integers
$\si_{xy}=\frac{8m}{2\pi}, \ \ m\in \Z$.}
This result is valid even if we
consider higher dimensional $K$-matrices.

It is interesting to see that the potential SPT invariants for bosonic
$U(1)$ SPT states are integers (the integrally quantized Hall conductances). But
the actual  SPT invariants are even integers.  Similarly,  the
potential SPT invariants for fermionic $U^f(1)$ SPT states are also
integers (the integrally quantized Hall conductances).  However, the actual
SPT invariants are 8 times integers.

\subsection{Fermionic $Z_2^f$ SPT phases}

Next, we consider fermionic $Z_2^f$ SPT phases in 3-dimensional space-time.  We
find that the 2+1D fermionic $Z_2^f$ SPT phases have two types of potential
SPT invariants. However, so far we cannot find any fermionic SPT phases
that give rise to non-trivial SPT invariants.  This suggests that there
is no non-trivial fermionic $Z_2^f$ SPT phases in 3-dimensional space-time.
Let us use $fSPT^d_{G_f}$ to denote the Abelian group that classifies the
fermionic SPT phases with full symmetry group $G_f$ in $d$-dimensional
space-time.  The above result can be written as $fSPT^3_{Z^f_2}=0$.

Let us discuss the first potential SPT invariant.  We again create two
identical $Z_2^f$ monodromy defects on a closed 2D space.  We then measure the
$P_f$ quantum number $(-)^q$ for ground state with the two identical
$Z_2^f$ monodromy defects.  So the potential  SPT invariants $q$ are
elements in $\Z_2$.  But what are the actual  SPT invariants?
Can we realize the non-trivial SPT invariant $q=1$?

We may view a fermion $U^f(1)$ SPT phase discussed above as a $Z_2^f$ SPT phase
by viewing the $\pi$ $U^f(1)$ rotation as $P_f$.  In this case the SPT
invariants $q_U$ for the $U^f(1)$ SPT phases become the SPT invariants
$q$ for $Z_2^f$ SPT phases: $q=q_U$ mod 2.  To see this result, we note that
$q_U$ is the induced $U^f(1)$-charge by $2\pi$ $U^f(1)$-flux. $2\pi$ $U^f(1)$
flux can be viewed as two identical $Z_2^f$ vortex (each has $\pi$ $U^f(1)$
flux).  So the induced $Z_2^f$-charge is $q=q_U$ mod 2.

Since $q_U=0$ mod 8. Therefore fermionic $U^f(1)$ SPT phases always correspond
to a trivial  $Z_2^f$ SPT phase. We fail to get any non-trivial fermionic
$Z_2^f$ SPT phases from the  fermionic $U^f(1)$ SPT phases.

We like to point out that the induced $P_f$ quantum numbers by  two identical
$Z_2^f$ monodromy defects are not the only type of SPT invariants.
There exist a new type of SPT invariants for fermion systems:
\topinv{two identical $Z_2^f$
monodromy defects may induce topological degeneracy,\cite{WNtop} with different
degenerate states carrying different $P_f$ quantum numbers.}
This new type of SPT invariants is realized by a $p+\ii p$ state where
$2N$ identical $Z_2^f$ monodromy defects induce $2^N$ topologically degenerate
ground states.  Those topologically degenerate ground states are described by
$2N$ Majorana zero modes which correspond to $N$  zero-energy orbitals for
complex fermions.\cite{RG0067,I0168} But the  $p+\ii p$ state have an intrinsic
topological order which is not a fermionic SPT state.  So far we cannot find
any fermionic SPT phases that give rise to non-trivial SPT invariants
of the second type.  Thus we believe that $fSPT^3_{Z^f_2}=0$.

In 0+1D, we have non-trivial  fermionic SPT phases $fSPT^1_{Z^f_2}=\Z_2$. The
two fermionic SPT phases correspond to 0-dimensional ground state with no
fermion and one fermion.  One can also show that $fSPT^2_{Z^f_2}=0$, \ie no
non-trivial  fermionic SPT phases in 1+1D.\cite{GW1248}

\section{SPT invariants of SPT states with symmetry $G=GG \times SG$}

\subsection{Bosonic $U(1)\times \tilde U(1)$ SPT phases in 2+1D}

In this section, we are going to discuss the SPT invariant for bosonic
$U(1)\times \tilde U(1)$ SPT states in $3$-dimensional
space-time.\cite{CGL1172,CGL1204,LV1219}  To construct the SPT
invariance, we first ``gauge'' the $U(1)\times \tilde U(1)$ symmetry, and then
put the boson system on a 2D closed space $M_2$ with a $U(1) \times \tilde
U(1)$ gauge configuration $(A_\mu,\t A_\mu)$ that carries a unit of the
$U(1)$-gauge flux $\int_{M_2} \frac{F}{2\pi}=1$.  We then measure the
$U(1)$-charge $c_{11}$ and the $\tilde U(1)$-charge $c_{12}$ of the ground
state.  Next, we put another $U(1) \times \tilde U(1)$ gauge configuration on
$M_2$ with a unit of the $\tilde U(1)$ gauge flux $\int_{M_2} \frac{\tilde
F}{2\pi}=1$,  then measure the $U(1)$-charge $c_{21}$ and the $\tilde U(1)$
charge $c_{22}$.  We can use $c_{ij}$ to form a two by two integer matrix $\v
C$.  So an integer matrix $\v C$ is a potential SPT invariant for
fermionic $U(1)\times \tilde U(1)$ SPT phases in $3$-dimensional space-time.

But what are the actual realizable SPT invariants?  To answer this
question, let us consider the following $U(1)\times U(1)$ Chern-Simons theory
that describe the bosonic $U(1)\times \tilde U(1)$ SPT state
\begin{align}
\label{CSbb}
\cL &=
\frac{1}{4\pi} K_{IJ} a_{I\mu}\prt_\nu a_{J\la}\eps^{\mu\nu\la}
\\
&\ \ \ \
+
\frac{1}{2\pi} q_{1,I} A_{\mu}\prt_\nu a_{I\la}\eps^{\mu\nu\la}
+\frac{1}{2\pi}  q_{2,I} \tilde A_{\mu}\prt_\nu a_{I\la}\eps^{\mu\nu\la}
+\cdots
\nonumber
\end{align}
with the $K$-matrix and \emph{two} charge vectors
$\v q_1$,
$\v q_2$:
\begin{align}
\label{Kq}
K &=\bpm
0 & 1 \\
1 & 0 \\
\epm,\ \ \ \
\v q_1=\bpm
k_1  \\
k_2  \\
\epm, \ \ \
\v q_2=\bpm
k_3  \\
k_4  \\
\epm,
\nonumber\\
&
k_i=\text{ integers}.
\end{align}
The SPT invariant $\v C$ is given by
\begin{align}
 \v C= \Big( \v q_i^T K^{-1} \v q_j \Big).
\end{align}

Since stacking two SPT states with SPT invariants $\v C_1$ and $\v C_2$
give us a SPT state  with a SPT invariant $\v C_1 +\v C_2$, so the
actual SPT invariants form a vector space.  We find that the actual
SPT invariants form a three-dimensional vector space with basis vectors
\begin{align}
 \v C_1=
\begin{pmatrix}
 2 & 0\\
 0 & 0\\
\end{pmatrix}
,\ \ \
 \v C_2=
\begin{pmatrix}
 0 & 0\\
 0 & 2\\
\end{pmatrix}
,\ \ \
 \v C_3=
\begin{pmatrix}
 0 & 1\\
 1 & 0\\
\end{pmatrix}
.
\end{align}
So the bosonic $U(1)\times \tilde U(1)$ SPT phases in $3$-dimensional
space-time are described by three integers $\v C= m_1 \v C_1 +m_2 \v C_2 +m_3
\v C_3$, which agrees with the group cohomology result $\cH^3[U(1)\times \tilde
U(1),\RZ]=\Z^3$.

\subsection{Fermionic $U(1)\times U^f(1)$ SPT phases in 2+1D}
\label{U1Uf1}

Now let us discuss the SPT invariant for fermionic SPT states in
$3$-dimensional space-time, which has a full symmetry group $G_f=U(1)\times
U^f(1)$ (with $Z_2^f$ as a subgroup where odd $U^f(1)$-charges are always
fermions).  To construct the SPT invariance, we again ``gauge'' the
$U(1)\times U^f(1)$ symmetry, and then put the fermion system on a 2D closed
space $M_2$ with a $U(1) \times U^f(1)$ gauge configuration that carries a unit
of the $U(1)$-gauge flux $\int_{M_2} \frac{F}{2\pi}=1$.  We then measure the
$U(1)$-charge $c_{11}$ and the $U^f(1)$-charge $c_{12}$ of the ground state on
$M_2$ induced by the $U(1)$-gauge flux.  Next, we put another $U(1) \times
U^f(1)$ gauge configuration on $M_2$ with a unit of the $U^f(1)$ gauge flux
$\int_{M_2} \frac{\tilde F}{2\pi}=1$,  then measure the $U(1)$-charge $c_{21}$
and the $U^f(1)$-charge $c_{22}$.  So an integer matrix $\v C$ formed by
$c_{ij}$ is a potential SPT invariant for fermionic $U(1)\times U^f(1)$
SPT phases in $3$-dimensional space-time.

But what are the actual SPT invariants?  Let us consider the following
$U(1)\times U(1)$ Chern-Simons theory that describe the fermionic $U(1)\times
U^f(1)$ SPT state
\begin{align}
\label{CSff}
\cL &=
\frac{1}{4\pi} K_{IJ} a_{I\mu}\prt_\nu a_{J\la}\eps^{\mu\nu\la}
\nonumber\\
&\ \ \ \
+
\frac{1}{2\pi} q_{1,I} A_{\mu}\prt_\nu a_{I\la}\eps^{\mu\nu\la}
+\frac{1}{2\pi}  q_{2,I} \tilde A_{\mu}\prt_\nu a_{I\la}\eps^{\mu\nu\la}
+\cdots
\end{align}
with the $K$-matrix and \emph{two} charge vectors
$\v q_1$,
$\v q_2$:
\begin{align}
\label{Kq}
K &=\bpm
1 & 0 \\
0 & -1 \\
\epm,\ \ \ \
\v q_1=\bpm
m_1  \\
m_2  \\
\epm, \ \ \
\v q_2=\bpm
m_3  \\
m_4  \\
\epm,
\nonumber\\
&
m_{3,4}=\text{odd integers}.
\end{align}
The requirement ``$m_{3,4}=\text{ odd integers}$'' comes from the fact that odd
$U^f(1)$-charges are always fermions.  The SPT invariant $\v C$ is
given by
\begin{align}
 \v C= \Big( \v q_i^T K^{-1} \v q_j \Big).
\end{align}
We find that the actual SPT invariants form a three-dimensional
vector space with basis vectors
\begin{align}
 \v C_1=
\begin{pmatrix}
 1 & 1\\
 1 & 0\\
\end{pmatrix}
,\ \ \
 \v C_2=
\begin{pmatrix}
 0 & 0\\
 0 & 8\\
\end{pmatrix}
,\ \ \
 \v C_3=
\begin{pmatrix}
 0 & 2\\
 2 & 0\\
\end{pmatrix}
.
\end{align}
So the fermionic $U(1)\times U^f(1)$ SPT phases in $3$-dimensional
space-time are also described by three integers $\Z^3$.

\subsection{A general discussion for the case $G=GG \times SG$}
\label{GGxSG}

With the above two simple examples to give us some intuitive pictures, here we
like to give a general discussion for $G=GG \times SG$ cases.  In the appendix,
we show that that (see \eqn{kunnU})
\begin{align}
\cH^d(G,\RZ) = \oplus_{k=0}^{d} \cH^{k}[SG,\cH^{d-k}(GG,\RZ)] .
\end{align}
This means that we can use $(m_0,\cdots,m_d)$ to label each element of
$\cH^d(G,\RZ)$ where $m_k \in  \cH^{k}[SG,\cH^{d-k}(GG,\RZ)]$.  Note that $m_k$
only involves the group cohomology of smaller groups, which may be simpler.
Using the similar set up in the  above two examples, here we like to discuss
how to physically measure each $m_k$?

First, we notice that $\cH^{d-k}(GG,\RZ)$ describes the bosonic SPT phases in
$(d-k)$-dimensional space-time.  To stress this point, we rewrite $\cH^{d-k}(GG,\RZ)$
as $bSPT^{d-k}_{GG}$, and rewrite above decomposition as
\begin{align}
 \cH^d(G,\RZ) = \oplus_{k=0}^{d} \cH^{k}[SG,bSPT^{d-k}_{GG}] .
\end{align}
Since $bSPT^{d-k}_{GG}$ is a direct sum of $\Z$'s and $Z_n$'s, $
\cH^{k}[SG,bSPT^{d-k}_{GG}]$ is direct sum of $ \cH^{k}[SG,\Z]$'s and $
\cH^{k}[SG,\Z_n]$'s.
Such a structure motivates the following construction of SPT invariants
that allow us to measure $m_k$: we first gauge the $SG$ symmetry and create
non-trivial gauge configurations  described by ``$H^{k}(SG,\Z)$''. Such gauge
configurations will induce SPT invariants whose ``value'' is in
$bSPT^{d-k}_{GG}=\cH^{d-k}(GG,\RZ)$.  Again, we like to stress that the gauge
potentials for $SG$ are treated as fixed classical background without any
fluctuations.

%In other words, the gauge field
%for $SG$ is a non-fluctuating probe field that couples to the $SG$ quantum
%numbers. We then, examine the properties of our model with such a
%non-fluctuating $SG$ gauge field as a background.

To create suitable gauge configurations, we may choose the space-time manifold
to have a form $M_k\times M_{d-k}$ where $M_k$ has  $k$ dimensions and
$M_{d-k}$ has  $d-k$ dimensions.  We assume the $SG$ gauge configuration to be
constant on $M_{d-k}$. Such a $SG$ gauge configuration can be viewed as a gauge
configuration on $M_k$.  Now we assume that $M_k$ is very small, and our system
can be viewed as a system on $M_{d-k}$ which has a $GG$ symmetry.  The ground
state of such a $GG$ symmetric system is $GG$ SPT state on $M_{d-k}$ which is
labeled by an element in $bSPT^{d-k}_{GG}=\cH^{d-k}(GG,\R/\Z)$.  This way, we
obtain a function $\t m_k$ that maps a $SG$ gauge configuration on $M_k$ to an
element in $\cH^{d-k}(GG,\R/\Z)$.  In the above, we have discussed how to
measure such an element physically when $GG=U(1), Z_n$.

We note that $m_k$ in $\cH^k[SG,\cH^{d-k}(GG,\R/\Z)]$ is a cocycle, which is
denoted as $\om_k$ in section \ref{nonhomo}. $\om_k$ maps a $SG$ gauge
configuration on a $k$-cell in $M_k$ to an element in $\cH^{d-k}(GG,\R/\Z)$.
In fact $\om_k$ (or $m_k$) is given by
\begin{align}
 \om_k( s_{01}, s_{12},\cdots,s_{k-1,k}) \in \cH^{d-k}(GG,\R/\Z),
\end{align}
where $s_{ij} \in SG$ live on the edges of the $k$-cell which describe a $SG$
gauge configuration on the $k$-cell.  If we sum over the contributions from all
the $k$-cells in $M_k$, we will obtain the above $\t m_k$ function
that maps an $SG$ gauge configuration on $M_k$ to an element in
$\cH^{d-k}(GG,\R/\Z)$.

The key issue is that whether the function $\t m_k$ allows
us to fully detect $m_k \in \cH^k[SG,\cH^{d-k}(GG,\R/\Z)]$, \ie whether
different $m_k$ always lead to different $\t m_k$.  We can show that this is
indeed the case using the classifying space.  Let $BSG$ be the classifying
space of $SG$.  We know that the group cocycles in
$\cH^k[SG,\cH^{d-k}(GG,\R/\Z)]$ can be one-to-one represented by the
topological cocycles in $H^k[BSG,\cH^{d-k}(GG,\R/\Z)]$.  We know that a
topological cocycle $m_k^B$ in  $H^k[BSG,\cH^{d-k}(GG,\R/\Z)]$ gives rise to a
function that maps all the $k$-cycles in $BSG$ to $\cH^{d-k}(GG,\R/\Z)$.  And
such a function can fully detect the cocycle $m_k^B$ (\ie different  cocycles
always lead to different mappings).  We also know that each  $k$-cycles in
$BSG$ can be viewed as an embedding map from a $k$-dimensional space-time $M_k$
to $BSG$, and each embedding map define a $SG$ gauge configuration on $M_k$.
Thus the topological cocycle $m_k^B$ is actually a
function that maps a  $SG$ gauge configuration in space-time to
$\cH^{d-k}(GG,\R/\Z)$, and such a mapping can fully detect $m_k^B$.  All the
$k$-cycles in $BSG$ can be continuously deformed into a particular type of $k$
cycles where all the vertices on the $k$-cycle occupy one point in $BSG$.  The
$m_k^B$ that maps the $k$-cycles to $\cH^{d-k}(GG,\R/\Z)$ is a constant under
such a deformation.  $m_k^B$, when restricted on the  $k$-cycles whose vertices
all occupy one point, become the map $\t m_k$.  This way, we show that the
function $\t m_k$ can fully detect the group cocycles $m_k$ in
$\cH^k[SG,\cH^{d-k}(GG,\R/\Z)]$.  This is how we fully measure $m_k$.

In the above we see that each  embedding map from  $k$-dimensional space-time
$M_k$ to $BSG$ define a  $SG$ gauge configuration on $M_k$.  This relation
tells us how to choose the $SG$ gauge configurations on $M_k$ so that we can
fully measure $m_k$.  We choose the  $SG$ gauge configurations on $M_k$ that
come from the  embedding maps from  $M_k$ to $BSG$ such that the images are the
non-trivial $k$-cycles in $BSG$.

\subsection{An example with $SG=U(1)$ and $GG=U(1)$ }

\subsubsection{2+1D}
\label{U1U13d}

Let us reconsider the bosonic SPT states with symmetry  $G=U^{SG}(1)\times
U^{GG}(1)$ (\ie $SG=U(1)\equiv U^{SG}(1)$ and $GG=U(1)\equiv U^{GG}(1)$) in 3
space-time dimensions.  Such SPT states are described by $\cH^3(G,\R/\Z)$ with
$G=U^{SG}(1)\times U^{GG}(1)$.  We have
\begin{align}
\label{SGGG}
& \cH^3(G,\R/\Z) =
\oplus_{k=0}^{3} \cH^{k}[U(1)^{SG},\cH^{3-k}(U(1)^{GG},\R/\Z)]
\nonumber\\ &
=
\cH^3(U^{GG}(1),\R/\Z) \oplus \cH^2[U^{SG}(1),\cH^{1}(U^{GG}(1),\R/\Z)]
\nonumber\\ &\ \ \ \ \
 \oplus
\cH^3(U^{SG}(1),\R/\Z) ,
\end{align}
 with
\begin{align}
\cH^3(U^{GG}(1),\R/\Z) &=\Z=\{m_0\},
\nonumber\\
\cH^2[U^{SG}(1),\cH^{1}(U^{GG}(1),\R/\Z)]  &=
\\
\cH^1[U^{SG}(1),\RZ]\otimes_\Z \cH^{1}[U^{GG}(1),\R/\Z]  &=\Z=\{m_2\},
\nonumber\\
\cH^3(U^{SG}(1),\R/\Z) &=\Z=\{m_3\}.
\nonumber
\end{align}
$m_0$ labels different 2+1D $U^{GG}(1)$ SPT states and $m_3$ labels different
2+1D $U^{SG}(1)$ SPT states.  We have discussed how to measure $m_0$ and $m_3$
in section \ref{bU1SPT}.  Here we will discuss how to measure $m_2$.  The
structure of the K\"unneth expansion directly suggests the way to construct the
SPT invariant.

We first choose the space-time manifold to be $S_2\times S_1$, where $S_n$ is a
$n$-dimensional sphere.  We gauge the $GG$ symmetry and consider a $GG$ gauge
configuration with 1 unit of flux on $S_2$.  The flux on $S_2$ correspond to an
element in $\cH^2[U^{GG}(1),\Z)=\cH^1[U^{GG}(1),\RZ]$.  In the small $S_2$
limit, our system becomes a 0+1D $U^{SG}(1)$ symmetric theory on $S_1$.  The
ground state of such a 0+1D theory is a SPT state described by
$\cH^1[U^{SG}(1),\RZ]=\Z$ which corresponds to the $U^{SG}(1)$-charge of the
ground state. Such a charge happen to be $m_2$ that we intend to measure.  

This example also suggests the following general construction of SPT
invariant (see section \ref{GGxSG}): \topinv{ \label{U1U13dTop} In order to
measure $m\in \cH^{k-1}(GG,\RZ)\otimes_\Z \cH^{d-k}(SG,\RZ)\subset
\cH^{d}(GG\times SG,\RZ)$, we choose the space-time to have a topology
$M_k\times M_{d-k}$.  Next, we construct a $GG$ gauge configuration on $M_k$
that corresponds to an element of $\cH^k(GG,\Z)$.  In the large $ M_{d-k}$
limit, the system can be viewed as having a $d-k$ space-time dimension with
$SG$ symmetry.  Such a $d-k$ dimensional system is described by an element in $
\cH^{d-k}(SG,\RZ)$, which is the response to the $GG$ gauge configuration.
Measuring the responses for all possible  $GG$ gauge configurations allow us
to measure  $m\in \cH^{k-1}(GG,\RZ)\otimes_\Z \cH^{d-k}(SG,\RZ)$.} In the above
example, we try to measure $m_2\in \cH^1[U^{GG}(1),\RZ]\otimes_\Z
\cH^1[U^{SG}(1),\RZ]$ by choosing the space-time to be $S_2\times M_1$.  ``The
$U^{GG}$ gauge configuration with 1 unit of flux on $S_2$'' corresponds to an
element in $ \cH^2[U^{GG}(1),\Z]=\cH^1[U^{GG}(1),\RZ]$.  ``The
$U^{SG}(1)$-charge of the ground state'' corresponds to an element in $
\cH^1[U^{SG}(1),\RZ]$.  This example illustrate the idea of using the K\"unneth
formula \eq{kunn1} to construct SPT invariants.

In fact, if we also gauge the $U^{GG}(1)$ symmetry and integrate out the matter
fields (described by $a_{I\mu}$'s) in \eqn{CSbb}, $m_2$ will correspond to an
induced topological Chern-Simons term in $U^{SG}(1) \times U^{GG}(1)$ gauge
theory
\begin{align}
\label{CSy2}
\cL=\frac{m_2}{2\pi} A_{SG} F_{GG}
\end{align}
where $A_{SG}$ is the gauge potential one-form for the $U^{SG}(1)$ gauge field
and $F_{GG}$ is the field strength two-form for the $U^{GG}(1)$ gauge field.
Similarly, $m_0$ and $m_3$ also correspond to
topological Chern-Simons terms in $U^{SG}(1) \times U^{GG}(1)$ gauge theory
\begin{align}
\label{CSy0y3}
\cL= \frac{m_0}{2\pi} A_{GG} F_{GG}
+\frac{m_3}{2\pi} A_{SG} F_{SG}
\end{align}
So the topological partition function
$Z_\text{top}(M_d,A_\mu)=\ee^{\ii \int \dd^d x \cL_\text{top}}$
is given by
\begin{align}
\label{CSy0y2y3}
\cL_\text{top}= \frac{m_0}{2\pi} A_{GG} F_{GG}
+\frac{m_2}{2\pi} A_{SG} F_{GG}
+\frac{m_3}{2\pi} A_{SG} F_{SG}.
\end{align}
We see a direct correspondence between the K\"unneth expansion of the group
cohomology and the gauge topological term.

If we turn on one unit of $U^{GG}(1)$-flux on $S_2$ (described by a background
field $\bar A_{GG}$),
the above topological terms become (with $A_{GG}=\del A_{GG} + \bar A_{GG}$):
\begin{align}
\cL_\text{top}= \frac{2m_0}{2\pi} \del A_{GG} \bar F_{GG}
+ O(  \del A_{GG}^2) + \cdots
\end{align}
which implies that one  unit of $U^{GG}(1)$-flux on $S_2$ will induce $2m_0$
unit of $U^{GG}(1)$-charge. The factor 2 agrees with the result of
even-integer-quantized Hall conductance obtained before.

\subsubsection{4+1D}
\label{U1U15d}

Next, we consider bosonic $U^{SG}(1)\times U^{GG}(1)$ SPT states in 4+1D.
The SPT states are described by
\begin{align}
& \cH^5(G,\R/\Z) =
\oplus_{k=0}^{5} \cH^{k}[U(1)^{SG},\cH^{5-k}(U(1)^{GG},\R/\Z)]
\\ &
=
\cH^5(U^{GG}(1),\R/\Z)
\oplus \cH^2[U^{SG}(1),\cH^{3}(U^{GG}(1),\R/\Z)]
\nonumber\\ &\ \
\oplus \cH^4[U^{SG}(1),\cH^{1}(U^{GG}(1),\R/\Z)]
 \oplus
\cH^5(U^{SG}(1),\R/\Z) ,
\nonumber
\end{align}
 with
\begin{align}
\cH^5(U^{GG}(1),\R/\Z) &=\Z=\{m_0\},
\nonumber\\
\cH^2[U^{SG}(1),\cH^{3}(U^{GG}(1),\R/\Z)]  &=\Z=\{m_2\},
\nonumber\\
\cH^4[U^{SG}(1),\cH^{1}(U^{GG}(1),\R/\Z)]  &=\Z=\{m_4\},
\nonumber\\
\cH^5(U^{SG}(1),\R/\Z) &=\Z=\{m_5\}.
\end{align}
The topological terms labeled by $m_k$ are the Chern-Simons terms:
\begin{align}
\label{AGGASG}
 \cL_\text{top} &=
 \frac{m_0}{(2\pi)^2} A_{GG}F_{GG}^2
+\frac{m_2}{(2\pi)^2}  A_{SG}F_{GG}^2
\nonumber\\
&\ \ \ \
+\frac{m_4}{(2\pi)^2}  A_{GG}F_{SG}^2
+\frac{m_5}{(2\pi)^2} A_{SG}F_{SG}^2 .
\end{align}
which gives rise to the topological partition function
$Z_\text{top}(M_d,A_\mu)=\ee^{\ii \int \dd^d x \cL_\text{top}}$.

Why the topological terms must take the above form? Here we give an argument by
considering the following general topological terms with $\ka$ $U(1)$-gauge
fields
\begin{align}
 \cL_\text{top} =  \sum_{1\leq I\leq J \leq M \leq \ka}
\frac{K_{IJM}}{(2\pi)^2}  A^IF^JF^M.
\end{align}
First we assume $K_{IJM}$ are real numbers.  Then we like to show that, when
$I\neq J\neq M$,  $K_{IJM}$ must be quantized as integers. Otherwise, a gauge
configuration of $\int_{M_4} \frac{1}{(2\pi)^2}F^JF^M =1$ in the 4D space $M_4$
will induce a fractional $A^I$-charge.  Also, the quantization conditions on
$K_{IJM}$ should be invariant under the $SL(\ka,\Z)$ transformation $A^I \to
U_{IJ}A^J,\ U\in SL(\ka,\Z)$.  In this case, an integral $K_{IJM}$ for $I\neq
J\neq M$ will generate  integral $K_{IJM}$ for general $I,J,M$.  This leads us
to believe that $K_{IJM}$ are quantized as integers for general $I,J,M$.  So
the topological terms must take the form as in \eqn{AGGASG}.

Now let us go back to the $U^{SG}(1)\times U^{GG}(1)$ topological terms
\eq{AGGASG}.  We have discussed the measurement of $m_0$ and $m_5$ before in
our discussion of $U(1)$ SPT states.  To measure $m_2$, we choose a space-time
manifold of a form $M_2\times M'_2\times S_1$ (where $S_1$ is the time
direction).  We put  a $SG$ gauge field on space $M_2$ such that $\int_{M_2}
\frac{1}{2\pi}  F_{SG}=1$.  In the small $M_2$ limit, our theory reduces to a
$GG$-gauge theory on $M'_2\times S_1$ described by $m_2$ in
$\cH^3[U^{GG}(1),\R/\Z]$.  We can then put  a $GG$ gauge field on space $M'_2$
such that $\int_{M'_2} \frac{1}{2\pi}  F_{GG}=1$.  Such a configuration will
induce $2m_2$ unit of $U^{GG}(1)$-charges.  In other words,  a $SG$ gauge field
on space $M_2$ such that $\int_{M_2} \frac{1}{2\pi} F_{SG}=1$ and a  $GG$ gauge
field on space $M'_2$ such that $\int_{M'_2} \frac{1}{2\pi}  F_{GG}=1$ will
induce $2m_2$ units of $U^{GG}(1)$-charges.

The $m_4$ term can be measured by putting a $SG$ gauge field on space $M_4$
such that $\int_{M_4} \frac{1}{8\pi^2}  F_{SG}^2=1$.  Such a $SG$ gauge
configuration will induce $2m_4$ units of the $U^{GG}(1)$-charges.  The $SG$
gauge configuration will also induce $6m_5$ units of the $U^{SG}(1)$-charges.

\subsection{Bosonic $Z_{n_1}\times Z_{n_2}$ SPT states}

\subsubsection{2+1D}
\label{bZ2Z23d}

Next, let us consider SPT states with symmetry $G=Z_{n_1}\times Z_{n_2}$ in 2+1
dimensions. Such a theory was studied in \Ref{HW1227,LV1219,HW1227,LV1334}
using $U(1)$ Chern-Simons theory.  The $Z_{n_1}\times Z_{n_2}$ SPT states are
described by $\cH^3(Z_{n_1}\times Z_{n_2},\R/\Z)$, which has the following
decomposition (see \eqn{kunnU})
\begin{align}
& \cH^3(Z_{n_1}\times Z_{n_2},\R/\Z) =
\oplus_{k=0}^{3} \cH^{k}[Z_{n_1},\cH^{3-k}(Z_{n_2},\R/\Z)]
\nonumber\\ &
=
\cH^3(Z_{n_1},\R/\Z) \oplus \cH^2[Z_{n_1},\cH^{1}(Z_{n_2},\R/\Z)]
\nonumber\\ &\ \ \ \ \
 \oplus
\cH^3(Z_{n_2},\R/\Z) ,
\end{align}
 with
\begin{align}
  \cH^3(Z_{n_2},\R/\Z) &=\Z_{n_2}=\{m_0\},
\nonumber\\
\cH^2[Z_{n_1},\cH^{1}(Z_{n_2},\R/\Z)]  &=
\nonumber\\
\cH^1(Z_{n_1},\RZ)\otimes_\Z \cH^{1}(Z_{n_2},\R/\Z)  &=\Z_{\<n_1,n_2\>}=\{m_2\},
\nonumber\\
\cH^3(Z_{n_1},\R/\Z) &=\Z_{n_1}=\{m_3\}.
\end{align}
where $\<n_1,n_2\>$ is the greatest common divider of $n_1$ and $n_2$.  $m_0$
labels different 2+1D $Z_{n_2}$ SPT states and $m_3$ labels different 2+1D
$Z_{n_1}$ SPT states.  To measure $m_k$, we may create two identical $Z_{n_1}$
monodromy defects on a closed 2D space. We then measure the induced
$Z_{n_1}$-charge, which measures $2m_3$. We can also measure the induced
$Z_{n_2}$-charge, which measures $m_2$.

To understand why measuring the induced $Z_{n_1}$-charges and $Z_{n_2}$ charges
allow us to measure $2m_3$ and $m_2$, let us start with the dual gauge theory
description of the $Z_{n_1}\times Z_{n_2}$ SPT state.  The total Lagrangian has
a form
\begin{align}
 \cL+W_\text{top}=\frac{1}{4\pi} K_{IJ} a_{I\mu}\prt_\nu a_{J\la} + ...
\end{align}
with
\begin{align}
\label{csz2z2}
 K=\bpm
-2m_3 &n_1& -m_2 &0\\
n_1&0&0&0\\
 -m_2 &0&-2m_0 &n_2\\
0&0&n_2&0\\
\epm .
\end{align}
Note that, here, $a_{I\mu}$ are dynamical gauge fields. They are not
fixed probe gauge fields which are denoted by capital letter $A_\mu$.
Two $K$-matrices $K_1$ and $K_2$ are equivalent $K_1\sim K_2$ (\ie give rise to
the same theory) if $K_1=U^TK_2U$ for an integer matrix with det$(U)=\pm 1$. We
find $K(m_3 , m_2 ,m_0 ) \sim K(m_3 +n_1, m_2 ,m_0 ) \sim K(m_3 , m_2 +\< n_1,n_2\>,m_0 )
\sim K(m_3 , m_2 ,m_0 +n_2)$.  Thus only $\text{mod}(m_3,n_1)$,
$\text{mod}(m_2,\<n_1,n_2\>)$, $\text{mod}(m_0,n_2)$ give rise to
nonequivalent $K$-matrices.

A particle carrying $l_I$ $a^I_\mu$-charge will have a statistics
\begin{align}
 \th_l=\pi l_I (K^{-1})^{IJ}l_J .
\end{align}
A particle carrying $l_I$ $a^I_\mu$-charge will have a mutual statistics with a
particle carrying $\t l_I$ $a^I_\mu$-charge:
\begin{align}
\label{mutual}
 \th_{l,\t l}=2\pi l_I (K^{-1})^{IJ}\t l_J .
\end{align}

A particle with a unit of $Z_{n_1}$-charge is described by a particle with a
unit $a^1_\mu$-charge.  A particle with a unit of $Z_{n_2}$-charge is
described by a particle with a unit $a^3_\mu$-charge.  Using
\begin{align}
 K^{-1}=\bpm
0&\frac{1}{n_1}&0&0\\
\frac{1}{n_1}&\frac{2m_3}{n_1^2} &0& \frac{m_2}{n_1n_2} \\
0&0&0&\frac{1}{n_2}\\
0& \frac{m_2}{n_1n_2} &\frac{1}{n_2}&\frac{2m_0}{n_2^2} \\
\epm ,
\end{align}
we find that the $Z_{n_1}$-charge (the unit $a^1_\mu$-charge) and the
$Z_{n_2}$-charge (the unit $a^3_\mu$-charge) are always bosonic.

The $Z_{n_1}$ monodromy defect in the original theory corresponds to
$2\pi/n_1$-flux in $a^1_\mu$, since the unit $a^1_\mu$-charge corresponds to the
$Z_{n_1}$-charge in the original theory. We note that a particle carry $l_I$
$a^I_\mu$-charge created a $2l_2\pi/n_1$ flux in $a^1_\mu$.  So a unit
$a^2_\mu$-charge always represent a $Z_{n_1}$ monodromy defect.  Similarly,
a unit $a^4_\mu$-charge always represent a $Z_{n_2}$ monodromy defect.

Since a $Z_{n_1}$ monodromy defect corresponds to $2\pi/n$-flux in $a^1_\mu$,
by itself, a single monodromy defect is not an allowed excitation.  However,
$n_1$ identical $Z_{n_1}$ monodromy defects (\ie $n_1$ particles that each
carries a  unit $a^2_\mu$-charge) correspond to $2\pi$-flux in $a^1_\mu$, which
is an allowed excitation.  We note that $n$ units of $a^2_\mu$-charges can be
viewed as a bound state of a particle with
$(l_1,l_2,l_3,l_4)=(-2m_3,n_1,-m_2,0)$ $a^I_\mu$-charges and a particle with
$(l_1,l_2,l_3,l_4)=(2m_3,0,m_2,0)$ $a^I_\mu$-charges.  The  particle with
$(l_1,l_2,l_3,l_4)=(-2m_3,n_1,-m_2,0)$ $a^I_\mu$-charges is a trivial
excitation that carry zero $(Z_{n_1},Z_{n_2})$ charges.  The particle
with$(l_1,l_2,l_3,l_4)=(2m_3,0,m_2,0)$ $a^I_\mu$-charges carries $2m_3$
$Z_{n_1}$ charges and $m_2$ $Z_{n_2}$ charges.  Thus, \topinv{In a 2+1D
$Z_{n_1}\times Z_{n_2}$ bosonic SPT state labeled by $(m_0,m_2,m_3)$, $n_1$
identical elementary $Z_{n_1}$ monodromy defects have $2m_3$ total $Z_{n_1}$
charges and $m_2+ \<n_1,n_2\>\times$integer total $Z_{n_2}$ charges.  Similarly,
$n_2$ identical elementary $Z_{n_2}$ monodromy defects have $2m_0$ total
$Z_{n_2}$ charges and $m_2+ \<n_1,n_2\>\times$integer total $Z_{n_1}$ charges.  }
We see that to probe $m_2 \in \cH^2[Z_{n_1},\cH^{1}(Z_{n_2},\R/\Z)]
=\cH^1(Z_{n_1},\RZ) \otimes_\Z \cH^{1}(Z_{n_2},\R/\Z) $, we first turn on a
$Z_{n_1}$ gauge configuration of $n_1$ identical $Z_{n_1}$ monodromy defects
which corresponds to an element in $\cH^1(Z_{n_1},\RZ)=\cH^2(Z_{n_1},\Z)$.  We
then measure the induced $Z_{n_2}$ charge which corresponds to an element in
$\cH^1(Z_{n_2},\RZ)$.

We note that, some times, the above SPT invariants cannot fully detect
$m_0$ and $m_3$.
More complete SPT invariants can be obtained from the statistics of the
monodromy defects.  Let $\th_{11}$ be the statistic angle of the elementary
$Z_{n_1}$ monodromy defect and $\th_{22}$ be the statistic angle of the
elementary $Z_{n_2}$ monodromy defect.  Note that a generic  elementary
$Z_{n_1}$ monodromy defect is describe by a particle with
$(l_1,l_2,l_3,l_4)=(l^{Z_{n_1}}_1,1,l^{Z_{n_1}}_3,0)$ $a^I_\mu$-charges and a
generic elementary $Z_{n_2}$ monodromy defect is describe by a particle with
$(l_1,l_2,l_3,l_4)=(l^{Z_{n_2}}_1,0,l^{Z_{n_2}}_3,1)$ $a^I_\mu$-charges, where
$l^{Z_{n_1}}_{1,3}$ and  $l^{Z_{n_2}}_{1,3}$ describe different $Z_{n_1.n_2}$
charges that a generic monodromy defect may carry.
We find that an elementary
$Z_{n_1}$ monodromy defect has a statistics
\begin{align}
 \th_{11} &=2\pi \Big(\frac{m_3}{n_1^2}
+ \frac{l^{Z_{n_1}}_1}{n_1}\Big)
\end{align}
So $\th_{11}$
mod $  \frac{2\pi}{n_1}= 2\pi \frac{m_3}{n_1^2}$ is a SPT
invariance.  Similarly,  $\th_{22}$ mod $\frac{2\pi}{n_2}=2\pi
\frac{m_0}{n_2^2}$ is also a SPT invariance.  Let $\th_{12}$ be the
mutual statistical angle between an elementary $Z_{n_1}$ monodromy defect and
an  elementary $Z_{n_2}$ monodromy defect.  We find that $\th_{12}$ mod $
\frac{2\pi}{\{n_1,n_2\}}= 2\pi \frac{m_2}{n_1n_2}$ is a SPT invariance.
Here $\{n,m\}$ is the smallest common  multiple of $n$ and $m$.
Therefore, the  statistic of the monodromy defects give us the following
SPT invariants
\begin{align}
\v\Th &=\begin{pmatrix}
\th_{11} \text{ mod }   \frac{2\pi}{n_1}, & \th_{12}\text{ mod }  \frac{2\pi}{\{n_1,n_2\}} \\
\th_{12} \text{ mod }  \frac{2\pi}{\{n_1,n_2\}}, & \th_{22}\text{ mod }  \frac{2\pi}{n_2}  \\
\end{pmatrix}
\nonumber\\
&=
\begin{pmatrix}
\frac{2\pi m_3}{n_1^2} \text{ mod }   \frac{2\pi}{n_1} ,
& \frac{2\pi m_2}{n_1n_2} \text{ mod }   \frac{2\pi}{\{n_1,n_2\}}\\
 \frac{2\pi m_2}{n_1n_2} \text{ mod }   \frac{2\pi}{\{n_1,n_2\}} , &
\frac{2\pi m_0}{n_1^2} \text{ mod }   \frac{2\pi}{n_2}\\
\end{pmatrix}
\end{align}
We note that if we stack two SPT states with SPT invariants $(\v C,
\v\Th)$ and $(\v C', \v\Th')$, we obtain a new SPT state with SPT
invariants
\begin{align}
(\v C'', \v\Th'')
=(\v C, \v\Th)+(\v C', \v\Th')
.
\end{align}
\topinv{In a 2+1D $Z_{n_1}\times Z_{n_2}$ bosonic SPT state labeled by
$(m_0,m_2,m_3)$, the statistics/mutual-statistics matrix $\v\Th$ can fully
detect $m_0$, $m_2$, and $m_3$.}

Just like the bosonic $U^{SG}(1)\times U^{GG}(1)$ SPT states can be
characterized by the $U^{SG}(1)\times U^{GG}(1)$ Chern-Simons topological term
(see \eqn{CSy0y2y3}) after we gauge the global symmetry $U^{SG}(1)\times
U^{GG}(1)$, the bosonic $Z_{n_1}\times Z_{n_2}$ SPT states can also be
characterized by a $Z_{n_1}\times Z_{n_2}$ gauge topological term after we
gauge the  global $Z_{n_1}\times Z_{n_2}$ symmetry.  The  $Z_{n_1}\times
Z_{n_2}$ gauge topological term is obtained by integrating out the matter
fields in a back ground of $Z_{n_1}\times Z_{n_2}$ gauge configuration.  In
terms of the discrete differential forms (see appendix \ref{dform}), the
$Z_{n_1}\times Z_{n_2}$ gauge topological term can be written as
\begin{align}
 \cL_\text{top}=
 2\pi m_0 \om^{Z_{n_2}}_3
+2\pi m_2 \om^{Z_{n_1},Z_{n_2}}_{2,1}
+2\pi m_3 \om^{Z_{n_1}}_3
\end{align}
where $\om^{Z_{n_2}}_3 \in \cH^3(Z_{n_2},\R/\Z)$, $\om^{Z_{n_1}}_3 \in
\cH^3(Z_{n_1},\R/\Z)$, and $\om^{Z_{n_1},Z_{n_2}}_{2,1} \in
\cH^2[Z_{n_1},\cH^{1}(Z_{n_2},\R/\Z)]$.
Compare to \eqn{CSy0y2y3}, the above can be viewed as
discrete Chern-Simons terms for
$Z_{n_1}\times Z_{n_2}$ gauge fields.

\subsubsection{1+1D}
\label{Zn1Zn22d}

In the above examples, we see that measuring topological responses give rise to
a complete set of SPT invariants which fully characterize the SPT
states. We believe this is true in general.  Next we will use this idea to
study the $Z_{n_1}\times Z_{n_2}$ SPT states in 1+1D and 3+1D.

The 1+1D bosonic $G=Z_{n_1}\times Z_{n_2}$ SPT states are described by
$\cH^2(G,\R/\Z)$, which has the following decomposition (see \eqn{kunnU} and
\eqn{ucfGG})
\begin{align}
& \cH^2(G,\R/\Z) =
\oplus_{k=0}^{2} \cH^{k}[Z_{n_1},\cH^{2-k}(Z_{n_2},\R/\Z)]
\nonumber\\ &
= \cH^1[Z_{n_1},\cH^{1}(Z_{n_2},\R/\Z)]
\\ &
= \cH^1(Z_{n_1},\RZ) \boxtimes_\Z \cH^{1}(Z_{n_2},\R/\Z)
=\Z_{\<n_1,n_2\>}=\{m_1\}
\nonumber
\end{align}
To measure $m_1$, we choose the space to be $S_1$ and create a twist boundary
condition on $S_1$ generated by $g^{(1)}= \ee^{\ii 2\pi/n_1} \in Z_{n_1}$
(which corresponds to the generating element in $\cH^1(Z_{n_1},\Z_n)$).  Then
we measure the induced $Z_{n_2}$-charge on $S_1$ (which is
$\cH^1(Z_{n_2},\RZ)$).  The physical meaning of the above decomposition is that
the induced $Z_{n_2}$-charge mod $\<n_1,n_2\>$ is $m_1$ (for details see
section \ref{GGxSG}).  Thus, \topinv{In a 1+1D $Z_{n_1}\times Z_{n_2}$ SPT state
labeled by $m_1 \in \cH^2(Z_{n_1}\times Z_{n_2},\RZ)$, a twist boundary
condition on the space $S_1$ generated by $g^{(1)}= \ee^{\ii 2\pi/n_1} \in
Z_{n_1}$ will induce a $Z_{n_2}$-charge $m_1+\<n_1,n_2\>\times$integer in the
ground state.} 

This example also suggests the following general construction of SPT
invariant (see section \ref{GGxSG}): \topinv{ \label{Zn1Zn22dTop} In order to
measure $m\in \cH^k(GG,\RZ)\boxtimes_\Z \cH^{d-k}(SG,\RZ)\subset
\cH^{d}(GG\times SG,\RZ)$, we choose the space-time to have a topology
$M_k\times M_{d-k}$.  Next, we construct a $GG$ gauge configuration on $M_k$
that corresponds to an element of $\cH^k(GG,\RZ)$.  In the large $ M_{d-k}$
limit, the system can be viewed as having a $d-k$ space-time dimension with
$SG$ symmetry.  Such a $d-k$ dimensional system is described by an element in $
\cH^{d-k}(SG,\RZ)$, which is the response of the $GG$ gauge configuration.
Measuring the responses for all possible $GG$ gauge configurations allow us to
measure  $m\in \cH^k(GG,\RZ)\boxtimes_\Z \cH^{d-k}(SG,\RZ)$.} In the above
example, we try to measure $m_1\in \cH^1(Z_{n_1},\RZ)\boxtimes_\Z
\cH^1(Z_{n_2},\RZ)$ by choosing the space-time to be $S_1\times M_1$.  ``A
twist boundary condition on the space $S_1$ generated by $g^{(1)}= \ee^{\ii
2\pi/n_1} \in Z_{n_1}$'' corresponds to a $Z_{n_1}$ gauge configuration in
$\cH^1(Z_{n_1},\RZ)$.  ``The induced $Z_{n_2}$-charge in the ground state'' is
response in $\cH^1(Z_{n_2},\RZ)$.

\subsubsection{3+1D}
\label{Zn1Zn24d}

The 3+1D bosonic $G=Z_{n_1}\times Z_{n_2}$ SPT states are described by
$\cH^3(G,\R/\Z)$ with the following decomposition (see \eqn{kunnU})
\begin{align}
&\ \ \ \ \cH^4(G,\R/\Z) =
\oplus_{k=0}^{4} \cH^{k}[Z_{n_1},\cH^{4-k}(Z_{n_2},\R/\Z)]
\\ &
=
\cH^3[Z_{n_1},\cH^{1}(Z_{n_2},\R/\Z)]
\oplus \cH^1[Z_{n_1},\cH^{3}(Z_{n_2},\R/\Z)]
\nonumber
\end{align}
 with
\begin{align}
\cH^1[Z_{n_1},\cH^{3}(Z_{n_2},\R/\Z)]  &=
\nonumber\\
\cH^1(Z_{n_1},\RZ)\boxtimes_\Z \cH^3(Z_{n_2},\R/\Z)]  &=\Z_{\<n_1,n_2\>}=\{m_1\},
\nonumber\\
\cH^3[Z_{n_1},\cH^{1}(Z_{n_2},\R/\Z)]  &=
\\
\cH^3(Z_{n_1},\RZ)\boxtimes_\Z \cH^1(Z_{n_2},\R/\Z)]  &=\Z_{\<n_1,n_2\>}=\{m_3\},
\nonumber 
\end{align}
Motivated by the structure of the K\"unneth expansion, we can construct
SPT invariants in a similar way as what we did for the 1+1D SPT state.
For example, to measure $m_1$, we choose the space to be $S_1\times M_2$. We
then create a twist boundary condition on $S_1$ generated by $g^{(1)}= \ee^{\ii
2\pi/n_1} \in Z_{n_1}$ (which correspond to an element in
$\cH^1(Z_{n_1},\RZ)$).  In the small
$S_1$ limit, the SPT state on $S_1\times M_2$ reduces to a SPT state on $M_2$
which is described by $m_1+\<n_1,n_2\>\times \text{integer} \in
\cH^{3}(Z_{n_2},\R/\Z)$.  The element $m_1+\<n_1,n_2\>\times \text{integer}$ in
$\cH^{3}(Z_{n_2},\R/\Z)$ can be measured by the SPT invariants
discussed in section \ref{bZ2Z23d}.
%Similarly, if we create a twist boundary condition on $S_1$ generated by
%$g^{(1)}= \ee^{\ii 2\pi/n_2} \in Z_{n_2}$, then in the large $M_2$ limit, the
%$Z_{n_1}$ SPT state on $M_2$ will be labeled by an element in
%$\cH^{3}(Z_{n_1},\R/\Z)$ which is another SPT invariant.
To summarize,
\topinv{consider a 3+1D $Z_{n_1}\times Z_{n_2}$ SPT state labeled by $m_1,m_3$ on
a space with topology $M_2\times S_1$.  Adding the minimal $Z_{n_1}$-flux
through $S_1$ will reduce the 3+1D $Z_{n_1}\times Z_{n_2}$ SPT
state to a 2+1D $Z_{n_2}$ SPT state on $M_2$ labeled
by $m_1+\<n_1,n_2\>\times \text{integer}$  in $\cH^{3}(Z_{n_2},\R/\Z)$.  By
symmetry, adding the minimal $Z_{n_2}$-flux through $S_1$  will reduce the 3+1D
$Z_{n_1}\times Z_{n_2}$ SPT state to a 2+1D $Z_{n_1}$
SPT state on $M_2$ labeled by $m_3+\<n_1,n_2\>\times \text{integer}$  in
$\cH^{3}(Z_{n_1},\R/\Z)$.}

Just like the bosonic $U^{SG}(1)\times U^{GG}(1)$ SPT states can be
characterized by the $U^{SG}(1)\times U^{GG}(1)$ Chern-Simons topological term
(see \eqn{CSy0y2y3}) after we gauge the global symmetry $U^{SG}(1)\times
U^{GG}(1)$, the bosonic $Z_{n_1}\times Z_{n_2}$ SPT states can also be
characterized by a $Z_{n_1}\times Z_{n_2}$ gauge topological term. If we
gauge the  global $Z_{n_1}\times Z_{n_2}$ symmetry and integrating out the
matter fields, we will get a $Z_{n_1}\times Z_{n_2}$ gauge topological term in
3+1D:
\begin{align}
 \cL_\text{top}=
2\pi m_1 \om^{Z_{n_1},Z_{n_2}}_{1,3}
+2\pi m_3 \om^{Z_{n_1},Z_{n_2}}_{3,1}
\end{align}
where
$\om^{Z_{n_1},Z_{n_2}}_{1,3} \in \cH^1[Z_{n_1},\cH^{3}(Z_{n_2},\R/\Z)]$ and
$\om^{Z_{n_1},Z_{n_2}}_{3,1} \in \cH^3[Z_{n_1},\cH^{1}(Z_{n_2},\R/\Z)]$.
%Compare to \eqn{F2top}, the above can be viewed as
%discrete $F^2$-type topological term for
%$Z_{n_1}\times Z_{n_2}$ gauge fields.

\subsection{2+1D Bosonic $U(1)\times Z_2$ SPT phases}

In this section, we like to consider SPT states with symmetry $G=U(1)\times
Z_2$ in 2+1 dimensions.  The $U(1)\times Z_2$ SPT states are described by
$\cH^3(G,\R/\Z)$, which has the following decomposition (see \eqn{kunnU})
\begin{align}
& \cH^3(G,\R/\Z) =
\oplus_{k=0}^{3} \cH^{k}[Z_2,\cH^{3-k}(U(1),\R/\Z)]
\nonumber\\ &
=
\cH^3(U(1),\R/\Z) \oplus \cH^2[Z_2,\cH^{1}(U(1),\R/\Z)]
\nonumber\\ &\ \ \ \ \
 \oplus
\cH^3(Z_2,\R/\Z) ,
\end{align}
 with
\begin{align}
  \cH^3(U(1),\R/\Z) &=\Z=\{m_0\},
\nonumber\\
\cH^2[Z_2,\cH^{1}(U(1),\R/\Z)]  &=
\nonumber\\
\cH^1(Z_2,\RZ)\otimes_\Z \cH^{1}(U(1),\R/\Z)  &=\Z_2=\{m_2\},
\nonumber\\
\cH^3(Z_2,\R/\Z) &=\Z_2=\{m_3\}.
\end{align}
$m_0$ labels different 2+1D $U(1)$ SPT states and $m_3$ labels different 2+1D
$Z_2$ SPT states, whose measurement were discussed before.  \topinv{To measure
$m_2\in \cH^1(Z_2,\RZ)\otimes_\Z \cH^{1}(U(1),\R/\Z)$, we may create two identical $Z_2$ monodromy defects on a closed 2D
space.  We then measure the induced $U(1)$-charge  mod 2, which measures
$m_2$.} This result can be obtained by viewing the $U(1)\times Z_2$ SPT states
as $Z_2\times Z_2$ SPT states and use the result in section \ref{bZ2Z23d}.

%Thus the bosonic $U(1)\times Z_2$ SPT phases is described by $\Z\oplus \Z_2
%\oplus \Z_2$ in 2+1D.

If we gauge the  global $U(1)\times Z_2$ symmetry and integrating out
the matter fields, we will get a $U(1)\times Z_2$ gauge topological
term in 2+1D:
\begin{align}
 \cL_\text{top}=
\frac{m_0}{2\pi} AF
+2\pi m_2 \om^{Z_2,U(1)}_{2,1}
+2\pi m_3 \om^{Z_2}_{3}
\end{align}
where $\om^{Z_2,U(1)}_{2,1} \in \cH^2[Z_2,\cH^1(U(1),\R/\Z)]$ and
$\om^{Z_2}_{3} \in \cH^3[Z_2,\R/\Z]$.
Also $A$ and $F$ are the gauge potential one-form
and the field strength two-form for the $U(1)$-gauge field.
We can further rewrite the above
$U(1)\times Z_2$ gauge topological term as
\begin{align}
 \cL_\text{top}=
\frac{m_0}{2\pi} AF
+ m_2 \Om^{Z_2}_2 A
+2\pi m_3 \om^{Z_2}_{3}
\end{align}
where $\Om^{Z_2}_2 \in  \cH^2(Z_2,\Z)$ which is viewed a discrete differential
two-form (see appendix \ref{dform}). $\Om^{Z_2}_2 A = \Om^{Z_2}_2 \wedge A$ is
the wedge product of the differential forms.

%In the following, we like to show that it is valid to write $
%\om^{Z_2,U(1)}_{2,1}$ as $\Om^{Z_2}_2 \wedge A$.

\subsection{Bosonic $U(1)\times Z_2^T$ SPT phases}

In this section, we are going to consider bosonic $U(1)\times Z_2^T$ SPT
phases.  The $U(1)\times Z_2$ SPT phases can be realized by time reversal
symmetric spin systems where the spin rotation symmetry is partially broken.

\subsubsection{1+1D}
\label{U1Z2T2d}

We first consider SPT states with symmetry $G=U(1)\times Z_2^T$ in 1+1
dimensions, where $Z_2^T$ is the anti-unitary time reversal symmetry.  The
$U(1)\times Z_2^T$ SPT states are described by $\cH^2(G,\R/\Z)$, which has the
following decomposition (see \eqn{kunnU})
\begin{align}
& \cH^2(G,\R/\Z) =
\oplus_{k=0}^{2} \cH^{k}[U(1),\cH^{2-k}(Z_2^T,(\R/\Z)_T)]
\nonumber\\ &
=
\cH^2(Z_2^T,(\R/\Z)_T) \oplus \cH^2(U(1),\Z_2) ,
\end{align}
 with
\begin{align}
&\ \ \ \
  \cH^2(Z_2^T,(\R/\Z)_T) =\Z_2=\{m_0\},
\nonumber\\
&\ \ \ \ \cH^2(U(1),\Z_2) 
\\
&=
\cH^1[U(1),\RZ]\otimes_\Z \cH^0[Z_2^T,(\RZ)_T] =\Z_2=\{m_2\}.
\nonumber 
\end{align}
$m_0$ labels different 1+1D $Z_2^T$ SPT states and $m_2$ labels different 1+1D
$U(1)$ SPT states whose action amplitudes are real numbers (\ie $\pm 1$).  To
measure $m_k$, we put the system on a finite line $I_1$.  At an end of the
line, we get degenerate states that form a projective representation of
$U(1)\times Z_2^T$, which is classified by $\cH^2[U(1)\times
Z_2^T,\R/\Z]$.\cite{CGW1107,SPC1139,CGW1128}
We find that \topinv{ a 1+1D bosonic $U(1)\times Z_2^T$ SPT state labeled by
$(m_0,m_2)$ has a degenerate Kramer doublet at an open boundary if
$(m_0,m_2)=(1,0)$ or a degenerate doublet of $U(1)$ charge $\pm 1/2$ if
$(m_0,m_2)=(0,1)$.  The time
reversal transformation flips the sign of the $U(1)$-charge.}

Another way to probe $m_2$ is to gauge the $U(1)$ symmetry. The
$U(1)\times Z_2^T$ SPT states are described by the following gauge
topological term (induced by integrating out the matter fields)
\begin{align}
\label{Ftop2d1}
 \cL_\text{top}=\frac{m_2}{2} F
\end{align}
where $F$ is the field strength two form for the $U(1)$-gauge field.  Under
$Z_2^T$ transformation,
\begin{align}
 A_0\to -A_0, \ \ \
 A_i \to A_i, \ \ \ F\to -F.
\end{align}
(Note that  under $Z_2^T$, the $U(1)$-charge changes sign.) Since $\int_{M_2}
\frac{m_2}{2} F =m_2 \pi\times$ integers, on any closed 1+1D space-time
manifold $M_2$, the $Z_2^T$ symmetry requires $m_2$ to be quantized as an
integer.

If the space-time $M_2$ has a boundary, the above topological term naively
reduce to an effective Lagrangian on the boundary
\begin{align}
 \cL_{0+1D}=\frac{m_2}{2} A
\end{align}
where $A$ is the gauge potential one form.  This is nothing but a 1D $U(1)$
Chern-Simons term with a \emph{fractional} coefficient.  But such a  1D $U(1)$
Chern-Simons term breaks the $Z_2^T$ symmetry, since $A_0\to -A_0$ under the
time reversal transformation. So only if the $Z_2^T$ symmetry is broken at the
boundary, can the  topological term reduce to the above 1D Chern-Simons term on
the boundary. We find that \topinv{for a 1+1D $U(1)\times Z_2^T$ SPT state on a
open chain, if the $Z_2^T$ symmetry is broken at the boundary, the boundary
will carry a $U(1)$ charge $m_2/2$ or $-m_2/2$ mod 1.}

If the $Z_2$ symmetry is not broken, we have the following
effective boundary theory
\begin{align}
 \cL_{0+1D}=\frac{m_2 \si }{2} A +\cL(\si)
\end{align}
where the $\si(x)$ field only takes two values $\si=\pm 1$.  We see that if
$m_2=0$, the ground state of the 0+1D system is not degenerate
$|\text{ground}\>=|\si=1\>+|\si=-1\>$.  If $m_2=1$, the ground states of the
0+1D system is degenerate with $|\si=\pm 1\>$ states carrying fractional $\pm
1/2$ $U(1)$-charges. Such states form a projective representation
of $U(1)\times Z_2^T$.

\subsubsection{2+1D}

Next, we consider SPT states with symmetry $G=U(1)\times Z_2^T$ in 2+1
dimensions.  The $U(1)\times Z_2^T$ SPT states are described by
$\cH^3(G,\R/\Z)$, which has the decomposition (see \eqn{kunnU} and \eqn{U1Zn})
\begin{align}
& \cH^3(G,\R/\Z) =
\oplus_{k=0}^{3} \cH^{k}[U(1),\cH^{3-k}(Z_2^T,(\R/\Z)_T)]
\nonumber\\ &
= \cH^3(Z_2^T,(\R/\Z)_T) 
\oplus \cH^1[U(1), \cH^2(Z_2^T,(\R/\Z)_T)]
\\
&\ 
\oplus \cH^2[U(1), \cH^1(Z_2^T,(\R/\Z)_T)]
\oplus
\cH^3(U(1),\Z_2) = 0.
\nonumber 
\end{align}
Thus there is no non-trivial $U(1)\times Z_2^T$ SPT states in 2+1 dimensions.

\subsubsection{3+1D}

Now we consider $U(1)\times Z_2^T$ SPT states  in 3+1 dimensions,
which are described by $\cH^3(G,\R/\Z)$:
\begin{align}
& \cH^4(G,\R/\Z) =
\oplus_{k=0}^{4} \cH^{k}[U(1),\cH^{4-k}(Z_2^T,(\R/\Z)_T)]
\nonumber\\ &
=
\cH^4[Z_2^T,(\R/\Z)_T] \oplus \cH^2[U(1),\cH^{2}(Z_2^T,(\R/\Z)_T)]
\nonumber\\ &\ \ \ \ \
 \oplus \cH^4(U(1),\Z_2) ,
\end{align}
 with
\begin{align}
\cH^4(Z_2^T,(\R/\Z)_T) &=\Z_2=\{m_0\}.
\nonumber\\
\cH^2[U(1),\cH^{2}(Z_2^T,(\R/\Z)_T)] &=
\nonumber\\
\cH^1[U(1),\RZ]\otimes_\Z \cH^{2}[Z_2^T,(\R/\Z)_T] &=\Z_2=\{m_2\},
\nonumber\\
  \cH^4(U(1),\Z_2) &=
\nonumber\\
\cH^3[U(1),\RZ]\otimes_\Z \cH^0[Z_2^T,(\R/\Z)_T] &=\Z_2=\{m_4\}.
\end{align}
The elements in
$\cH^4(U(1)\times Z_2^T,(\R/\Z)_T)$ can also be labeled by a set of
$\{(m_0',m_1',m_2',m_3',m_4')\}$ (see appendix \ref{LHS}), where
\begin{align}
m_0' &\in \cH^0[Z_2^T,\cH^4[U(1),\RZ]_T) =\Z_1,
\nonumber\\
m_1' &\in \cH^1[Z_2^T,\cH^3[U(1),\RZ]_T) = \cH^1(Z_2^T,\Z_T) =\Z_2,
\nonumber \\
m_2' &\in \cH^2[Z_2^T,\cH^2[U(1),\RZ]_T) = \Z_1,
\\
m_3' &\in \cH^3[Z_2^T,\cH^1[U(1),\RZ]_T) = \cH^3(Z_2^T,\Z_T) =\Z_2,
\nonumber\\
m_4' &\in \cH^4[Z_2^T,\cH^0[U(1),\RZ]_T) = \cH^4(Z_2^T,(\R/\Z)_T) =\Z_2,
\nonumber
\end{align}
where $\cH^k[U(1),\RZ]_T$ means that $Z_2^T$ has a non-trivial action on
$\cH^k[U(1),\RZ]_T$.  Again, we see that $\cH^4(U(1)\times
Z_2^T,(\R/\Z)_T)=\Z_2^3$

$m_0$ labels different 3+1D $Z_2^T$ SPT states, and $m_4$ labels different
3+1D $U(1)$ SPT states whose action amplitudes are real numbers (\ie $\pm
1$).  To detect $m_2$, we consider a 3D space with topology
$M_2\times I_1$ where $M_2$ is closed 2D manifold.  We then put a $U(1)$-gauge
configuration that carries a unit of the $U(1)$-gauge flux $\int_{M_2}
\frac{F}{2\pi}=1$ on $M_2$.  In the large $I_1$ limit, we may view the system
as a 1+1D system on $I_1$ with the same $U(1)\times Z_2^T$ symmetry (note that
the $U(1)$ flux does not break the $Z_2^T$ time reversal symmetry).  The
resulting 1+1D $U(1)\times Z_2^T$ SPT state is classified by $\cH^2[U(1)\times
Z_2^T,\RZ]=\Z_2^2$ discussed in section \ref{U1Z2T2d}.  For such a set up,
a non-zero $m_2$ (and $m_0=m_4=0$) will give rise to a degenerate Kramer
doublet at each end of the line $I_1$ which carry no $U(1)$-charge.  We find
that \topinv{in a 3+1D bosonic  $U(1)\times Z_2^T$ SPT state labeled by
$(m_0,m_2,m_4)=(0,1,0)$, a $U(1)$ monopole of unit magnetic charge will carries
a $U(1)$-neutral degenerate Kramer doublet.}

From section \ref{U1Z2T2d}, we also know that the other kind of  1+1D
$U(1)\times Z_2^T$ SPT states is characterized by the degenerate doublet states
of $U(1)$-charge $\pm 1/2$ at each end of the line $I_1$.  One may wonder if a
non-zero $m_4$ (and $m_0=m_2=0$) will give rise to such a 1+1D $U(1)\times
Z_2^T$ SPT state on the line $I_1$?  In the following, we will argue that a
non-zero $m_4=1$ does not give rise to a non-trivial 1+1D $U(1)\times Z_2^T$ SPT
state.

As before, a way to probe $m_4$ is to gauge the $U(1)$ symmetry. We believe
that the $U(1)\times Z_2^T$ SPT states labeled by $(m_0,m_2,m_4)=(0,0,m_4)$ are
described by the following $U(1)$-gauge topological term
\begin{align}
\label{F2top}
 \cL_\text{top}=\frac{ m_4\pi }{(2\pi)^2} F^2
\end{align}
Under the $Z_2^T$ transformation, $F^2\to -F^2$ and $\ee^{\ii \int_{M_4}
\frac{m_4\pi}{(2\pi)^2} F^2}\to \ee^{-\ii \int_{M_4} \frac{m_4\pi}{2(2\pi)^2}
F^2}$. Because $\int_{M_4} \frac{m_4\pi}{(2\pi)^2} F^2 =\pi m_4 \times$
integers, on any closed 3+1D orientable space-time manifold $M_4$, the $Z_2^T$
symmetry is not broken due to the fact that $m_4$ is an integer. $m_4$=odd
describes the non-trivial  3+1D $U(1)\times Z_2^T$ SPT state, while $m_4$=even
describes the trivial SPT state.

If we put a $U(1)$-gauge configuration that carries a unit of the $U(1)$-gauge
flux $\int_{M_2} \frac{F}{2\pi}=1$ on $M_2$, the above 3+1D $U(1)$-gauge
topological term \eq{F2top} will reduce to a 1+1D $U(1)$-gauge topological
term:
\begin{align}
\label{Ftop2d}
 \cL_\text{top}=2\frac{ m_4\pi }{2\pi} F.
\end{align}
Compare to \eqn{Ftop2d1}, we see that even $m_4=1$ will
give rise to a trivial 1+1D
$U(1)\times Z_2^T$ SPT state.

To measure $m_4$, we need to use the statistical effect discussed in
\Ref{W7983,GMW8921,MKF1335}: \topinv{in a 3+1D bosonic  $U(1)\times Z_2^T$ SPT
state labeled by $(m_0,m_2,m_4)=(0,0,m_4)$, a dyon of the $U(1)$ gauge field
with ($U(1)$-charge, magnetic charge) = $(q,m)$ has a statistics
$(-)^{m(q-m_4)}$ (where $+ \to$ boson and $- \to$ fermion).}

If the space-time $M_4$ has a boundary, the topological term \eq{F2top}
reduces to an effective Lagrangian on the boundary
\begin{align}
 \cL_{2+1D}=\frac{m_4}{4\pi} AF ,
\end{align}
if  the $Z_2^T$ time-reversal symmetry is broken on the boundary.  The above is
nothing but a 2+1D $U(1)$ Chern-Simons term with a quantized Hall conductance
$\si_{xy}=m_4/2\pi$.  We note that if a 2+1D state with $U(1)$ symmetry has no
topological order,  a Hall conductance must be quantized as even integer
$\si_{xy}=\text{even}/2\pi$. Thus 
\topinv{in a 3+1D bosonic  $U(1)\times Z_2^T$ SPT
state labeled by $(m_0,m_2,m_4)$, 
the gapped time-reversal symmetry breaking boundary
has a Hall conductance $\si_{xy}=\frac{m_4}{2\pi}+\frac{\text{even}}{2\pi}$.
}

If the $Z_2^T$ symmetry is not broken, we
actually have the following effective boundary theory
\begin{align}
 \cL_{2+1D}=\frac{m_4 \si }{4\pi} AF +\cL(\si)
\end{align}
where the $\si(x)$ field only takes two values $\si=\pm 1$.  The gapless edge
states on the domain wall between $\si=1$ and $\si=-1$ regions may give rise to
the gapless boundary excitations on the 2+1D surface.

\subsection{Bosonic $Z_2\times Z_2^T$ SPT phases}

In this section, we are going to consider bosonic $Z_2\times Z_2^T$ SPT
phases.  The $Z_2\times Z_2^T$ SPT phases can be realized by time reversal
symmetric spin systems where the spin rotation symmetry is partially broken.

\subsubsection{1+1D}
\label{Z2Z2T2d}

We first consider SPT states with symmetry $G=Z_2\times Z_2^T$ in 1+1
dimensions.  The
$Z_2\times Z_2^T$ SPT states are described by $\cH^2(G,\R/\Z)$, which has the
following decomposition (see \eqn{kunnU})
\begin{align}
& \cH^2(G,\R/\Z) =
\oplus_{k=0}^{2} \cH^{k}[Z_2,\cH^{2-k}(Z_2^T,(\R/\Z)_T)]
\nonumber\\ &
=
\cH^2(Z_2^T,(\R/\Z)_T) \oplus \cH^2(Z_2,\Z_2) ,
\end{align}
 with
\begin{align}
  \cH^2(Z_2^T,(\R/\Z)_T)&= 
\cH^0(Z_2,\RZ)\boxtimes_\Z \cH^2(Z_2^T,(\R/\Z)_T) 
\nonumber\\
&=\Z_2=\{m_0\},
\nonumber\\
\cH^2(Z_2,\Z_2) &= 
\cH^1(Z_2,\RZ)\otimes_\Z \cH^0(Z_2^T,(\R/\Z)_T)
\nonumber\\
&=\Z_2=\{m_2\}.
\end{align}
$m_0$ labels different 1+1D $Z_2^T$ SPT states and $m_2$ labels different 1+1D
$Z_2$ SPT states whose action amplitudes are real numbers (\ie $\pm 1$).  To
measure $m_k$, we put the system on a finite line $I_1$.  At an end of the
line, we get degenerate states that form a projective representation of
$Z_2\times Z_2^T$, which is classified by $\cH^2[Z_2\times
Z_2^T,\R/\Z]$.\cite{CGW1107,SPC1139,CGW1128}
We find that \topinv{ a 1+1D bosonic $Z_2\times Z_2^T$ SPT state labeled by
$(m_0,m_2)$ has a degenerate Kramer doublet at an open boundary if
$(m_0,m_2)=(1,0)$ or a degenerate doublet of $Z_2$ charge $\pm 1/2$ mod 2 if
$(m_0,m_2)=(0,1)$.  The time
reversal transformation flips the sign of the $Z_2$-charge.}
The above result can also be understood by viewing the $U(1)\times Z_2^T$ SPT
phases discussed in the last section as $Z_2\times Z_2^T$ SPT phases.

\subsubsection{2+1D}

Next, we consider SPT states with symmetry $G=Z_2\times Z_2^T$ in 2+1
dimensions.  The $Z_2\times Z_2^T$ SPT states are described by
$\cH^3(G,\R/\Z)$, which has the decomposition (see \eqn{kunnU})
\begin{align}
& \cH^3(G,\R/\Z) =
\oplus_{k=0}^{3} \cH^{k}[U(1),\cH^{3-k}(Z_2^T,(\R/\Z)_T)]
\nonumber\\ &
= \cH^1[Z_2, \cH^2(Z_2^T,(\R/\Z)_T)] \oplus \cH^3[Z_2, \Z_2],
\end{align}
 with (see \eqn{ZmZn})
\begin{align}
\label{Z2Z2T3dm1m3}
&\ \ \ \  \cH^1(Z_2,\cH^2(Z_2^T,(\R/\Z)_T)) 
\nonumber\\
&=
  \cH^1(Z_2,\RZ)\boxtimes_\Z \cH^2[Z_2^T,(\R/\Z)_T]
=\Z_2=\{m_1\},
\nonumber\\
&\ \ \ \ \cH^3(Z_2,\Z_2) 
\\
&=
  \cH^3(Z_2,\RZ)\boxtimes_\Z \cH^0[Z_2^T,(\R/\Z)_T]
=\Z_2=\{m_3\}.
\nonumber 
\end{align}
We note that $m_3$ actually describe the $Z_2$ SPT order in 2+1D.
Such a non-trivial $Z_2$ SPT order can survive even if we break the
time-reversal symmetry. We can use the fractional statistics
of the $Z_2$ monodromy defects to detect $m_3$ (see section \ref{ZnSta}).

To detect/measure $m_1$ we can use the results in SPT invariant
\ref{Zn1Zn22dTop}.  We first choose the space-time topology to be $S_1\times
M_2$ We next add a $Z_2$ flux through $S_1$ which is an element in
$\cH^1(Z_2,\RZ)$. We then measure it response by measuring the induced $Z_2^T$
SPT state on $M_2$ which is an element in $ \cH^2[Z_2^T,(\R/\Z)_T]$.  Thus
\topinv{consider a 2+1D $Z_2\times Z_2^T$ SPT state.  The  $Z_2$ monodromy
defect will carry a degenerate Kramer doublet if $m_1=1$ and no Kramer doublet
if $m_1=0$.  }

\subsubsection{3+1D}

Now we consider $Z_2\times Z_2^T$ SPT states  in 3+1 dimensions,
which are described by $\cH^3(G,\R/\Z)$:
\begin{align}
& \cH^4(G,\R/\Z) =
\oplus_{k=0}^{4} \cH^{k}[Z_2,\cH^{4-k}(Z_2^T,(\R/\Z)_T)]
\nonumber\\ &
=
\cH^4[Z_2^T,(\R/\Z)_T] \oplus \cH^2[Z_2,\cH^{2}(Z_2^T,(\R/\Z)_T)]
\nonumber\\ &\ \ \ \ \
 \oplus \cH^4(Z_2,\Z_2) ,
\end{align}
 with
\begin{align}
\cH^4(Z_2^T,(\R/\Z)_T) &=\Z_2=\{m_0\}.
\nonumber\\
\cH^2[Z_2,\cH^{2}(Z_2^T,(\R/\Z)_T)]
  &=\Z_2=\{m_2\},
\nonumber\\
  \cH^4(Z_2,\Z_2) &=\Z_2=\{m_4\},
\end{align}
$m_0$ labels different 3+1D $Z_2^T$ SPT states, and $m_4$ labels different
3+1D $Z_2$ SPT states whose action amplitudes are real numbers (\ie $\pm
1$).  

To detect $m_2$, we consider a 3D space with topology $M_2\times I_1$ where
$M_2$ is closed 2D manifold and $I_1$ is an 1D segment.  In the large $I_1$
limit, we may view the system as 1+1D gapped state with $Z_2^T$ symmetry. Let
us assume that the end of $I_1$ does not carry degenerate Kramer doublet.  We
then put two \emph{identical} $Z_2$ monodromy defects on $M_2$.  In the large
$I_1$ limit, we may view the system as a 1+1D system on $I_1$ with the same
$Z_2\times Z_2^T$ symmetry (note that the  $Z_2$ monodromy defects do not break
the $Z_2^T$ time reversal symmetry).  The resulting 1+1D $U(1)\times Z_2^T$ SPT
state is classified by $\cH^2[Z_2\times Z_2^T,\RZ]=\Z_2^2$ discussed in section
\ref{Z2Z2T2d}.  For such a set up, a non-zero $m_2$ (and $m_0=m_4=0$) will give
rise to a degenerate Kramer doublet at each end of the line $I_1$ which carry
no $Z_2$-charge.  We find that \topinv{in a 3+1D bosonic  $Z_2\times Z_2^T$ SPT
state labeled by $(m_0,m_2,m_4)=(0,1,0)$,  two \emph{identical} $Z_2$ monodromy
defects on the surface of the sample will induce a $Z_2$-neutral degenerate
Kramer doublet.}

To detect $m_4$, let us view the $U(1)\times Z_2^T$ SPT state as a  $Z_2\times
Z_2^T$ SPT state, and assume that a $U(1)\times Z_2^T$ SPT state describe by
$m_4$ is a  $Z_2\times Z_2^T$ SPT state  describe by the same $m_4$.  (Note
that the $U(1)\times Z_2^T$ SPT states and the  $Z_2\times Z_2^T$ SPT states
are labeled by the same set of $m_i$ quantum numbers.) We have seen that if the
space-time $M_4$ has a boundary, the gapped boundary of 3+1D $U(1)\times Z^T_2$
SPT state can have a  quantized Hall conductance $\si_{xy}=m_4/2\pi$ if the
time-reversal symmetry is broken only at the boundary.  Let assume that we have
a fat $U(1)$ $\pi$-flux on the surface. Such a  fat $U(1)$ $\pi$-flux will
induce a $U(1)$ charge $m_4/2$.  Similarly, a  fat $U(1)$ $-\pi$-flux will
induce a $U(1)$ charge $-m_4/2$.  If we shrink the  fat $U(1)$ $\pm\pi$-flux to
a point, the $U(1)$ $\pm\pi$-flux will become the same  $Z_2$ monodromy defect,
which will have degenerate states with $U(1)$ charge $\pm m_4/2$. 
\topinv{Consider a gapped surface of $Z_2\times Z_2^T$ SPT state that break the
$Z_2^T$ symmetry.  If $m_4=1$, then a $Z_2$ monodromy line-defect that end on
the surface will have two degenerate states whose $Z_2$ charge differ by 1, or
the $Z_2$ monodromy line-defect in the bulk is gapless.}

We also know that for a $U(1)\times Z_2^T$ SPT state, a gapped time-reversal
symmetry breaking boundary has a Hall conductance $\si_{xy} = 1/2\pi$.  As a
result, the symmetry breaking domain wall will carry the edge excitations of
the 2+1D $U(1)$ STP state with a Hall conductance $\si_{xy} = 1/\pi$. We can
view the $U(1)\times Z_2^T$ SPT state as a $Z_2 \times Z_2^T$ SPT state, and
the edge excitations of the 2+1D $U(1)$ STP state as the edge excitations of
the 2+1D $Z_2$ STP state. This way, we find that \topinv{Consider a gapped
surface of $Z_2 \times Z_2^T$ SPT state that break the $Z_2^T$ symmetry. If
$m_4 = 1$, then the symmetry breaking domain wall will carry the edge
excitations of the 2+1D $Z_2$ STP state.  }

\subsection{2+1D fermionic $U(1)\times Z_2^f$ SPT phases}
\label{U1Z2f}

%\subsubsection{2+1D}

The fermionic $U(1)\times Z_2^f$ SPT phases can be realized by systems with two
types of fermions, one carry the $U(1)$-charge and the other is neutral.  To
construct the SPT invariants for the fermionic $U(1)\times Z_2^f$ SPT
states, we again ``gauge'' the $U(1)\times Z^f_2$ symmetry, and then put the
fermion system on a 2D closed space $M_2$ with a $U(1) \times Z^f_2$ gauge
configuration that carries a unit of the $U(1)$-gauge flux $\int_{M_2}
\frac{F}{2\pi}=1$.  We then measure the $U(1)$-charge $c_{11}$ and the
$Z^f_2$-charge $c_{12}$ of the ground state on $M_2$ induced by the $U(1)$
gauge flux.  Next, we put another $U(1) \times Z^f_2$ gauge configuration on
$M_2$ with no $U(1)$ flux but two \emph{identical} $\Z_2^f$ vortices, then
measure the $U(1)$-charge $c_{21}$ (mod 2) and the $Z^f_2$-charge $c_{22}$.  So
an integer matrix $\v C$ formed by $c_{ij}$
\begin{align}
 \v C=
\begin{pmatrix}
 c_{11} & c_{12}\text{ mod } 2 \\
 c_{21}\text{ mod } 2  &  c_{22}\text{ mod } 2\\
\end{pmatrix}
\end{align}
is a potential SPT invariant for fermionic $U(1)\times Z^f_2$ SPT phases
in $3$-dimensional space-time.

But which SPT invariants can be realized? What are the actual
SPT invariants?  One way to realize the fermionic $U(1)\times Z_2^f$
SPT phases is to view them as the fermionic $U(1)\times U^f(1)$ SPT phases discuss
in section \ref{U1Uf1}.  Using the $U(1)\times U(1)$ Chern-Simons theory for
the fermionic $U(1)\times U^f(1)$ SPT phases, we see that the following
SPT invariant
\begin{align}
 \v C_1=
\begin{pmatrix}
 1 & 1\\
 1 & 0\\
\end{pmatrix}
\end{align}
can be realized.

By binding the $U(1)$-charged fermion and neutral fermion to form a $U(1)$
charged boson, we can form other fermionic $U(1)\times Z_2^f$ SPT phases
through the bosonic $U(1)$  SPT phases of the above bosonic bound states.  This
allows us to realize the following  SPT invariant
\begin{align}
 \v C'_1=
\begin{pmatrix}
 2 & 0\\
 0 & 0\\
\end{pmatrix}
\end{align}
which is twice of $\v C_1$.  This suggests that the realizable SPT
invariants are $\v C_1 \times$integers.

%We may also assume that the fermionic $U(1)\times Z_2^f$ SPT phases are at
%least described by $m_k \in \cH^k[U(1), fSPT^{3-k}_{Z^f_2}]$ $k=0,1,2$, and
%$m_3 \in bSPT^3_{U(1)}$.  ($m_3 \in bSPT^3_{U(1)}$ because $U(1)$ does not
%contain $Z_2^f$ and is a bosonic symmetry for the fermion bound states
%discussed above.)  Using $fSPT^{1}_{Z^f_2}=Z_2$ and $fSPT^{k}_{Z^f_2}=0$ for
%$k>1$, we have
%\begin{align}
%& m_0 = 0, \ \ \ m_1 = 0,
%\nonumber\\
%& m_2 \in  \cH^2[U(1) fSPT^{1}_{Z^f_2}] =\cH^2[U(1),\Z_2]=\Z_2
%\nonumber\\
%& m_3 \in  bSPT^{3}_{U(1)} =\cH^3[U(1),\RZ]=\Z.
%\end{align}
%$m_2$ and $m_3$ can be probed by putting a $U(1) \times Z^f_2$ gauge configuration that
%carries a unit of the $U(1)$ gauge flux $\int_{M_2} \frac{F}{2\pi}=1$ on a
%closed 2D  space, and then measure the induced $U(1)$ charges and the fermion
%numbers (\ie the $Z_2^f$-charges).  $(m_2,m_3)=(1,1)$ may lead to the
%SPT invariant $C_1$ discussed above.
%%, while $(m_2,m_3)=(0,1)$
%%corresponds to the SPT invariant $C_1'$.

To summarize, some of the fermionic $U(1)\times Z_2^f$ SPT phases are
described by $\Z$ in $3$-dimensional space-time, whose SPT invariant
is $\v C_1$ times an integer.  It is not clear if those are all the  fermionic
$U(1)\times Z_2^f$ SPT phases.  The integer $\Z$ that label the fermionic
$U(1)\times Z_2^f$ SPT phases correspond to the integer Hall conductance.  This
result should to contrasted with the result for the fermionic $U^f(1)$ SPT
phases discussed in section \ref{Uf1}, where the  Hall conductance is quantized
as 8 times integer.

%\subsubsection{3+1D}
%
%Let us assume that the fermionic $U(1)\times Z_2^f$ SPT phases in 3+1D are
%described by $m_k \in \cH^k[U(1), fSPT^{4-k}_{Z^f_2}]$ $k=0,1,2,3$, and $m_4
%\in bSPT^4_{U(1)}$ (since $U(1)$ does not contain $Z_2^f$ and is a bosonic
%symmetry for the fermion bound states discussed above).
% Using
%$fSPT^{1}_{Z^f_2}=Z_2$ and $fSPT^{k}_{Z^f_2}=0$ for $k>1$, we have
%\begin{align}
%& m_0 = 0, \ \ \ m_1 = 0,\ \ \ m_2=0,
%\nonumber\\
%& m_3 \in  \cH^3[U(1), fSPT^{1}_{Z^f_2}] =\cH^3[U(1),\Z_2]=0
%\nonumber\\
%& m_4 \in  bSPT^{4}_{U(1)} =\cH^4[U(1),\RZ]=0.
%\end{align}
%This suggests that the  fermionic $U(1)\times Z_2^f$ SPT phases in 3+1D are
%always trivial.

\subsection{2+1D fermionic $Z_2\times Z^f_2$ SPT states}

Now, let us consider fermionic SPT states with full symmetry $Z_2\times Z^f_2$
in 2+1 dimensions.  This kind of fermionic SPT states were studied in
\Ref{GW1248} using group super-cohomology theory where four fermionic
$Z_2\times Z^f_2$ SPT states (including the trivial one) were constructed.
They were also studied in \Ref{RZ1232,GL1369} where 8 SPT states were obtained
(see also \Ref{Q1283,YR1205}).  To construct SPT invariants for the
fermionic $Z_2\times Z^f_2$ SPT states, we may create two identical $Z_2$
monodromy defects on a closed 2D space. We then measure the induced
$Z_2$-charge $c_{11}$ and the $Z_2^f$-charge $c_{12}$.  We then  create two
identical $Z_2^f$ monodromy defects, and measure the induced $Z_2$-charge
$c_{21}$ and the $Z_2^f$-charge $c_{22}$.  Note that $c_{ij}=c_{ji}=0,1$.  Thus
there are 8 potential different SPT invariants described by 2 by 2
symmetric integer matrix
\begin{align}
 \v C=\begin{pmatrix}
 c_{11} & c_{21}\\
 c_{12} & c_{22}\\
\end{pmatrix} \text{ mod 2}.
\end{align}

More general SPT invariants can be obtained from the statistics of the
monodromy defects.  Let $\th_{11}$ mod $\pi$ be the statistic angle of the
$Z_2$ monodromy defect and $\th_{22}$ mod $2\pi$ be the statistic angle of the
$Z_2^f$ monodromy defect.  Note that adding a $Z_2$ neutral fermion to a  $Z_2$
monodromy defect will change its statistical angle by $\pi$. So $\th_{11}$ is
only well defined mod $\pi$.  Adding a fermion to a  $Z_2^f$ monodromy defect
will not change its statistic since a fermion always carries a non-trivial
$Z_2^f$ charge. So $\th_{22}$ is well defined mod $2\pi$. Also Moving a
$Z_2$-monodromy defect around a $Z_2^f$ monodromy defect gives us a mutual
statistics angle $\th_{12}$ mod $\pi$.  Note that  adding a  fermion to a
$Z_2$ monodromy defect will change the mutual statistics angle $\th_{12}$ by
$\pi$, and thus $\th_{12}$ is well defined mod $\pi$.
So the  statistic of the monodromy defects give us the following
SPT invariants
\begin{align}
\v\Th=\begin{pmatrix}
\th_{11} \text{ mod } \pi & \th_{12}\text{ mod } \pi\\
\th_{12} \text{ mod } \pi& \th_{22}\text{ mod } 2\pi\\
\end{pmatrix}
\end{align}
%We note if we stack two SPT states with SPT invariants $(\v C, \v\Th)$
%and $(\v C', \v\Th')$, we obtain a new SPT state with SPT invariants
%\begin{align}
%(\v C'', \v\Th'')
%=(\v C, \v\Th)+(\v C', \v\Th')
%.
%\end{align}

But which values of the above SPT invariants can be realized by actual
fermion systems?  We may view the 2+1D fermionic $U(1)\times U^f(1)$ SPT states
discussed in section \ref{U1Uf1} as  fermionic $Z_2\times Z^f_2$ SPT states.
The different  $U(1)\times U^f(1)$ SPT states can be obtained by stacking a
fermion system where the $Z_2$-charged fermions form a $\nu=1$ integer quantum
Hall state and the $Z_2$-neutral fermions form a $\nu=-1$ integer quantum Hall
state.  Such a $(\nu=1)/(\nu=-1)$ double integer quantum Hall state can realize the
SPT invariants
\begin{align}
&\v C_1=\begin{pmatrix}
 1 & 1\\
 1 & 0\\
\end{pmatrix} \text{ mod 2},
\\
&
\v\Th_1=\begin{pmatrix}
\th_{11} & \th_{12}\\
\th_{12} & \th_{22}\\
\end{pmatrix}
=
\begin{pmatrix}
 \pi/4 \text{ mod } \pi& \pi/2\text{ mod } \pi\\
 \pi/2 \text{ mod } \pi& 0\text{ mod } 2\pi\\
\end{pmatrix} .
\nonumber
\end{align}
This because a monodromy defect of $Z_2$ in the  $(\nu=1)/(\nu=-1)$ double integer
quantum Hall state carries a $(Z_2,Z_2^f)$-charge $(1/2,1/2)+$integer  and a
statistics $\th_{11}=\pi/4$ mod $\pi$, while  a monodromy defect of $Z_2^f$ in
the $(\nu=1)/(\nu=-1)$ double integer quantum Hall state carries a
$(Z_2,Z_2^f)$-charge $(1/2,0)+$integer  and a statistics $\th_{22}=0$.  Also,
moving a $Z_2$-monodromy defect around a $Z_2^f$ monodromy defect gives us a
mutual statistics $\th_{12}=\pi/2$ mod $2\pi$.

If we assume that the fermions form bound states, we will get a bosonic system
with $Z_2$ symmetry.  Such  a bosonic system can realize  a SPT
invariant
\begin{align}
&\v C_2=\begin{pmatrix}
 2 & 0\\
 0 & 0\\
\end{pmatrix} \text{ mod 2},
\\
&
\v\Th_2=\begin{pmatrix}
\th_{11} & \th_{12}\\
\th_{12} & \th_{22}\\
\end{pmatrix}
=
\begin{pmatrix}
 \pi/2 \text{ mod } \pi & 0 \text{ mod } \pi\\
 0 \text{ mod } \pi & 0\text{ mod } 2\pi\\
\end{pmatrix} .
\nonumber
\end{align}
The calculation of $\v C_2$ was discussed in section \ref{bZ2} and the
calculation $\th_{11}$ was given by \eqn{thMZn}.  The other entries of
$\v\Th_2$ are obtained by noting the the $Z_2^f$ monodromy defect is trivial
since the $Z_2^f$ symmetry acts trivially.  We note that $(2\v C_1,
2\v\Th_1) = (\v C_2, \v\Th_2)$.  So it is possible that
the bosonic $Z_2$ SPT state is the same SPT state obtained by
stacking two $(\nu=1)/(\nu=-1)$ double integer quantum Hall states.

As we have mentioned that the SPT invariant $(\v C_1, \v\Th_1)$ is
realized by a fermion system where the $Z_2$-charged fermions form a $\nu=1$
integer quantum Hall state and the $Z_2$-neutral fermions form a $\nu=-1$
integer quantum Hall state.  We can have a new  SPT invariant which is
realized by a fermion system where the $Z_2$-charged fermions form a $p+\ii p$
superconducting state and the $Z_2$-neutral fermions form a $p-\ii p$
superconducting state.\cite{RG0067,I0168}  We note that the $Z_2$
monodromy defects in the $(p+\ii p)/(p-\ii p)$ superconducting state will have
non-Abelian statistics.\cite{RG0067} We can not simply use $\v\Th_1/2$ to
described their statistics.  We also note that two $Z_2$ monodromy defects
in the $(p+\ii p)/(p-\ii p)$ superconducting state have topological
degeneracy,\cite{RG0067,I0168} where the two degenerate states carry different
$Z_2$ and $Z_2^f$ quantum numbers.  We can not simply use $\v C_1/2$ to
describe the induced $Z_2$ and  $Z_2^f$ charges either.

Stacking four $(\nu=1)/(\nu=-1)$ double integer quantum Hall states (or eight
$(p+\ii p)/(p-\ii p)$ superconducting states) will give us a trivial fermionic
$Z_2\times Z^f_2$ SPT state since $(4\v C_1, 4\v\Th_1)$ is trivial.  This
agrees with the result obtained in \Ref{RG0067}.

Let us examine the assumption that the fermionic $Z_2\times Z_2^f$ SPT phases
are described by $m_k \in \cH^k[Z_2, fSPT^{3-k}_{Z^f_2}]$ $k=0,1,2$, and $m_3
\in bSPT^3_{Z_2}$ (note that $Z_2$ does not contain $Z_2^f$ and is a
symmetry for the bosonic two-fermion bound states discussed above).  Using
$fSPT^{1}_{Z^f_2}=Z_2$ and $fSPT^{k}_{Z^f_2}=0$ for $k>1$, we have
\begin{align}
& m_0 = 0, \ \ \ m_1 = 0,
\nonumber\\
& m_2 \in  \cH^2[Z_2, fSPT^{1}_{Z^f_2}] =\cH^2[Z_2,\Z_2]=\Z_2
\nonumber\\
& m_3 \in  bSPT^{3}_{Z_2} =\cH^3[Z_2,\RZ]=\Z_2.
\end{align}
%$m_2$ can be measured by putting two identical $Z_2$-monodromy defects on on a
%closed 2D space, and then measure the induced fermion numbers (\ie the $Z_2^f$
%charges).  The possible induced fermion numbers are $0$ and $1$, but there is
%another possibility where there are two degenerate ground states: one with no
%fermion and the other with one fermion.  Let us denote the later possibility as
%$m_2=1/2$.  We see that $(m_2,m_3)=(1,0)$ corresponds to the SPT
%invariant $C_1$ discussed above, $(m_2,m_3)=(1/2,0)$ corresponds to the
%SPT invariant $C_1/2$, and $(m_2,m_3)=(0,1)$ corresponds to the
%SPT invariant $C_2$.  So the assumption that $m_2 \in  \cH^2[Z_2,
%fSPT^{1}_{Z^f_2}]$ is not correct. It should be generalized to $m_2 \in
%\cH^2[Z_2, fSPT^{1}_{Z^f_2}] + $extra.
The above only give us 4 different SPT states.  So not all fermionic $Z_2\times
Z_2^f$ SPT phases can be described by $m_k \in \cH^k[Z_2, fSPT^{3-k}_{Z^f_2}]$
$k=0,1,2$, and $m_3 \in bSPT^3_{Z_2}$.

\section{Gapless boundary excitations or degenerate boundary states as
experimentally measurable SPT invariants}

In the above, we have discussed many SPT invariants for SPT states.
However, those SPT invariants are designed for numerical calculations
and can be probe by numerical calculations. They are hard to measure in real
experiments.  In this section, we like to argue that \topinv{a non-trivial SPT
state with symmetry $G$, must have gapless boundary excitations or degenerate
boundary states that transform non-trivially under the symmetry
transformations, even when the symmetry is not spontaneously broken at the
boundary.} Those low energy states can be probed by perturbations that break
the symmetry.

The above result is proven for 2+1D SPT states in \Ref{CLW1141} which has a
stronger form \topinv{a non-trivial 2+1D SPT state with symmetry $G$, must have
gapless boundary excitations that transform non-trivially under the symmetry
transformations, even if the symmetry is not spontaneously broken at the
boundary.} This is due to the fact there are no (intrinsic) topological orders
in 1+1D.  In the following, we will present some arguments for the above result
through a few simple examples, The new arguments are valid for higher
dimensions.

\begin{figure}[tb]
\begin{center}
\includegraphics[scale=0.3]{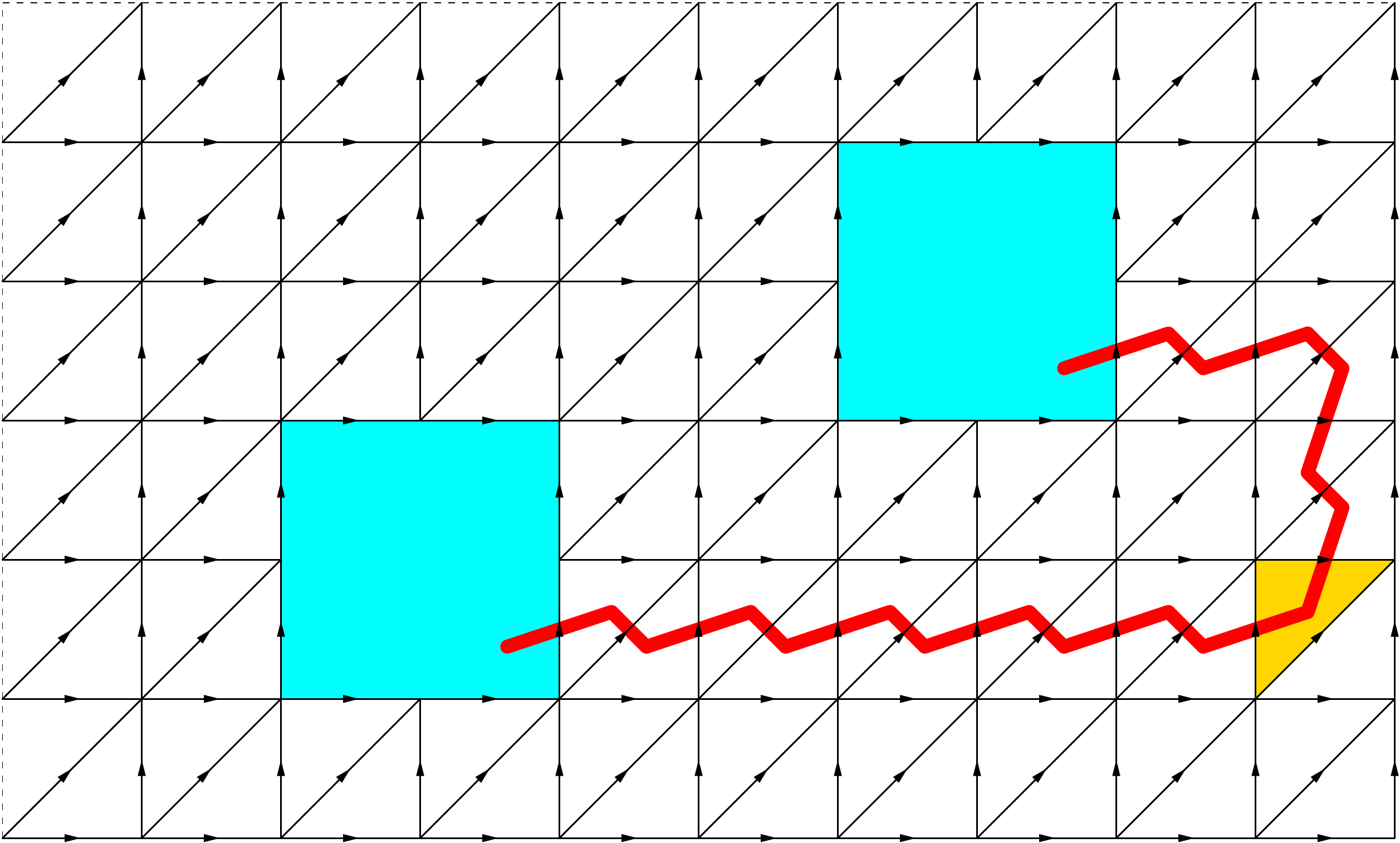} \end{center}
%Fig. 4
\caption{ (Color online)
A $Z_2$-gauge configuration with two \emph{identical} holes on a torus that
contains a $Z_2$-monodromy defect in each hole.  Such a $Z_2$-gauge
configuration has $U(-1)=-1$ (each yellow triangle contributes a factor $-1$)
(see Fig. \ref{z2gauge}).
}
\label{z2gaugeH}
\end{figure}

\subsection{Bosonic $Z_n$ SPT state in 2+1D}

We have shown that, in a non-trivial 2+1D $Z_n$ SPT state labeled by $m \in
\cH^3(Z_n,\RZ),\ \ m\neq 0$, $n$ \emph{identical} $Z_n$  monodromy defects will
carry a total $Z_n$-charge $2m$ mod $n$ (see section \ref{Zn3D}).  We may
realize the $n$ identical $Z_n$  monodromy defects through $n$ large holes in
the 2D space (see Fig. \ref{z2gaugeH}).  Let us assume that the $Z_n$ symmetry
is not spontaneously broken at the edge of the holes. Then depending on if a
hole contains a $Z_n$-monodromy defect or not, the $Z_n$-charge of the hole
will be $2m/n$ or $0$.

In the large hole limit, adding a  monodromy defect to a hole correspond to
twisting the boundary condition as we go around the edge of the hole.  Such a
twist of boundary condition costs zero energy in the large hole limit,  (since
the branch cut of a  monodromy defect costs no energy).  If twisting the
boundary condition around the edge change the $Z_n$-charge on the edge by
$2m/n$, then we will change the $Z_n$-charge on the edge by $2m$ if we make $n$
\emph{identical} twists of the boundary condition around the edge.  Since
twists cost zero energy and $n$ twists are equivalent to no twist, this way, we
show that \topinv{the edge of 2+1D $Z_n$ SPT state labeled by $m \in
\cH^3(Z_n,\RZ)$ contains nearly degenerate ground states that carry different
$Z_n$-charges (by $2m$) in the large edge limit.} According to the above
result, when $n$=odd, there will be (at least) $n$-fold degenerate edge states,
and when $n$=even, there will be (at least) $n/2$-fold degenerate edge states.

So the edge states of the holes must be gapless or degenerate, at least when
$n>2$.  Also the gapless low energy excitations or the degenerate states must
transform non-trivially under the the $Z_n$ symmetry transformations.  In
\Ref{LG1220,L1355}, using the non-trivial statistics of the monodromy defects,
one can argue more generally that edge states of the holes must be gapless or
degenerate even for $n=2$ case.

\subsection{Bosonic $Z_n$ SPT state in 4+1D}

Next, we consider bosonic $Z_n$ SPT state in 4+1D, labeled by $m \in
\cH^5(Z_n,\RZ),\ \ m\neq 0$. We assume the space to have a topology $M_2\times
M_2'$.  We have shown that, $n$ \emph{identical} $Z_n$  monodromy defects in
$M_2'$ will will induce a 2+1D $Z_n$ SPT state on $M_2$, labeled by $3m \in
\cH^3(Z_n,\RZ)$ (see section \ref{Zn5dsec}).  Again, we can realize the $n$
identical $Z_n$  monodromy defects through $n$ large holes on $M_2'$ and assume
that the $Z_n$ symmetry is not spontaneously broken at the edge of the holes.
Then depending on if each hole contains a $Z_n$-monodromy defect or not, the
2+1D $Z_n$ SPT state on $M_2$ will be labeled by $3m$ or $0$ in
$\cH^3(Z_n,\RZ)$.  We see that twisting the boundary condition around the edges
of the $n$ holes change the 2+1D $Z_n$ SPT state on $M_2$.  Since each twist
costs no energy in the large hole limit, the edge states of a hole must be
gapless or degenerate, at least when mod$(3m,n)\neq 0$.

\subsection{$U(1)$ SPT state in 2+1D and beyond}

\begin{figure}[tb]
\begin{center}
\includegraphics[scale=0.4]{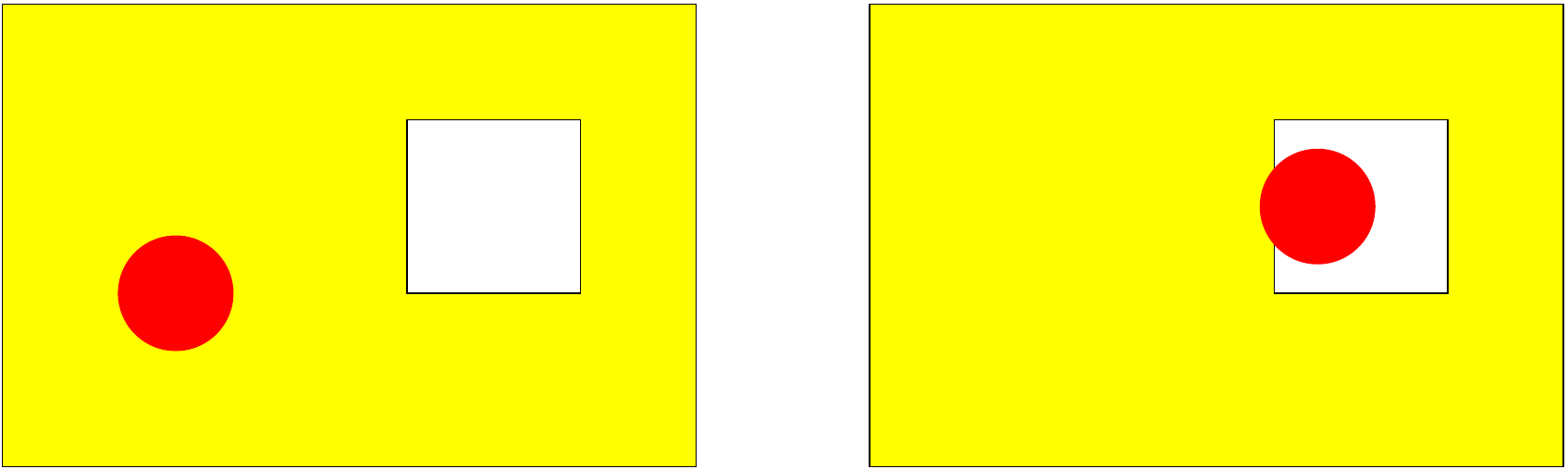} \end{center}
%Fig. 5
\caption{ (Color online) The circle represents $2\pi n$ flux which induce an
integer $U(1)$-charge.  As we move the  flux into the hole, the induced  $U(1)$
charge disappears.
}
\label{U1flux}
\end{figure}

We have discussed bosonic and fermionic  $U(1)$ SPT states in 2+1D.  Those
$U(1)$ SPT states are characterized by a non-zero Hall conductance.  In
\Ref{L8132,W9125}, it was shown that a non-zero  Hall conductance implies
gapless edge excitations.  Here we will review the argument.

We consider a 2D space with a hole and $2\pi n$ flux far away from the hole.
We assume that there is no $U(1)$ symmetry breaking.  The $2\pi n$ flux will
induce a non-zero charge $Q=n m$, $m \in \Z$.  As we move the  $2\pi n$ flux
into the hole, the  induced $U(1)$-charge will become the charge on the edge.
Since $2\pi n$ flux  in the hole do not change the boundary condition, the
induced $U(1)$-charge is an excitations of the edge.

If the $2\pi n$ flux is generate by a weak field, moving the  $2\pi n$ flux
into the hole represents a weak perturbation.  Since the weak perturbation
causes a finite change in the induced charge and also since there are infinite
many  weak perturbations cause infinite many different change in the induced
charges, the excitations on the edge of the hole is gapless.

We can also use a similar argument to show that \topinv{non-trivial bosonic and
fermionic  $U(1)$ SPT states have gapless boundary excitations in any
dimensions.}

\section{SPT invariants of SPT states with symmetry $G=GG \rtimes SG$}

In this section, we will discuss some examples of SPT states where the symmetry
group has a form $G=GG \times SG$.  

\subsection{Bosonic $U(1)\rtimes Z_2$ SPT phases}

Let us first consider bosonic $U(1)\rtimes Z_2$ SPT phases.
We note that $U(1)\rtimes Z_2$ is a subgroup of $SO(3)$.  So the $U(1)\rtimes
Z_2$ SPT phases can be realized by spin systems where the spin rotation
symmetry is partially broken.

\subsubsection{1+1D}

The SPT states with a non-Abelian symmetry $U(1)\rtimes Z_2$
in 1+1 dimensions  are described by
$\cH^2[U(1)\rtimes Z_2,\R/\Z]=\Z_2$, whose elements can be labeled by a subset
of $\{(m_0,m_1,m_2)\}$, according to the result in appendix \ref{LHS}:
\begin{align}
m_0 &\in \cH^2(Z_2,\R/\Z) =\Z_1,
\nonumber\\
m_1 &\in \cH^{1}(Z_2,\cH^1[U(1),\R/\Z]_{Z_2})
=\cH^1(Z_2,\Z_{Z_2}) =\Z_2,
\nonumber\\
m_2 &\in \cH^2(Z_2,\RZ) =\Z_1.
\end{align}
The second equation in the above is obtained by noting that the nonhomogenous
cocycle $\om_1(\th) \in \cH^1[U(1),\R/\Z] \cong \Z$ has a form $\om_1(\th)= m
\frac{\th}{2\pi}$, $m\in \Z$ [\ie $\ee^{\ii 2\pi \om_1(\th)}$ forms a 1D
representation of the $U(1)$].  Under the $Z_2$ transformation $g$,
$\om_1(\th)$ transforms as $\om_1(\th) \to \om_1(g\th g^{-1})=-\om_1(\th)$ or
$m\to -m$, since $g\th g^{-1} =-\th$.  Therefore, $Z_2$ has a non-trivial
action on $\cH^1[U(1),\R/\Z]=\Z$. We rewrite $\Z$ as $\Z_{Z_2}$ and
$\cH^1[U(1),\R/\Z])$ as $\cH^1[U(1),\R/\Z]_{Z_2}$ to indicate such a
non-trivial action.

Note that $\cH^1(Z_2,\Z)=\Z_1$ while $\cH^1(Z_2,\Z_{Z_2})=\Z_2$.  This is
because the cocycle condition for $\cH^1(Z_2,\Z_{Z_2})$ is
\begin{align}
 (\dd \om_1)(g_0,g_1)
&=g_0\cdot \om_1(g_1) - \om_1(g_0g_1) + \om_1(g_0)
=0,
\nonumber\\
 g_0,g_1 &\in Z_2=\{1,-1\}.
\end{align}
Using $g_0\cdot \om_1(g_1) = \pm \om_1(g_1)$ when $g_0=\pm 1$,
we can reduce the above to
\begin{align}
 \om_1(1)=0, \ \ \ -\om_1(-1)-\om_1(1)+ \om_1(-1)=0.
\end{align}
So the cocycles are given by
\begin{align}
 \om_1(1)=0,\ \ \ \om_1(-1)=\text{integer}.
\end{align}
The 1-coboundaries are given by
\begin{align}
 (\dd \om_0)(g_0)=g_0\cdot \om_0 - \om_0
\end{align}
or
\begin{align}
  (\dd \om_0)(1)=0,\ \ \
 (\dd \om_0)(-1)=\text{even integer}.
\end{align}
We see that $\cH^1(Z_2,\Z_{Z_2})=\Z_2$.

We also note that every elements in $\cH^2[U(1)\rtimes Z_2,\R/\Z]$ can  be
labeled by at least one $(m_0,m_1,m_2)$, but it is possible that not every
$(m_0,m_1,m_2)$ labels an element in  $\cH^2[U(1)\rtimes Z_2,\R/\Z]$.  In other
word, the two sets, $\{(m_0,m_1,m_2)\}$ and $\cH^2[U(1)\rtimes Z_2,\R/\Z]$, are
related by a sequence
\begin{align}
 \{(m_0,m_1,m_2)\} \to \cH^2[U(1)\rtimes Z_2,\R/\Z] \to 0.
\end{align}
In this particular case, since $\{m_0\}=\{m_2\}=\Z_1$, we know that
$\{m_1\}=\Z_2$ and $\cH^2[U(1)\rtimes Z_2,\R/\Z]$ has an one-to-one
correspondence.

To measure $m_1$, we put the system on a finite line $I_1$.  At an end of the
line, we get degenerate states that form a projective representation of
$U(1)\rtimes Z_2$,\cite{CGW1107,SPC1139,CGW1128} if $m_1\neq 0$.  If we view
$U(1)\rtimes Z_2$ as a subgroup of $SO(3)$, the projective representations of
$U(1)\rtimes Z_2$ are simply half-integer spin representations of $SO(3)$.

One way to understand such a result is to gauge the $U(1)\rtimes Z_2$ symmetry,
the $U(1)\rtimes Z_2$ SPT states are described are described by the following
gauge topological term (induced by integrating out the matter fields)
\begin{align}
 \cL_\text{top}=\frac{m_1}{2} F
\end{align}
where $F$ is the field strength two form for the $U(1)$-gauge field.  Under
$Z_2$ transformation, $F\to -F$.  Since $\int_{M_2} \frac{m_1}{2} F =m_1 \pi$
on any closed 1+1D space-time manifold $M_2$, $\cL_\text{top}$ respects the
$Z_2$ symmetry, since $m_1$ is an integer.

If the space-time $M_2$ has a boundary, the above topological term
naively reduce to an effective Lagrangian on the boundary
\begin{align}
 \cL_{0+1D}=\frac{m_1}{2} A
\end{align}
where $A$ is the gauge potential one form.  This is nothing but a 1D $U(1)$
Chern-Simons term with a \emph{fractional} coefficient.  But such a  1D $U(1)$
Chern-Simons term breaks the $Z_2$ symmetry. So only if the $Z_2$ symmetry is
broken at the boundary, can the  topological term reduce to the  1D
Chern-Simons term on the boundary. If the $Z_2$ symmetry is not broken, we have
the following effective boundary theory
\begin{align}
 \cL_{0+1D}=\frac{m_1 \si }{2} A +\cL(\si)
\end{align}
where the $\si(x)$ field only takes two values $\si=\pm 1$.  We see that if
$m_1=0$, the ground state of the 0+1D system is not degenerate
$|\text{ground}\>=|\si=1\>+|\si=-1\>$.  If $m_1=1$, the ground states of the
0+1D system is degenerate, which are described by $|\si=\pm 1\>$ states
carrying fractional $\pm 1/2$ $U(1)$-charges. Such states form a projective
representation of $U(1)\rtimes Z_2$.

We can also view the  $U(1)\rtimes Z_2$ SPT states as $Z_2\times Z_2$ SPT
states. Using the results in SPT invariant \ref{Zn1Zn22dTop}, we find that \topinv{Consider
a 1+1D bosonic $U(1)\rtimes Z_2$ SPT state labeled by $m_1 \in
\cH^2(U(1)\rtimes Z_2,\RZ)$. If we put the SPT state on a circle $S_1$, adding
$\pi$-flux of $U(1)$ through $S_1$ will induce a $Z_2$-charge $m_1$, and adding
$\pi$-flux of $Z_2$ through $S_1$ will induce a $U(1)$-charge $m_1+$even
integers in the ground state.  }

\subsubsection{2+1D}

The $U(1)\rtimes Z_2$ SPT states
in 2+1 dimensions  are described by
$\cH^3[U(1)\rtimes Z_2,\R/\Z]=\Z_2$, whose elements can be labeled by a subset
of $\{(m_0,m_1,m_2,m_3)\}$ (see appendix \ref{LHS}), where
\begin{align}
m_0 &\in \cH^3[U(1),\R/\Z] =\Z,
\nonumber\\
m_1 &\in \cH^{1}(Z_2,\cH^2[U(1),\R/\Z])
=\cH^1(Z_2,\Z_1) =\Z_1,
\nonumber\\
m_2 &\in \cH^{2}(Z_2,\cH^1[U(1),\R/\Z]_{Z_2})
=\cH^2(Z_2,\Z_{Z_2}) =\Z_1,
\nonumber\\
m_3 &\in \cH^2(Z_2,\RZ) =\Z_2.
\end{align}
We see that the 2+1D $U(1)\rtimes Z_2$ SPT states can be viewed as 2+1D $U(1)$
SPT states (described by $ \cH^3[U(1),\R/\Z] =\Z$) or 2+1D $Z_2$ SPT states
(described by $ \cH^3[Z_2,\R/\Z] =\Z_2$). Their SPT invariants have
been discussed before.

\subsubsection{3+1D}
\label{U1rZ24d}

The 3+1D $U(1)\rtimes Z_2$ SPT states
  are described by
$\cH^4[U(1)\rtimes Z_2,\R/\Z]=\Z_2$, whose elements can be labeled by a subset
of $\{(m_0,m_1,m_2,m_3,m_4)\}$ (see appendix \ref{LHS}), where
\begin{align}
m_0 &\in \cH^4[U(1),\R/\Z] =\Z_1,
\nonumber\\
m_1 &\in \cH^1(Z_2,\cH^3[U(1),\R/\Z])
=\cH^1(Z_2,\Z) =\Z_1,
\nonumber\\
m_2 &\in \cH^{2}(Z_2,\cH^2[U(1),\R/\Z])
=\cH^2(Z_2,\Z_1) =\Z_1,
\nonumber\\
m_3 &\in \cH^{3}(Z_2,\cH^1[U(1),\R/\Z]_{Z_2})
=\cH^3(Z_2,\Z_{Z_2}) =\Z_2,
\nonumber\\
m_4 &\in \cH^4(Z_2,\RZ) =\Z_1.
\end{align}
Note that $Z_2$ has a trivial action on $\cH^3[U(1),\R/\Z]$.
To construct the SPT invariants that probe $m_3$, we can view the
$U(1)\rtimes Z_2$ SPT states as  $Z_2\times Z_2$ SPT states and use the result
in section \ref{Zn1Zn24d}.  This is because, as we replace $U(1)$ by $Z_2$,
$\cH^{3}(Z_2,\cH^1[U(1),\R/\Z]_{Z_2})$ becomes
$\cH^{3}[Z_2,\cH^1(Z_2,\R/\Z)]=\Z_2$.  In  section \ref{Zn1Zn24d}, we have
discussed how to measure $\cH^{3}[Z_2,\cH^1(Z_2,\R/\Z)]$. The same set up also
measure $\cH^{3}(Z_2,\cH^1[U(1),\R/\Z]_{Z_2})$. This allow us to obtain the
following result.  \topinv{Consider a 3+1D bosonic $U(1)\rtimes Z_2$ SPT state
labeled by $m_3 \in \cH^4(U(1)\rtimes Z_2,\RZ)$. If we put the SPT state on a
space with topology $S_1\times M_2$, adding $\pi$-flux of $U(1)$ through $S_1$
will induce a bosonic $Z_2$ SPT state in the 2D space labeled by $m_3$ in
$\cH^3(Z_2,\RZ)$.  This also implies that a $\pi$-flux vortex line in $U(1)$
will carry the gapless/degenerate edge states\cite{CW1217} of the 2+1D bosonic
$Z_2$ SPT state labeled by $m_3$ in $\cH^3(Z_2,\RZ)$.}

\subsection{Bosonic $U_c(1)\times [U_s(1)\rtimes Z_2]$ SPT states}

After the preparation of the last section, in this section, we will use the
tools (\ie the SPT invariants) developed so far to study a more
complicated example: bosonic $U_c(1)\times [U_s(1)\rtimes Z_2]$ SPT states in
various dimensions.  We note that $U_s(1)\rtimes Z_2$ is a subgroup of $SO(3)$.
So the results obtained here apply to integer-spin boson gas with boson number
conservation.  For this reason, we will call $U_c(1)$ the charge $U(1)$ and
$U_s(1)$ the spin $U(1)$.

\subsubsection{1+1D}

The different $U_c(1)\times [U_s(1)\rtimes Z_2]$ bosonic SPT states in 1+1D are
described by $\cH^2[U_c(1)\times [U_s(1)\rtimes Z_2],\RZ]$ According to the
K\"unneth formula (see appendix \ref{HBGRZ})
\begin{align}
&\ \ \ \
 \cH^2[U_c(1)\times [U_s(1)\rtimes Z_2],\RZ]
\nonumber\\
& =
\cH^0(U_c(1), \cH^2[U_s(1)\rtimes Z_2,\RZ])
\nonumber\\ &
=
\cH^2[U_s(1)\rtimes Z_2,\RZ]
 = \Z_2 =\{m_0\},
\end{align}
We see that there are two $U_c(1)\times [U_s(1)\rtimes Z_2]$ bosonic SPT states in
1+1D (including the trivial one), labeled by $m_0=0,1$.  The SPT states involve
only the $U_s(1)\rtimes Z_2$ symmetry.  The non-trivial 1D SPT state carries a
projective representation of $U_s(1)\rtimes Z_2$ at each end if the 1D SPT state
form an open chain.\cite{CGW1107,SPC1139,CGW1128}
This state was discussed in the last section.

\subsubsection{2+1D}

\paragraph{\bf Group cohomology description:}
The different $U_c(1)\times [U_s(1)\rtimes Z_2]$ bosonic SPT states in 1+1D are
described by $\cH^3[U_c(1)\times [U_s(1)\rtimes Z_2],\RZ]$:
\begin{align}
&\ \ \ \
 \cH^3[U_c(1)\times [U_s(1)\rtimes Z_2],\RZ]
\nonumber\\
& =
\cH^0(U_c(1), \cH^3[U_s(1)\rtimes Z_2,\RZ])
\oplus
\nonumber\\ & \ \ \ \
\cH^1(U_c(1), \cH^2[U_s(1)\rtimes Z_2,\RZ])
\oplus
\nonumber\\ & \ \ \ \
\cH^2(U_c(1), \cH^3[U_s(1)\rtimes Z_2,\RZ])
\oplus
\nonumber\\ & \ \ \ \
\cH^3(U_c(1), \cH^0[U_s(1)\rtimes Z_2,\RZ]),
\end{align}
where
\begin{align}
&\ \ \ \ \cH^0(U_c(1), \cH^3[U_s(1)\rtimes Z_2,\RZ])
\nonumber\\
&=
 \cH^3[U_s(1)\rtimes Z_2,\RZ]=\Z\oplus\Z_2=\{m_0,m_0'\} ,
\end{align}
\begin{align}
&\ \ \ \ \cH^1(U_c(1), \cH^2[U_s(1)\rtimes Z_2,\RZ])
\nonumber\\
&=
 \cH^1[U_c(1),\Z_2]=0 ,
\end{align}
\begin{align}
&\ \ \ \ \cH^2(U_c(1), \cH^1[U_s(1)\rtimes Z_2,\RZ])
\nonumber\\
&=
 \cH^2[U_c(1),\Z_2]=\Z_2=\{m_2\} ,
\end{align}
\begin{align}
&\ \ \ \ \cH^3(U_c(1), \cH^0[U_s(1)\rtimes Z_2,\RZ])
\nonumber\\
&=
 \cH^3[U_c(1),\RZ]=\Z=\{m_3\} .
\end{align}
We see that $U_c(1)\times [U_s(1)\rtimes Z_2]$ bosonic SPT states in 2+1D are
labeled by $m_0,m_3 \in \Z$ and by $m_0',m_2 \in \Z_2$.

\paragraph{\bf The $(m_0,m_0',m_2,m_3)=(m_0,0,0,0)$ SPT states:}
We note that a $(m_0,0,0,0)$ SPT state is still non-trivial if we break the
$Z_2$ symmetry and the charge $U(1)$ symmetry since $\cH^3[U_c(1),\RZ]=\Z$ for the
spin $U(1)$ symmetry.  Thus, if we
probe the $(m_0,0,0,0)$ SPT state by a non-dynamical  $U(1)$-gauge field $A_{\mu}$, after we
integrate out the matter fields, we will obtain the following \emph{quantized}
gauge topological term in 3+1D:\cite{HW1267}
\begin{align}
\label{fUUZ}
\cL_{2+1D}=
\frac{2m_0}{4\pi} A_{\mu} \prt_\nu A_{\la} \eps^{\mu\nu\la},
\end{align}
which characterize the $(m_0,0,0,0)$ SPT state.  The Hall conduce for the
charge $U(1)$ symmetry is quantized as an even integer
$\si_{xy}=\frac{2m_0}{2\pi}$, which is the SPT invariant that fully
characterizes the  $(m_0,0,0,0)$ SPT states.

\paragraph{\bf The $(m_0,m_0',m_2,m_3)=(0,m_0',0,0)$ SPT states:}
Again, the $(0,m_0',0,0)$ SPT states only involves the $U_s(1)\rtimes Z_2$
symmetry.  The charge $U(1)$ is not relevant here. So we will drop it in the
following discussion.  To probe the $(0,m_0',0,0)$ SPT states, we create two
identical monodromy defects of the spin $U(1)$ symmetry, each with a $\pi$
twist. Such  monodromy defects do not break the $U_s(1)\rtimes Z_2$ symmetry.
The SPT invariant for the  $(0,m_0',0,0)$ SPT states is the total $Z_2$
charge of the two monodromy defects, which is given by $m_0'$.
Such a SPT invariant  fully characterizes the  $(0,m_0',0,0)$
SPT states.

In fact, we can view the 2+1D $U_s(1)\rtimes Z_2$ SPT states as $Z_2\times Z_2$
SPT states. Then the above SPT invariant is one of those discussed in
section \ref{bZ2Z23d}.

\subsubsection{3+1D}

\paragraph{\bf Group cohomology description:}
The different $U_c(1)\times [U_s(1)\rtimes Z_2]$ bosonic SPT states in 3+1D
are described by $\cH^4[U_c(1)\times [U_s(1)\rtimes Z_2],\RZ]$.
According to the K\"unneth formula (see appendix \ref{HBGRZ})
\begin{align}
&\ \ \ \
 \cH^4[U_c(1)\times [U_s(1)\rtimes Z_2],\RZ]
\nonumber\\
& =
\cH^2(U_c(1), \cH^2[U_s(1)\rtimes Z_2,\RZ]) \oplus
\nonumber\\ & \ \ \ \
\cH^0(U_c(1), \cH^4[U_s(1)\rtimes Z_2,\RZ])
\end{align}
where we have only kept the non-zero terms, and
\begin{align}
&\ \ \ \
\cH^2(U_c(1), \cH^2[U_s(1)\rtimes Z_2,\RZ])
\nonumber\\
&=
\cH^2[U_c(1), \Z_2] = \Z_2 =\{m_2\},
\end{align}
\begin{align}
&\ \ \ \
\cH^0(U_c(1), \cH^4[U_s(1)\rtimes Z_2,\RZ])
\nonumber\\
&=
\cH^4[U_s(1)\rtimes Z_2,\RZ] = \Z_2 =\{m_0\} .
\end{align}
We see that there are four $U_c(1)\times [U_s(1)\rtimes Z_2]$ bosonic SPT states in
3+1D (including the trivial one), labeled by $m_0=0,1$ and $m_2=0,1$.  The SPT
state $(m_0,m_2)=(1,0)$ involves only the $U_s(1)\rtimes Z_2$ symmetry, which is
discussed in section \ref{U1rZ24d}.  On the other hand, the $(m_0,m_2)=(0,1)$
SPT state involves the full  $U_c(1)\times [U_s(1)\rtimes Z_2]$ symmetry and is new.

\paragraph{\bf The $(m_0,m_2)=(0,1)$ SPT state:}
One way to probe the $(m_0,m_2)=(0,1)$ SPT state is to couple the the $U_c(1)$
and $U_s(1)$
charges to non-dynamical gauge fields $A_{c\mu}$ and $A_{s\mu}$.  After we
integrate out the matter fields, we will obtain the following \emph{quantized}
gauge topological term in 3+1D:\cite{HW1267}
\begin{align}
\label{fUUZ}
\cL_{3+1D}=
\frac{\pi}{(2\pi)^2} \prt_\mu A_{c\nu} \prt_\la A_{s\ga} \eps^{\mu\nu\la\ga}
\end{align}
The structure of the above quantized gauge topological term is consistent with
corresponding group cohomology class $\cH^2(U_c(1), \cH^2[U_s(1)\rtimes Z_2,\RZ])$.

To understand the physical properties (\ie the SPT invariants)  of the
$(m_0,m_2)=(0,1)$ SPT state, let us assume that the 3+1D space-time has a
topology $M_2\times M_2'$.  We also assume that the $A_{c\mu}$ gauge field has
$2\pi$ flux on $M_2'$.  In the large $M_2$ limit, the Lagrangian \eq{fUUZ}
reduces to an effective Lagrangian on  $M_2$ which has a form
\begin{align}
\cL_{M_2}=
\frac{\pi}{2\pi} \prt_\mu A_{s\nu} \eps^{\mu\nu}
.
\end{align}
We note that the  $A_{c\mu}$ gauge configuration preserve the  $U_c(1)\times(
U_s(1) \rtimes Z_2)$ symmetry. The above Lagrangian is the effective
Lagrangian of the $U_c(1)\times [U_s(1)\rtimes Z_2]$ symmetric theory on $M_2$
probed by the $A_{s\mu}$ gauge field. Such an effective  Lagrangian
implies that the $U_s(1)\rtimes Z_2$ symmetric theory on $M_2$ describe a
non-trivial $U_s(1)\rtimes Z_2$ SPT state labeled by the non-trivial element
$m_2=1$ in $\cH^2[U_c(1) \rtimes Z_2,\RZ]=\Z_2$.  (The charge $U(1)$ does not
play a role here.)

The non-trivial  1+1D $U_s(1)\rtimes Z_2$ SPT state on $M_2$ has the following
property: Let $M_2=R_t\times I$, where $R_t$ is the time and $I$ is a spatial
line segment. Then the excitations at the end of the line are degenerate, and
the degenerate end-states form a projective representation of $U_s(1)\rtimes
Z_2$,\cite{CGW1107,SPC1139,CGW1128,PBT1039} provided that the $A_{c\mu}$ gauge
field has $2\pi$ flux on $M_2'$.

The above result has another interpretation.  Let the 3+1D space-time has a
topology $R_t\times I\times M_2'$.  Such a space-time has two boundaries.  Each
boundary has a topology $R_t\times M_2'$.  The above result implies that the
excitations on $M_2'$ form a linear representation of $U_c(1)\times
[U_s(1)\rtimes Z_2]$, if the $A_{c\mu}$ gauge field is zero on $M_2'$.
However, the excitations on $M_2'$ will form a projective representation of
$U_c(1)\times [U_s(1)\rtimes Z_2]$, if the $A_{c\mu}$ gauge field has $2\pi$
flux on $M_2'$.  If we shrink the boundary $ M_2'$ to a point, we see that
\topinv{the monopole of $A_{c\mu}$ gauge field in the  3+1D $U_c(1)\times
[U_s(1)\rtimes Z_2]$ SPT state with $(m_0,m_2)=(0,1)$ will carries a projective
representation of $U_s(1)\rtimes Z_2$.} Note that the  monopole of charge
$A_{c\mu}$ gauge field does not break the $U_s(1)\rtimes Z_2$ symmetry.  If we
view $U_s(1)\rtimes Z_2$ as a subgroup of $SO(3)$, we may say that the monopole
of $A_{c\mu}$ gauge field carries a half-integer spin.

Adding $2\pi$ flux of $A_{c\mu}$ gauge field is a weak perturbation if $M_2'$
is large.  Such a  perturbation changes the $U_s(1)\rtimes Z_2$ representation
of all the low energy boundary excitations from linear to projective, which
implies that \emph{the 2+1D boundary $R_t\times M_2'$ has gapless excitations
or degenerate ground states}.  In other words, we can prove the basic
conjecture regarding to the boundary of SPT phases, for the case of
$U_c(1)\times [U_s(1)\rtimes Z_2]$ SPT state.

\subsection{Bosonic $U(1)\rtimes Z_2^T$ SPT phases}

In this section, we are going to study bosonic $U(1)\rtimes Z_2^T$ SPT phases.
Those SPT phases can be realized by charged bosons with time reversal symmetry.

\subsubsection{1+1D}

Let us first consider 1+1D SPT states with symmetry $U(1)\rtimes Z_2^T$, which
are described by $\cH^2(U(1)\rtimes Z_2^T,(\R/\Z)_T)= \Z_2$.  According to the
result in appendix \ref{LHS}, the elements in $\cH^2(U(1)\rtimes
Z_2^T,(\R/\Z)_T)$ can be labeled by a subset of $\{(m_0,m_1,m_2)\}$, where
\begin{align}
m_0 &\in \cH^0[Z_2^T,\cH^2[U(1),\RZ]) =\Z_1,
\\
m_1 &\in \cH^1[Z_2^T,\cH^1[U(1),\RZ]) = \cH^1(Z_2^T,\Z) =\Z_1,
\nonumber\\
m_2 &\in \cH^2[Z_2^T,\cH^0[U(1),\RZ]_T) = \cH^2(Z_2^T,(\R/\Z)_T) =\Z_2.
\nonumber
\end{align}
where we have use the fact that $Z_2^T$ has a trivial action on
$\cH^1[U(1),\RZ]$.
We see that $m_2=0,1$ describes the two 1+1D $U(1)\rtimes Z_2^T$ SPT states.
The $U(1)$ symmetry is irrelevant here.  Therefore, \topinv{a 1+1D bosonic
$U(1)\rtimes Z_2^T$ SPT state labeled by $m_2=1$ has a degenerate Kramer
doublet at an open boundary.  }

\subsubsection{2+1D}

Next, we consider the $U(1)\rtimes Z_2^T$ SPT states in 2+1 dimensions, which
are described by $\cH^3(U(1)\rtimes Z_2^T,(\R/\Z)_T)= \Z\oplus \Z_2$.  The
elements in $\cH^3(U(1)\rtimes Z_2^T,(\R/\Z)_T)$ can be labeled by a subset of
$\{(m_0,m_1,m_2,m_3)\}$ (see appendix \ref{LHS}), where
\begin{align}
m_0 &\in \cH^0[Z_2^T,\cH^3[U(1),\RZ]_T) =\cH^0(Z_2^T,\Z_T)=\Z_1,
\nonumber\\
m_1 &\in \cH^1[Z_2^T,\cH^2[U(1),\RZ]) = \Z_1,
\\
m_2 &\in \cH^2[Z_2^T,\cH^1[U(1),\RZ]) 
\nonumber\\
&= \cH^1(Z_2^T,\RZ)\otimes_\Z \cH^1[U(1),\RZ]) =\Z_2,
\nonumber\\
m_3 &\in \cH^3[Z_2^T,\cH^0[U(1),\RZ]_T) = \cH^3(Z_2^T,(\R/\Z)_T) =\Z_1,
\nonumber
\end{align}
where we have use the fact that $Z_2^T$ has a trivial action on
$\cH^1[U(1),\RZ]$ and a non-trivial action on $\cH^3[U(1),\RZ]$.
We see that the $U(1)\rtimes Z_2^T$ SPT states are described by $m_2=0,1$.

%From the structure of $\cH^2[Z_2^T,\cH^1[U(1),\RZ])$, we propose the following
%construction of SPT invariant. We put the 2+1D  $U(1)\rtimes Z_2^T$ SPT
%state on space-time with topology $M_2\times S_1$ and twist the boundary
%condition around $S_1$ by the $\pi$ rotation in $U(1)$.  Such a configuration
%does not break the $U(1)\rtimes Z_2^T$ symmetry.  In the large $M_2$ limit, we
%view the 2+1D  $U(1)\rtimes Z_2^T$ SPT state on  $M_2\times S_1$ as a 1+1D
%$U(1)\rtimes Z_2^T$ SPT state on  $M_2$. Such a 1+1D $U(1)\rtimes Z_2^T$ SPT
%state is characterized by the elements in $\cH^2(U(1)\rtimes
%Z_2^T,(\R/\Z)_T)=\Z_2$ which measures $m_2$.  

To measure $m_2$, we note that if we break the $U(1)$ symmetry down to $Z_2$,
we the non-trivial SPT state with $m_2=1$ is still non-trivial, since
$\cH^2[Z_2^T,\cH^1[U(1),\RZ]) = \cH^2[Z_2^T,\cH^1(Z_2,\RZ)_T]=Z_2$.  If we also
break the $Z_2^T$ symmetry at the same time, the $U(1)\rtimes Z_2^T$ SPT state
described by $m_2=1$ will become a trivial  $Z_2$ SPT state.  So the
$U(1)\rtimes Z_2^T$ SPT state described by $m_2=1$ is also a non-trivial
$Z_2\times Z_2^T$ SPT state.  The above consideration suggest that such a
$Z_2\times Z_2^T$ SPT state is described by a non-trivial $m_1$ in
\eqn{Z2Z2T3dm1m3}.  So we can use the SPT invariant that detect $m_1$
of the $Z_2\times Z_2^T$ SPT states to detect $m_2$ of the $U(1)\rtimes Z_2^T$
SPT states.  Thus \topinv{Consider a 2+1D bosonic $U(1)\rtimes Z_2^T$ SPT state
labeled by $m_2=1$ in $\cH^3(U(1)\rtimes Z_2^T,(\R/\Z)_T)$ .  If we put the
state on a cylinder $I \times S_1$, then the states on one boundary will form
Kramer doublets, if we twist the boundary condition around $S_1$ by the $\pi$
rotation in $U(1)$.  This also implies that a $U(1)$ monodromy defect generated
by $\pi$ rotation carries a degenerate Kramer doublet.}

\newcommand{\tbox}[2]{\parbox[t]{#1}{#2}}
\begin{table*}[tb]
 \caption{ Symmetry-protected topological invariants for various bosonic SPT
states. Here the flux, the monodromy defects and the monopoles are always the minimal ones.
$\{n_1,n_2\}$ is the smallest common  multiple of $n_1,n_2$.
$\<n_1,n_2\>$ is the largest common divisor of $n_1,n_2$.
Also $\cH^d(G)\equiv \cH^d(G,\R/\Z)$.
The previously known results are indicated by the references.
} \label{TopInv}
 \centering
 \begin{tabular}{ |c|c|c|p{3.8in}| }
 \hline
symmetry & dim. & labels & ~~~~~~~~~~
symmetry-protected topological invariants \\
\hline
\hline
$Z_n$ & 2+1D & $m\in \cH^3(Z_n)=\Z_n$ &
$\bullet$ $n$ identical monodromy defects have a total $Z_n$ charge $2m$.
\hfill\break
$\bullet$ A monodromy defect has a
statistics $\th=2\pi (\frac{m}{n^2} +\frac{\text{integer}}{n})$.\cite{LG1220,MR1235,HW1227}
\\
\hline
$U(1)$ & 2+1D & $m\in \cH^3[U(1)]=\Z$ &
$\bullet$
Even-integer quantized Hall conductance $\si_{xy}=2m/2\pi$.\cite{LV1219,LW1224,CW1217,SL1204}
\\
\hline
 & 1+1D & $m_1\in \cH^1(Z_{n_1}) \boxtimes_\Z \cH^1(Z_{n_2})=\Z_{\<n_1,n_2\>}$ &
$\bullet$ The degenerate states at an end of chain form the $m_1^\text{th}$
projective representation of $Z_{n_1}\times Z_{n_2}$.
\hfill\break
$\bullet$  The $Z_{n_1}$-flux through the 1D circular space
induces a $Z_{n_2}$-charge $m+\<n_1,n_2\>\times$ integer.
\\
\cline{2-4}
$Z_{n_1}\times Z_{n_2}$ & 2+1D &
\tbox{2.25in}{$m_0\in \cH^3[Z_{n_2}]=\Z_{n_2}$\\
$m_2\in \cH^1(Z_{n_1})\otimes_\Z \cH^1(Z_{n_2})=\Z_{\<n_1,n_2\>}$\\
$m_3\in \cH^3(Z_{n_1})=\Z_{n_1}$}
&
$\bullet$ The statistics of $Z_{n_1}$  monodromy defects: $\th_{11}$ mod
$\frac{2\pi}{n_1} = \frac{2\pi m_3}{n_1^2}$.  The statistics of $Z_{n_2}$
monodromy defects: $\th_{22}$ mod $\frac{2\pi}{n_2} = \frac{2\pi m_0}{n_2^2}$.
The mutual statistics between $Z_{n_1}$ and $Z_{n_2}$  monodromy defects:
$\th_{12}$ mod $\frac{2\pi}{\{n_1,n_2\}} = \frac{2\pi
m_2}{n_1n_2}$.\cite{LG1220,MR1235,HW1227}
\hfill\break
$\bullet$ $n_1$ identical $Z_{n_1}$ monodromy defects carry $2m_3$ $Z_{n_1}$
charges and $m_2+\<n_1,n_2\>\times$ integer $Z_{n_2}$ charges.
$n_2$ identical $Z_{n_2}$ monodromy defects carry $2m_0$ $Z_{n_2}$
charges and $m_2+\<n_1,n_2\>\times$ integer $Z_{n_1}$ charges.
\\
\cline{2-4}
 & 3+1D &
\tbox{2.25in}{$m_1\in \cH^1(Z_{n_1})\boxtimes_\Z,\cH^3(Z_{n_2})=\Z_{\<n_1,n_2\>}$\\
$m_3\in \cH^3(Z_{n_1})\boxtimes_\Z\cH^1(Z_{n_2})=\Z_{\<n_1,n_2\>}$}
&
$\bullet$ A $Z_{n_1}$ monodromy line-defect will carry
gapless/degenerate edge states of  $m_1^\text{th}$ 2+1D bosonic $Z_{n_2}$
SPT states.
A $Z_{n_2}$ monodromy line-defect will carry
gapless/degenerate edge states of  $m_3^\text{th}$ 2+1D bosonic $Z_{n_1}$
SPT states.
\\
\hline
 & 1+1D & $m_1\in \cH^1(Z_2,\cH^1[U(1)]_{Z_2})=\Z_2$ &
$\bullet$ A degenerate $U(1)$-charge $\pm 1/2$ doublet at a boundary, if $m_1=1$.
\\
\cline{2-4}
$U(1)\rtimes Z_2$ & 2+1D & \tbox{2.25in}{$m_0\in \cH^3[U(1)]=\Z$\\
$m_3\in \cH^3(Z_2)=\Z_2$}
&
$\bullet$ Same as the $U(1)$ or the $Z_2$ SPT states in 2+1D.
\\
\cline{2-4}
 & 3+1D &
$m_3\in \cH^3(Z_2,\cH^1[U(1)]_{Z_2})=\Z_2$
&
$\bullet$ A $\pi$-flux line of the $U(1)$ will carry gapless/degenerate
edge states of the 2+1D bosonic $Z_2$ SPT state, if $m_3=1$.
\\
\hline
 & 1+1D & $m_2\in \cH^2(Z_2^T,\cH^0[U(1)]_T)=\Z_2$ &
$\bullet$ A neutral Kramer doublet at a boundary, if $m_2=1$.
\\
\cline{2-4}
& 2+1D & $m_2\in \cH^1(Z_2^T)\otimes_\Z \cH^1[U(1)]=\Z_2$ &
$\bullet$ A
monodromy defect generated by $U(1)$ $\pi$-rotation carries a degenerate Kramer
double, if $m_2=1$.
\\
\cline{2-4}
$U(1)\rtimes Z_2^T$ 
 & 3+1D &
\tbox{2.25in}{$m_1\in \cH^1(Z_2^T,\cH^3[U(1)]_T)=\Z_2$\\
$m_4\in \cH^4[Z_2^T,(\RZ)_T]=\Z_2$
}
&
$\bullet$ A dyon of (electric,magnetic) charge ($q$,$m$) has
a statistics $(-)^{m(q-m_1 m)}$.\cite{MKF1335}
\hfill\break
$\bullet$ 
A gapped time-reversal symmetry breaking boundary has a Hall conductance
$\si_{xy}=\frac{m_1}{2\pi}+\frac{\text{even}}{2\pi}$.\cite{VS1258}
\\
\hline
 & 1+1D & \tbox{2.25in}{
$m_0\in \cH^2[Z_2^T,(\RZ)_T]=\Z_2$\\
$m_2\in \cH^1[U(1)]\otimes_\Z \cH^0[Z_2^T,(\RZ)_T]=\Z_2$
}
&
$\bullet$ A neutral Kramer doublet at a boundary, if $(m_0,m_2)=(1,0)$.
\hfill\break
$\bullet$ A degenerate boundary charge-$\pm\frac12$ doublet, if $(m_0,m_2)=(0,1)$.
\\
\cline{2-4}
$U(1)\times Z_2^T$ & 3+1D &
\tbox{2.25in}{ $m_0\in \cH^4[Z_2^T,(\RZ)_T]=\Z_2$\\
$m_2\in \cH^1[U(1)]\otimes_\Z \cH^2[Z_2^T,(\RZ)_T]=\Z_2$\\
$m_4\in \cH^3[U(1)]\otimes_\Z \cH^0[Z_2^T,(\RZ)_T]=\Z_2$ }
& $\bullet$ A $U(1)$ monopole 
will carries a neutral degenerate Kramer doublet,
if $(m_1,m_2,m_4)=(0,1,0)$.
\hfill\break
$\bullet$ A dyon of (electric,magnetic) charge ($q$,$m$) has
a statistics $(-)^{m(q-m_4 m)}$, if $(m_1,m_2,m_4)=(0,0,m_4)$.
\hfill\break
$\bullet$ 
A gapped time-reversal symmetry breaking boundary
has a Hall conductance $\si_{xy}=\frac{m_4}{2\pi}+\frac{\text{even}}{2\pi}$.
\\
\hline
 & 1+1D & \tbox{2.25in}{
$m_0\in \cH^2[Z_2^T,(\RZ)_T]=\Z_2$\\
$m_2\in \cH^1[U(1)]\otimes_\Z \cH^0[Z_2^T,(\RZ)_T]=\Z_2$
}
&
$\bullet$ A neutral Kramer doublet at a boundary, if $(m_0,m_2)=(1,0)$.
\hfill\break
$\bullet$ A degenerate boundary charge-$\pm\frac12$ doublet, if $(m_0,m_2)=(0,1)$.
\\
\cline{2-4}
& 2+1D &
\tbox{2.25in}{$m_1\in \cH^1(Z_2)\boxtimes_\Z \cH^2[Z_2^T,(\RZ)_T]=\Z_2$\\
$m_3\in \cH^3[Z_2]\boxtimes_\Z \cH^0[Z_2^T,(\RZ)_T]=\Z_2$ }
& $\bullet$ A $Z_2$ monodromy defect will carries a degenerate Kramer doublet,
if $(m_1,m_3)=(1,0)$.  
\hfill\break
$\bullet$ $m_3$ can be measured as in the 2+1D $Z_2$ SPT states.
\\
\cline{2-4}
$Z_2\times Z_2^T$ 
& 3+1D &
\tbox{2.25in}{ $m_0\in \cH^4[Z_2^T,(\RZ)_T]=\Z_2$\\
$m_2\in \cH^1(Z_2)\otimes_\Z \cH^2[Z_2^T,(\RZ)_T]=\Z_2$\\
$m_4\in \cH^3(Z_2)\otimes_\Z \cH^0[Z_2^T,(\RZ)_T]=\Z_2$ }
& $\bullet$ 
When $(m_0,m_2,m_4)=(0,0,1)$,
the symmetry breaking domain wall
in a gapped time-reversal symmetry breaking boundary will carry the
edge excitations of the 2+1D $Z_2$ STP state. Also the $Z_2$ monodromy
line-defect will induce two-fold degenerate states with different Z2
charges at its intersection with a gapped time-reversal symmetry
breaking boundary, unless the line-defect is gapless.
\hfill\break
$\bullet$ 
When $(m_0,m_2,m_4)=(0,1,0)$,
two identical ends of the line-defects on
a compact  surface will induce a
degenerate Kramer doublet.
\\
\hline
\hline
 \end{tabular}
\end{table*}

\subsubsection{3+1D}

Last, we consider the $U(1)\rtimes Z_2^T$ SPT states in 3+1 dimensions.
Several SPT invariants for such states were discussed in
\Ref{VS1258,MKF1335}.  The $U(1)\rtimes Z_2^T$ SPT states are described by
$\cH^4(U(1)\rtimes Z_2^T,(\R/\Z)_T)= \Z_2\oplus Z_2$.  The elements in
$\cH^4(U(1)\rtimes Z_2^T,(\R/\Z)_T)$ can be labeled by a subset of
$\{(m_0,m_1,m_2,m_3,m_4)\}$ (see appendix \ref{LHS}), where
\begin{align}
m_0 &\in \cH^0[Z_2^T,\cH^4[U(1),\RZ]) =\Z_1,
\nonumber\\
m_1 &\in \cH^1[Z_2^T,\cH^3[U(1),\RZ]_T) = \cH^1(Z_2^T,\Z_T) =\Z_2,
\nonumber \\
m_2 &\in \cH^2[Z_2^T,\cH^2[U(1),\RZ]) = \Z_1,
\\
m_3 &\in \cH^3[Z_2^T,\cH^1[U(1),\RZ]) = \cH^3(Z_2^T,\Z) =\Z_1,
\nonumber\\
m_4 &\in \cH^4[Z_2^T,\cH^0[U(1),\RZ]_T) = \cH^4(Z_2^T,(\R/\Z)_T) =\Z_2,
\nonumber
\end{align}
We see that the 3+1D $U(1)\rtimes Z_2^T$ SPT states are labeled by $m_1=0,1$
and $m_4=0,1$.  $m_4$ labels different 3+1D $Z_2^T$ SPT states where the $U(1)$
symmetry is irrelevant.

To probe $m_1$, we may gauge the $U(1)$ symmetry. We believe
that the $U(1)\rtimes Z_2^T$ SPT states labeled by $(m_1,m_4)=(m_1,0)$ are
described by the following $U(1)$-gauge topological term
\begin{align}
\label{F2topTI}
 \cL_\text{top}=\frac{ m_1\pi }{(2\pi)^2} F^2
\end{align}
Under the $Z_2^T$ time-reversal transformation, $F^2\to -F^2$ and $\ee^{\ii
\int_{M_4} \frac{m_1\pi}{(2\pi)^2} F^2}\to \ee^{-\ii \int_{M_4}
\frac{m_1\pi}{2(2\pi)^2} F^2}$. Since $\int_{M_4} \frac{m_1\pi}{(2\pi)^2} F^2
=\pi m_1 \times$ integers, on any closed 3+1D orientable space-time manifold
$M_4$, the $Z_2^T$ symmetry is preserved since $m_1$ is an integer. $m_1$=odd
describes the non-trivial 3+1D $U(1)\rtimes Z_2^T$ SPT state, while $m_1$=even
describes the trivial SPT state.  Now we see that $m_1$ can be measured by the
statistical effect discussed in \Ref{W7983,GMW8921,MKF1335}: \topinv{in a 3+1D
bosonic  $U(1)\rtimes Z_2^T$ SPT state labeled by $(m_1,m_4)=(m_1,0)$, a dyon
of the $U(1)$ gauge field with ($U(1)$-charge, magnetic charge) = $(q,m)$ has a
statistics $(-)^{m(q-m_1)}$ (where $+ \to$ boson and $- \to$ fermion).}

If the space-time $M_4$ has a boundary, the topological term \eq{F2topTI}
reduces to an effective Lagrangian on the boundary
\begin{align}
 \cL_{2+1D}=\frac{m_1}{4\pi} AF ,
\end{align}
if  the $Z_2^T$ time-reversal symmetry is broken on the boundary.  The above is
nothing but a 2+1D $U(1)$ Chern-Simons term with a quantized Hall conductance
$\si_{xy}=m_1/2\pi$.\cite{VS1258}  Thus 
\topinv{in a 3+1D bosonic  $U(1)\rtimes Z_2^T$ SPT
state labeled by $(m_1,m_4)$, 
the gapped time-reversal symmetry breaking boundary
has a Hall conductance $\si_{xy}=\frac{m_1}{2\pi}+\frac{\text{even}}{2\pi}$.
}

If the $Z_2^T$ symmetry is not broken, we
actually have the following effective boundary theory
\begin{align}
 \cL_{2+1D}=\frac{m_1 \si }{4\pi} AF +\cL(\si)
\end{align}
where the $\si(x)$ field only takes two values $\si=\pm 1$.  The gapless edge
states on the domain wall between $\si=1$ and $\si=-1$ regions may give rise to
the gapless boundary excitations on the 2+1D surface.

\section{Summary}

\begin{table*}[tb]
\caption{Symmetry-protected topological invariants for some fermionic SPT
states.} \label{TopInvF}
 \centering
 \begin{tabular}{ |c|c|c|p{5.3in}| }
 \hline
symmetry & dim. & labels & ~~~~~~~~~~~~~
 symmetry-protected topological invariants \\
\hline
\hline
$U^f(1)$ & 2+1D & $m\in \Z$ &
$\bullet$
8-times-integer quantized Hall conductance $\si_{xy}=8m/2\pi$.
\\
\hline
$U(1)\times Z_2^f$ & 2+1D & $m\in \Z$ &
$\bullet$
Integer quantized Hall conductance $\si_{xy}=m/2\pi$.
\\
\hline
$Z_2\times Z_2^f$ & 2+1D & $m\in \Z_8$ &
$\bullet$ Abelian/non-Abelian statistics and mutual statistics of the $Z_2$
and $Z_2^f$ monodromy defects.
\hfill\break
$\bullet$ The mutual induced  $Z_2$ (or $Z_2^f$) charge by two identical
$Z_2^f$ (or $Z_2$) monodromy defects.
\\
\hline
\hline
 \end{tabular}
\end{table*}

\begin{table}[tb]
 \caption{ The fermionic SPT phases with the full symmetry $G_f$.  Here $0$
means that there is only trivial SPT phase.  $\Z_n$ means that the non-trivial
SPT phases plus the trivial phase are labeled by the elements in $\Z_n$.  }
\label{tb}
 \centering
 \begin{tabular}{ |c||c|c|c|c| }
 \hline
 $G_f$ & $0+1D$ & $1+1D$ & $2+1D$ & $3+1D$  \\
\hline
\hline
$Z_2^f$ & $\Z_2$  & $0$ & $0$ & $0$    \\
$U^f(1)$ & $\Z$  & $0$ & $\Z$ & $0$    \\
$U(1)\times U^f(1)$ & $\Z\oplus \Z$  & $0$ & $\Z\oplus \Z\oplus \Z$ & $0$    \\
$U(1)\times Z^f_2$ & $\Z\oplus \Z_2$  & $0$ & $\Z$ & 0    \\
$Z_2\times Z^f_2$ & $\Z_2\oplus \Z_2$  & $\Z_2$ & $\Z_8$ & ?    \\
\hline
 \end{tabular}
\end{table}

It has been shown that the SPT states and some of the SET states can be
described by the cocycles in the group cohomology class
$\cH^d(G,\RZ)$.\cite{CGL1172,GW1248} In this paper, we construct many
SPT invariants which allow us to physically measure the cocycles in
$\cH^d(G,\RZ)$ fully.  The constructed SPT invariants allow us to
physically or numerically detect and characterized the SPT states and some of
the SET states.

The SPT invariants are constructed by putting
the SPT states on a space-time with a topology $M_k\times M_{d-1-k}$ and
gauging a subgroup $GG$ of the symmetry group $G$.  We then put a non-trivial
$GG$ gauge configuration on the closed manifold $M_k$.  When $k=1$, the gauge
configuration can be a gauge flux through the ring.  When $k=2$, the gauge
configuration can be a gauge flux through $M_2$ if $GG$ is continuous or a few
\emph{identical}  gauge flux through $M_2$ (if $GG$ is discrete), \etc.

When $M_{d-1-k}$ is large, the SPT states on $M_k\times M_{d-1-k}$ can be
viewed as a SPT state on $M_{d-1-k}$ with a symmetry $SG$, where $SG$ is a
subgroup of $G$ that commute with $GG$.  The $SG$ SPT state on $M_{d-1-k}$ is
described by $\cH^{d-k}(SG,\RZ)$. This way, we can measure the the cocycles in
$\cH^d(G,\RZ)$ by  measuring the the cocycles in $\cH^{d-k}(SG,\RZ)$.  When
$d-k=1$, the cocycles in $\cH^{1}(SG,\RZ)$ can be measured by measuring the
$SG$ quantum number of the ground state.  When $d-k=2$, we can choose the
space-time $ M_{d-1-k}$ to have a space described by a finite line.  Then the
cocycles in $\cH^{2}(SG,\RZ)$ can be measured by measuring the projective
representation of $SG$ at one end of the line.

In Table \ref{TopInv}, we list the SPT
invariants for some simple bosonic SPT phases.  In Table \ref{TopInvF}, we list
the SPT invariants for a few fermionic  SPT
phases.  More SPT invariants are described by the \textbf{SPT
invariant} statements in the paper.  Those SPT invariants also allow us
to understand some of the SPT states for interacting fermions.  We list those
results in Table \ref{tb}.\cite{RZ1232,GL1369}

I like to thank Xie Chen, Zheng-Cheng Gu, Max Metlitski, and Juven Wang for
many helpful discussions.  This research is supported by NSF Grant No.
DMR-1005541, NSFC 11074140, and NSFC 11274192.  Research at Perimeter Institute
is supported by the Government of Canada through Industry Canada and by the
Province of Ontario through the Ministry of Research.

\appendix

\section{Group cohomology theory}
\label{gcoh}

\subsection{Homogeneous group cocycle}

In this section, we will briefly introduce group cohomology.  The group
cohomology class $\cH^d(G,\M)$ is an Abelian group constructed from a group $G$
and an Abelian group $\M$.   We will use ``+'' to represent the multiplication
of the Abelian groups.  Each elements of $G$ also induce a mapping $\M\to \M$,
which is denoted as
\begin{eqnarray}
g\cdot m = m', \ \ \ g\in G,\ m,m'\in \M.
\end{eqnarray}
The map $g\cdot$ is a group homomorphism:
\begin{eqnarray}
g\cdot (m_1+m_2)= g\cdot m_1 +g \cdot m_2.
\end{eqnarray}
The  Abelian group $\M$ with such a $G$-group homomorphism, is call a
$G$-module.

A homogeneous $d$-cochain
is a function $\nu_d: G^{d+1}\to \M$, that satisfies
\begin{align}
\nu_d(g_0,\cdots,g_d)
=g\cdot \nu_d(gg_0,\cdots,gg_d), \ \ \ \ g,g_i \in G.
\end{align}
We denote the set of $d$-cochains as $\cC^d(G,\M)$. Clearly $\cC^d(G,\M)$ is an
Abelian group.
homogeneous group cocycle

Let us define a mapping $\dd$ (group homomorphism) from
$\cC^d(G,\M)$ to $\cC^{d+1}(G,\M)$:
\begin{align}
  (\dd \nu_d)( g_0,\cdots, g_{d+1})=
  \sum_{i=0}^{d+1} (-)^i \nu_d( g_0,\cdots, \hat g_i
  ,\cdots,g_{d+1})
\end{align}
where $g_0,\cdots, \hat g_i ,\cdots,g_{d+1}$ is the sequence $g_0,\cdots, g_i
,\cdots,g_{d+1}$ with $g_i$ removed.
One can check that $\dd^2=0$.
The homogeneous $d$-cocycles are then
the homogeneous $d$-cochains that also satisfy the cocycle condition
\begin{eqnarray}
 \dd \nu_d =0.
\end{eqnarray}
We denote the set of $d$-cocycles as
$\cZ^d(G,\M)$. Clearly $\cZ^d(G,\M)$ is an Abelian subgroup of $\cC^d(G,\M)$.

Let us denote  $\cB^d(G,\M)$ as the image of the map $\dd: \cC^{d-1}(G,\M) \to
\cC^d(G,\M)$ and $\cB^0(G,\M)=\{0\}$.
The elements in $\cB^d(G,\M)$ are called $d$-coboundary.
Since $\dd^2=0$, $\cB^d(G,\M)$  is a subgroup of $\cZ^d(G,\M)$:
\begin{eqnarray}
\cB^d(G,\M) \subset \cZ^d(G,\M).
\end{eqnarray}
The group cohomology class $\cH^d(G,\M)$ is then defined as
\begin{eqnarray}
\cH^d(G,\M) =  \cZ^d(G,\M)/ \cB^d(G,\M) .
\end{eqnarray}
We note that the $\dd$ operator and the cochains $\cC^d(G,\M)$ (for all values
of $d$) form a so called cochain complex,
\begin{align}
\cdots
\stackrel{\dd}{\to}
\cC^d(G,\M)
\stackrel{\dd}{\to}
\cC^{d+1}(G,\M)
\stackrel{\dd}{\to}
\cdots
\end{align}
which is denoted as $C(G,\M)$.  So we may also write the group cohomology
$\cH^d(G,\M)$ as the standard cohomology of the cochain complex $H^d[C(G,\M)]$.

\subsection{Nonhomogeneous group cocycle}
\label{nonhomo}

The above definition of group cohomology class can be rewritten in terms of
nonhomogeneous group cochains/cocycles.  An nonhomogeneous group $d$-cochain is
a function $\om_d: G^d \to M$. All $\om_d(g_1,\cdots,g_d)$ form $\cC^d(G,\M)$.
The nonhomogeneous group cochains and the homogeneous group cochains are
related as
\begin{eqnarray}
\nu_d(g_0,g_1,\cdots,g_d)=
\om_d( g_{01},\cdots, g_{d-1,d}),
\end{eqnarray}
with
\begin{align}
g_0=1,\ \
g_1=g_0 g_{01}, \ \
g_2=g_1 g_{12}, \ \ \cdots \ \
g_d=g_{d-1} g_{d-1,d}.
\end{align}
Now the $\dd$ map has a form on $\om_d$:
\begin{align}
&
(\dd\om_d)(g_{01},\cdots, g_{d,d+1})=
 g_{01}\cdot \om_d( g_{12},\cdots,g_{d,d+1})
\nonumber\\
& \ \ \
+\sum_{i=1}^d (-)^i \om_d(g_{01},\cdots, g_{i-1,i}g_{i,i+1},\cdots, g_{d,d+1})
\nonumber\\
& \ \ \
+(-)^{d+1}\om_d(g_{01},\cdots,\t g_{d-1,d})
\end{align}
This allows us to define the nonhomogeneous group $d$-cocycles which satisfy
$\dd \om_d=0$ and  the nonhomogeneous group $d$-coboundaries which have a form
$\om_d = \dd \mu_{d-1}$.  In the following, we are going to use nonhomogeneous
group cocycles to study group cohomology.  Geometrically, we may view $g_i$ as
living on the vertex $i$, while $g_{ij}$ as living on the edge connecting the
two vertices $i$ to $j$.

\subsection{``Normalized'' cocycles}

We know that each elements in $\cH^d(G,\R/\Z)$ can be represented by many
cocycles.  In the following, we are going describe a way to simplify the
cocycles, so that the simplified cocycles can still represent all the elements
in $\cH^d(G,\R/\Z)$.

The simplification is obtained by considering ``normalized''
cochains,\cite{HS5310} which satisfy
\begin{align}
\om_d(g_1,\cdots, g_d)=0, \text{ if one of } g_i=1.
\end{align}
One can check that the $\dd$-operator maps a ``normalized'' cochain to a
``normalized'' cochain.  The group cohomology classes obtained from the
ordinary cochains is isomorphic to the group cohomology classes obtained from
the ``normalized'' cochains.  Let us use $\bar\cC^d(G,\M)$, $\bar\cZ^d(G,\M)$,
and $\bar\cB^d(G,\M)$ to denote the ``normalized'' cochains, cocycles, and
coboundaries.  We have $\cH^d(G,\M)=\bar\cZ^d(G,\M)/\bar\cB^d(G,\M)$.

\subsection{A ``differential form'' notation for group cocycles}
\label{dform}

We know that a cocycle $\om_d$ in $\cH^d(G,\RZ)$ is a linear map that map a
$d$-dimensional complex $M$, with $g_i$ on the vertices or $g_{ij}$ on the
edges, to a mod-1 number in $\RZ$.  Let us use a ``differential form'' notation
to denote such a map:
\begin{align}
 \int_M \om_d(g_{ij}) \in \RZ .
\end{align}
In the above, we have regarded $\om_d(g_{ij})$ as a function of $g_{ij}$ on the
edges.  We may also view $\om_d$  as a function of $g_{i}$ on the vertices by
replacing $g_{ij}$ by $g_jg_i^{-1}$: $\om_d(g_jg_i^{-1})$.  A differential form
$F$ is a linear map from a complex (or a manifold) to a real number:
\begin{align}
 \int_M F \in \R .
\end{align}
In fact, we can use a differential form $F_d(g_{ij})$ (that depends on
$g_{ij}$'s on the edges) to represent $\om_d(g_{ij})$:
\begin{align}
 \int_M \om_d(g_{ij}) = \int_M F_d(g_{ij}) \text{ mod } 1.
\end{align}
So we can treat $\om_d(g_{ij})$ as a differential form, or more precisely, a
\emph{discretized differential form}.  In fact, the cocycle is an analogue of
closed form.

In this paper, we will use such a
notation to described the fixed-point (or the ideal) Lagrangian for the SPT
states.  The  ideal fixed-point actions for SPT states contain only a pure
topological term which always has a form
\begin{align}
\label{Stop}
 S_\text{top}=2\pi \int_M \om_d(g_{ij})
\end{align}
where $\om_d$ is a cocycle in $\cH^d(G,\RZ)$ and $M$ is the space-time complex.
The factor $2\pi$ is needed to make the action amplitude $\ee^{\ii 2\pi \int_M
\om_d(g_{ij})}$ well defined.  The expression \eq{Stop} reflects the direct
connection between the SPT phases and cocycles in $ \cH^d(G,\RZ)$.

\section{Relation between $H^{d+1}(BG,\Z)$ and $\cH_B^d(G,\R/\Z)$}
\label{HBGZHcGRZ}

We can show that the topological cohomology of the classifying space,
$H^{d+1}(BG,\Z)$, and the Borel-group cohomology, $\cH_B^d(G,\R/\Z)$, are
directly related
\begin{align}
\label{HdHd1b}
   H^{d+1}(BG,\Z) \simeq  \cH_B^d(G,\R/\Z) .
\end{align}
This result is obtained from \Ref{WW1104}.  On page 16 of \Ref{WW1104}, it is mentioned
in Remark IV.16(3) that $\cH^d_B(G,\R)= \Z_1$ (there, $\cH^d_B(G,M)$ is denoted
as $\cH^d_\text{Moore}(G,M)$ which is equal to $\cH^d_\text{SM}(G,M)$).  It is
also shown in Remark IV.16(1) and in Remark IV.16(3) that
$\cH^d_\text{SM}(G,\Z)=H^{d}(BG,\Z)$ and
$\cH^d_\text{SM}(G,\R/\Z)=H^{d+1}(BG,\Z)$,
%On page 14, it is shown that\\ (3)
%$\cH^d_\text{loc,c}(G,\R/\Z)=\cH^{d}_\text{SM}(G,\R/\Z)$\\ Thus
%$\cH^d_B(G,\R/\Z)=\cH^{d+1}_B(G,\Z)= H^{d+1}(BG,\Z)$
(where $G$ can have a non-trivial action on $\R/\Z$ and $\Z$, and
$H^{d+1}(BG,\Z)$ is the usual topological cohomology on the classifying space
$BG$ of $G$).
%and $\cH^d_\text{loc,c}(G,\R/\Z)$ is the group cohomology where the cochain
%$\nu_d(g_0,...,g_d)$ is a continuous function near $g_i=1$.)
Therefore,
%we can use $\cH^d_\text{loc,c}(G,\R/\Z)$ to calculate $\cH^d_B(G,\R/\Z)$.
we have
\begin{align}
\label{HdR}
& \cH^d_B(G,\R/\Z)=\cH^{d+1}_B(G,\Z)=H^{d+1}(BG,\Z),
\nonumber\\
& \cH^d_B(G,\R)=\Z_1,\ \ d>0.
\end{align}
%These results will be useful later.
These results are valid for both continuous groups and discrete groups, as well
as for $G$ having a non-trivial action on the modules $\R/\Z$ and $\Z$.
%For finite groups, we also have
%\begin{align}
%\label{HdHd1}
%  H^d(BG,\R/\Z) \simeq H^{d+1}(BG,\Z) \simeq  \cH_B^d(G,\R/\Z) .
%\end{align}

\section{The K\"unneth formula} \label{HBGRZ}

The K\"unneth formula is a very helpful formula that allows us to calculate the
cohomology of chain complex $X\times X'$ in terms of  the cohomology of chain
complex $X$ and chain complex $X'$.
The K\"unneth formula is expressed in terms of
the tensor-product operation $\otimes_R$ and the torsion-product operation
$\boxtimes_R\equiv \text{Tor}_1^R$, which have the following properties:
\begin{align}
\label{tnprd}
& \M \otimes_\Z \M' \simeq \M' \otimes_\Z \M ,
\nonumber\\
& \Z \otimes_\Z \M \simeq \M \otimes_\Z \Z =\M ,
\nonumber\\
& \Z_n \otimes_\Z \M \simeq \M \otimes_\Z \Z_n = \M/n\M ,
\nonumber\\
& \Z_n \otimes_\Z \RZ \simeq \RZ \otimes_\Z \Z_n = 0,
\nonumber\\
& \Z_m \otimes_\Z \Z_n  =\Z_{\<m,n\>} ,
\nonumber\\
&  (\M'\oplus \M'')\otimes_R \M = (\M' \otimes_R \M)\oplus (\M'' \otimes_R \M)   ,
\nonumber\\
& \M \otimes_R (\M'\oplus \M'') = (\M \otimes_R \M')\oplus (\M \otimes_R \M'')   ;
\end{align}
and
\begin{align}
\label{trprd}
& \text{Tor}_1^R(\M,\M') \equiv \M\boxtimes_R \M'  ,
\nonumber\\
& \M\boxtimes_R \M' \simeq \M'\boxtimes_R \M  ,
\nonumber\\
& \Z\boxtimes_\Z  \M = \M\boxtimes_\Z  \Z = 0,
\nonumber\\
& \Z_n\boxtimes_\Z \M = \{m\in \M| nm=0\},
\nonumber\\
& \Z_n\boxtimes_\Z \RZ = \Z_n,
\nonumber\\
& \Z_m\boxtimes_\Z \Z_n = \Z_{\<m,n\>} ,
%\nonumber\\
%& \text{Tor}_1^\Z(U(1), U(1)) = 0 ,
\nonumber\\
& \M'\oplus \M''\boxtimes_R\M = \M'\boxtimes_R \M\oplus\M''\boxtimes_R \M,
\nonumber\\
& \M\boxtimes_R\M'\oplus \M'' = \M\boxtimes_R\M'\oplus\M\boxtimes_RB
,
\end{align}
where $\<m,n\>$ is the greatest common divisor of $m$ and $n$.  These
expressions allow us to compute the tensor-product $\otimes_R$ and  the
torsion-product $\boxtimes^R$.  Here $R$ is a ring and $\M,\M',\M''$ are
$R$-modules.  A $R$-module is like a vector space over $R$ (\ie we can
``multiply'' a vector by an element of $R$.)

The K\"unneth formula itself is given by (see
\Ref{Spa66} page 247)
\begin{align}
\label{kunn}
&\ \ \ \ H^d(X\times X',\M\otimes_R \M')
\nonumber\\
&\simeq \Big[\oplus_{k=0}^d H^k(X,\M)\otimes_R H^{d-k}(X',\M')\Big]\oplus
\nonumber\\
&\ \ \ \ \ \
\Big[\oplus_{k=0}^{d+1}
H^k(X,\M)\boxtimes_R H^{d-k+1}(X',\M')\Big]  .
\end{align}
Here $R$ is a principle ideal domain and $\M,\M'$ are $R$-modules such that
$\M\boxtimes_R \M'=0$.  We also require that $\M'$ and $H^d(X',\Z)$ are
finitely generated, such as $\M'=\Z\oplus \cdots \oplus \Z\oplus
\Z_n\oplus\Z_m\oplus\cdots$.

For more details on principal ideal domain and
$R$-module, see the corresponding Wiki articles.  Note that $\Z$ and $\R$ are
principal ideal domains, while $\R/\Z$ is not.  Also, $\R$ and $\R/\Z$ are not
finitely generate $R$-modules if $R=\Z$.  The K\"unneth formula works for
topological cohomology where $X$ and $X'$ are treated as topological spaces.
The K\"unneth formula also works for group cohomology, where $X$ and $X'$ are
treated as groups, $X=G$ and $X'=G'$, provided that $G'$ is a finite group.
However, the above K\"unneth formula  does not apply for Borel-group cohomology
when $X'=G'$ is a continuous group, since in that case $\cH_B^d(G',\Z)$ is not
finitely generated.

%The K\"unneth formula works for topological cohomology where $X$ and $X'$ are
%treated as topological spaces.  However, the above K\"unneth formula  does not
%apply to Boral-group cohomology where $X$ and $X'$ are treated as groups.

As the first application of K\"unneth formula, we like to use it to calculate
$H^*(X',\M)$ from $H^*(X',\Z)$,  by choosing $R=\M'=\Z$. In this case, the
condition $\M\boxtimes_R\M'=\M\boxtimes_\Z \Z=0$ is always
satisfied. So we have
\begin{align}
\label{kunnZ}
&\ \ \ \ H^d(X\times X',\M)
\nonumber\\
&\simeq \Big[\oplus_{k=0}^d H^k(X,\M)\otimes_{\Z} H^{d-k}(X',\Z)\Big]\oplus
\nonumber\\
&\ \ \ \ \ \
\Big[\oplus_{k=0}^{d+1}
H^k(X,{\M})\boxtimes_\Z H^{d-k+1}(X',\Z)\Big]  .
\end{align}
The above is valid for topological cohomology.
It is also valid for group  cohomology:
\begin{align}
\label{kunnZG1}
&\ \ \ \ \cH^d(G\times G',\M)
\nonumber\\
&\simeq \Big[\oplus_{k=0}^d \cH^k(G,\M)\otimes_{\Z} \cH^{d-k}(G',\Z)\Big]\oplus
\nonumber\\
&\ \ \ \ \ \
\Big[\oplus_{k=0}^{d+1}
\cH^k(G,\M)\boxtimes_\Z \cH^{d-k+1}(G',\Z)\Big]  .
\end{align}
provided that $G'$ is a finite group.
Using \eqn{HdR}, we can rewrite the above as
\begin{align}
\label{kunnZG2}
&\ \ \ \ \cH^d(G\times G',\M)
\simeq
\cH^d(G,\M)\oplus
\nonumber\\
&\ \ \ \ \ \
\Big[\oplus_{k=0}^{d-2} \cH^k(G,\M)\otimes_{\Z} \cH^{d-k-1}(G',\RZ)\Big]\oplus
\nonumber\\
&\ \ \ \ \ \
\Big[\oplus_{k=0}^{d-1}
\cH^k(G,\M)\boxtimes_\Z \cH^{d-k}(G',\RZ)\Big]  ,
\end{align}
where we have used
\begin{align}
 \cH^1(G',\Z)=0.
\end{align}
If we further choose $\M=\RZ$, we obtain
\begin{align}
\label{kunnZG}
&\ \ \ \ \cH^d(G\times G',\RZ)
\nonumber\\
&\simeq
\cH^d(G,\RZ)\oplus
\cH^d(G',\RZ)\oplus
\nonumber\\
&\ \ \ \ \ \
\Big[\oplus_{k=1}^{d-2} \cH^k(G,\RZ)\otimes_{\Z} \cH^{d-k-1}(G',\RZ)\Big]\oplus
\nonumber\\
&\ \ \ \ \ \
\Big[\oplus_{k=1}^{d-1}
\cH^k(G,\RZ)\boxtimes_\Z \cH^{d-k}(G',\RZ)\Big]  ,
\end{align}
where $G'$ is a finite group.

We can further choose $X$ to be the space of one point (or the trivial group of
one element) in \eqn{kunnZ} or \eqn{kunnZG1}, and use
\begin{align}
H^{d}(X,\M))=
\begin{cases}
\M, & \text{ if } d=0,\\
0, & \text{ if } d>0,
\end{cases}
\end{align}
to reduce \eqn{kunnZ} to
\begin{align}
\label{ucf}
 H^d(X,\M)
&\simeq  \M \otimes_{\Z} H^d(X,\Z)
\oplus
\M\boxtimes_\Z H^{d+1}(X,\Z)  .
\end{align}
where $X'$ is renamed as $X$.  The above is a form of the universal coefficient
theorem which can be used to calculate $H^*(X,\M)$ from $H^*(X,\Z)$ and the
module $\M$.  The  universal coefficient theorem works for topological
cohomology where $X$ is a topological space.  The  universal coefficient
theorem also works for group cohomology where $X$ is a finite group.

Using the universal
coefficient theorem, we can rewrite \eqn{kunnZ} as
\begin{align}
\label{kunnH}
H^d(X\times X',\M) \simeq \oplus_{k=0}^d H^k[X, H^{d-k}(X',\M)] .
\end{align}
The above is valid for topological cohomology.
It is also valid for group  cohomology:
\begin{align}
\label{kunnG}
\cH^d(G\times G',\M) \simeq \oplus_{k=0}^d \cH^k[G, \cH^{d-k}(G',\M)] ,
\end{align}
provided that both $G$ and $G'$ are finite groups.

We may apply the above to the classifying
spaces of group $G$ and $G'$. Using
$B(G\times G')=BG\times BG'$, we find
\begin{align*}
%\label{kunnHBG}
H^d[B(G\times G'),\M] \simeq \oplus_{k=0}^d H^k[BG, H^{d-k}(BG',\M)] .
\end{align*}
Choosing $\M=\RZ$ and using \eqn{HdR}, we have
\begin{align}
\label{kunnGGp}
&\ \ \ \
 \cH_B^d(G\times G',\RZ)=
 H^{d+1}[B(G\times G'),\Z]
\nonumber\\
&= \oplus_{k=0}^{d+1} H^k[BG, H^{d+1-k}(BG',\Z)]
\nonumber\\
&=\cH_B^d(G,\RZ)\oplus \cH_B^d(G',\RZ)\oplus
\nonumber\\
&\ \ \ \ \ \ \ \
\oplus_{k=1}^{d-1} H^k[BG, \cH_B^{d-k}(G',\RZ)]
\end{align}
where we have used $H^1(BG',\Z)=0$.
Using
\begin{align}
\label{HdcHdZ}
 H^d(BG,\Z)=\cH_B^d(G,\Z),\ \ \ \
 H^d(BG,\Z_n)=\cH_B^d(G,\Z_n),
\end{align}
we can rewrite the above as
\begin{align}
\label{kunnU}
\cH^d(GG & \times SG,\RZ) = \oplus_{k=0}^{d} \cH^{k}[SG,\cH^{d-k}(GG,\RZ)]
\nonumber\\
 &= \oplus_{k=0}^{d} \cH^{k}[GG,\cH^{d-k}(SG,\RZ)]
.
\end{align}
Eqn.  \ref{kunnU} is valid for any groups $GG$ and $SG$.

Choosing $X=BG$, $\M=\Z_n$, \eqn{ucf} becomes
\begin{align}
\label{ucfG}
 \cH^d(G,\Z_n)
&\simeq  \Z_n \otimes_{\Z} \cH^d(G,\Z)
\oplus
\Z_n\boxtimes_\Z \cH^{d+1}(G,\Z)  ,
\end{align}
where we have used \eqn{HdcHdZ}.
Eq. \ref{ucfG} is valid for any compact group $G$.
Using \eqn{ucfG}, we find that
\begin{align}
\label{ucfGG}
 \cH^d[G,&\cH^{d'}(G',\R/\Z)]
\simeq  \cH^{d'}(G',\R/\Z) \otimes_{\Z} \cH^{d-1}(G,\R/\Z)
\oplus
\nonumber\\
&
\cH^{d'}(G',\R/\Z)\boxtimes_\Z \cH^{d}(G,\R/\Z)  ,
\end{align}
Using \eqn{ucfG}, we also find that
\begin{align}
\label{ZmZn}
\cH^{d}(Z_m,\Z_n)=
\begin{cases}
\Z_n, & \text{ if } d=0,\\
\Z_{\<m,n\>}.  , & \text{ if } d>0,
\end{cases}
\end{align}
and
\begin{align}
\label{U1Zn}
\cH^{d}(U(1),\Z_n)=
\begin{cases}
\Z_n, & \text{ if } d=\text{even},\\
0, & \text{ if } d=\text{odd}.
\end{cases}
\end{align}

\section{Lyndon-Hochschild-Serre spectral sequence}
\label{LHS}

The Lyndon-Hochschild-Serre spectral sequence (see \Ref{L4871} page 280,291,
and \Ref{HS5310}) allows us to understand the structure of $\cH^d(GG\gext
SG,\RZ)$ to a certain degree. (Here $GG\gext SG\equiv PSG$ is a group extension
of $SG$ by $GG$: $SG=PSG/GG$.) We find that $\cH^d(GG\gext SG,\RZ)$, when
viewed as an Abelian group, contains a chain of subgroups
\begin{align}
\{0\}=H_{d+1}
\subset H_d
\subset \cdots
\subset H_0
=
 \cH^d(GG\gext SG,\RZ)
\end{align}
such that $H_k/H_{k+1}$ is a subgroup of a factor
group of $\cH^k[SG,\cH^{d-k}(GG,\RZ)_{SG}]$,
\ie $\cH^k[SG,\cH^{d-k}(GG,\RZ)_{SG}]$
contains a   subgroup $\Ga^k$, such that
\begin{align}
 H_k/H_{k+1} &\subset \cH^k[SG,\cH^{d-k}(GG,\RZ)_{SG}]/\Ga^k,
\nonumber\\
k&=0,\cdots,d.
\end{align}
Note that
$SG$ may have a non-trivial action on $\cH^{d-k}(GG,\RZ)$ as determined by the
structure $1\to GG \to GG\gext SG \to SG \to 1$.
We add the subscript $SG$ to  $\cH^{d-k}(GG,\RZ)$ to stress this point.
  We also have
\begin{align}
 H_0/H_{1} &\subset \cH^0[SG,\cH^{d}(GG,\RZ)_{SG}],
\nonumber\\
 H_d/H_{d+1}&=H_d = \cH^d(SG,\RZ)/\Ga^d.
\end{align}
In other words, all the elements in $\cH^d(GG\gext SG,\RZ)$ can be one-to-one labeled
by $(x_0,x_1,\cdots,x_d)$ with
\begin{align}
 x_k\in H_k/H_{k+1} \subset \cH^k[SG,\cH^{d-k}(GG,\RZ)_{SG}]/\Ga^k.
\end{align}

The above discussion implies that we can also use $(m_0,m_1,\cdots,m_d)$ with
\begin{align}
 m_k\in \cH^k[SG,\cH^{d-k}(GG,\RZ)_{SG}]
\end{align}
to  label all the elements in $\cH^d(G,\RZ)$. However, such a labeling scheme
may not be one-to-one, and it may happen that only some of
$(m_0,m_1,\cdots,m_d)$ correspond to  the  elements in $\cH^d(G,\RZ)$.  But, on
the other hand, for every element in $\cH^d(G,\RZ)$, we can find a
$(m_0,m_1,\cdots,m_d)$ that corresponds to it.

%For the special case $GG\times SG$, $(m_0,m_1,\cdots,m_d)$ will give us
%an one-to-one labeling of the elements in $\cH^d(GG\times SG,\RZ)$. In fact
%(see \eqn{kunnG})
%\begin{align}
%\label{kunnU}
%\cH^d(GG & \times SG,\RZ) = \oplus_{k=0}^{d} \cH^{k}[SG,\cH^{d-k}(GG,\RZ)]
%\nonumber\\
% &= \oplus_{k=0}^{d} \cH^{k}[GG,\cH^{d-k}(SG,\RZ)]
%.
%\end{align}
%Now, $G$ and $G'$ do not have to be finite groups.

\section{A duality relation between the SPT and the SET phases}
\label{dual}

There is a duality relation between the SPT and the SET phases described by
weak-coupling gauge field.\cite{LG1220,MR1235,HW1227} We first review a simple
formal description of such a  duality relation.  Then we will review an exact
description for finite gauge groups.

\subsection{A simple formal description}

To understand the duality between the SPT and the SET phases, we note that
a SPT state with symmetry $G$ in $d$-dimensional space-time $M$ can be
described by a non-linear $ \sigma $-model with $G$ as the target space
\begin{align}
 S= \int_M \dd^d x\ \Big[\frac{1}{ \lambda _s} [\prt g(x^\mu)]^2
+ \ii W_\text{top}(g) \Big].
\end{align}
in large $ \lambda _s$ limit.  Here we triangulate the $d$-dimensional
space-time manifold $M$ to make it a lattice or a $d$-dimensional complex, and
$g(x^\mu)$ live on the vertices of the complex: $g(x^\mu) = \{g_i\}$ where $i$
labels the vertices (the lattice sites).  So $\int \dd^d x$ is in fact a sum
over lattice sites and $ \partial $ is the lattice difference operator.  The
above action $S$ actually defines a lattice theory.  $ W_\text{top}[g(x^\mu)]$
is a lattice topological term which satisfy
\begin{align}
\label{cocyeq}
  \int_M \dd^d x\  W_\text{top}(\{g_i\}) &=
  \int_M \dd^d x\  W_\text{top}(\{g g_i\})
\in \R, \  g, g_i\in G,
\nonumber\\
  \int_M \dd^d x\  W_\text{top}[g(x^\mu)] &= 0 \text{ mod } 2\pi,
\text{ if $M$ has no boundary.}
\end{align}
We have rewritten $W_\text{top}[g(x^\mu)]$ as $W_\text{top}(\{g_i\})$ to stress
that the topological term is defined on lattice.  $W_\text{top}(\{g_i\})$
satisfying \eq{cocyeq} are the group cocycles.  Thus the lattice topological term
$W_\text{top}(\{g_i\})$ is defined and described by the elements (the
cocycles) in $\cH^d(G,\R/\Z)$.\cite{CGL1172,CGL1204} This is why the bosonic
SPT states are described by $\cH^d(G,\R/\Z)$.

If $G$ contains a normal subgroup $GG \subset G$,
we can ``gauge'' $GG$  to obtain a gauge theory in the bulk
\begin{align}
\label{Sset}
 S= \int \dd^d x\ \Big[\frac{[(\prt-\ii A) g]^2}{ \lambda _s} +
\frac{\Tr(F_{\mu\nu})^2}{ \lambda }
+ \ii W^\text{gauge}_\text{top}(g,A) \Big],
\end{align}
where $A$ is the $GG$ gauge potential. When $\la$ is small the above theory
is a weak-coupling gauge theory with a gauge group $GG$ and a global symmetry
group $SG=G/GG$.

The topological term $ W^\text{gauge}_\text{top}(g,A)$ in the gauge theory is a
generalization of the Chern-Simons term,\cite{DW9093,HW1267,HWW1295} which is obtained
by ``gauging'' the topological term $ W_\text{top}(g)$ in the non-linear $
\sigma $-model. The two topological terms $W^\text{gauge}_\text{top}(g,A)$ and
$W_\text{top}(g)$ are directly related when $A$ is a pure gauge:
\begin{align}
&W^\text{gauge}_\text{top}(g,A)= W_\text{top}[h(x)g(x)],
\nonumber\\
&\text{ where } A=h^{-1} \partial  h, \ h\in GG.
\end{align}
(A more detailed description of the two topological terms $W_\text{top}(g)$ and
$W^\text{gauge}_\text{top}(g,A)$ on lattice can be found in
\Ref{HW1267,HWW1295}. See also the next section.) So the topological term $
W^\text{gauge}_\text{top}(g,A)$ in the gauge theory is also classified by  same
$\cH^d(G,\R/\Z)$ that classifies $W_\text{top}(g)$.  (We like to remark that
although both  topological terms $W_\text{top}(g)$ and
$W^\text{gauge}_\text{top}(A)$ are classified by the same $\cH^d(G,\R/\Z)$,
when $\cH^d(G,\R/\Z)=\Z$, the correspondence can be tricky: for a  topological
term $W_\text{top}(g)$ that corresponds to an integer $k$ in $\cH^d(G,\R/\Z)$,
its corresponding topological term $W^\text{gauge}_\text{top}(g,A)$ may
correspond to an integer $nk$ in $\cH^d(G,\R/\Z)$.  However, for finite group
$G$, the correspondence is one-to-one.)

When the space-time dimensions $d=3$ or when $d>3$ and $GG$ is a finite group,
the theory \eq{Sset} is gapped in $\la_s \to \infty$ and $\la \to 0$ limit,
which describe a SET phase with symmetry group $SG$ and gauge group $GG$.  Such
SET phase are described by  $\cH^d(G,\R/\Z)$.

\subsection{Exactly soluble gauge theory with a finite gauge group
$GG$ and a global symmetry group $SG$}
\label{dispath}

To understand the above formal results more rigorously, we would like to review
the exactly soluble models of weak-coupling gauge theories with a \emph{finite}
gauge group $GG$ and a global symmetry group $SG$.  The exactly soluble models
were introduced in \Ref{LG1220,MR9162,ZW,HW1227}.  The  exactly soluble models is
defined on a space-time lattice, or more precisely, a triangulation of the
space-time.  So we will start by describing such a triangulation.

\subsubsection{Discretize space-time}
\label{disltgauge}

\begin{figure}[tb]
\begin{center}
\includegraphics[scale=0.6]{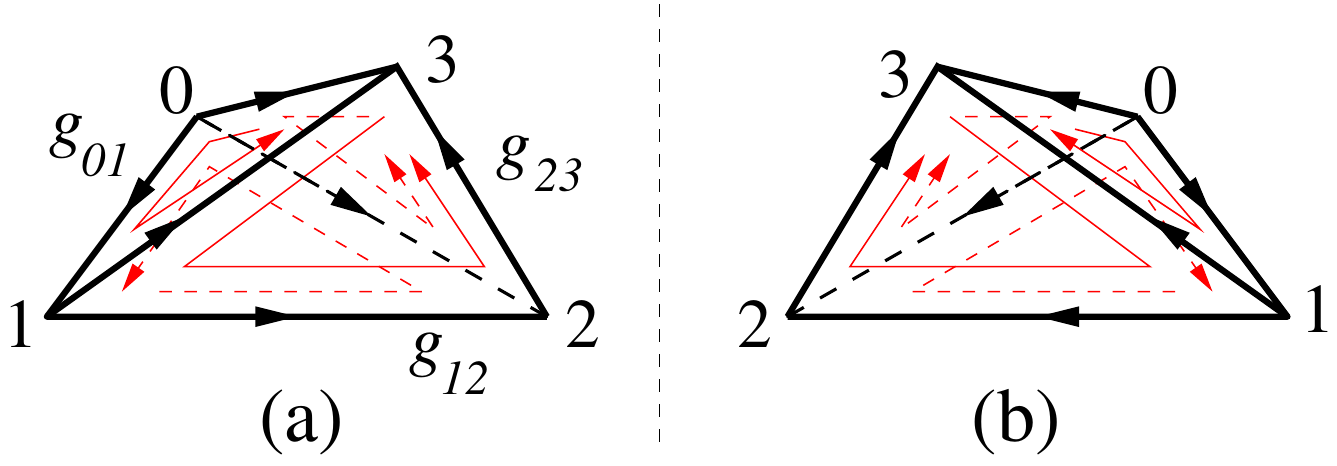} \end{center}
%Fig. 6
\caption{ (Color online) Two branched simplices with opposite orientations.
(a) A branched simplex with positive orientation and (b) a branched simplex
with negative orientation.  }
\label{mir}
\end{figure}

Let $M_\text{tri}$ be a triangulation of the $d$-dimensional space-time.  We
will call the triangulation $M_\text{tri}$ as a space-time complex, and a cell
in the complex as a simplex.  In order to define a generic lattice theory on
the space-time complex $M_\text{tri}$, it is important to give the  vertices of
each simplex a local order.  A nice local scheme to order  the  vertices is
given by a branching structure.\cite{C0527,CGL1172,CGL1204} A branching
structure is a choice of orientation of each edge in the $d$-dimensional
complex so that there is no oriented loop on any triangle (see Fig. \ref{mir}).

The branching structure induces a \emph{local order} of the vertices on each
simplex.  The first vertex of a simplex is the vertex with no incoming edges,
and the second vertex is the vertex with only one incoming edge, \etc.  So the
simplex in  Fig. \ref{mir}a has the following vertex ordering: $0,1,2,3$.

The branching structure also gives the simplex (and its sub simplexes) an
orientation denoted by $s_{ij \cdots k}=\pm 1$.  Fig. \ref{mir} illustrates two
$3$-simplices with opposite orientations $s_{0123}=1$ and $s_{0123}=*$.  The
red arrows indicate the orientations of the $2$-simplices which are the
subsimplices of the $3$-simplices.  The black arrows on the edges indicate the
orientations of the $1$-simplices.

\subsubsection{Lattice gauge theory with a global symmetry}

To define a lattice gauge theory with a gauge group $GG$ and  a global
symmetry group $SG$, let $G$ be an extension
of $SG$ by $GG$: $G = GG \gext SG$.
Here we will assume $GG$ to be a finite group.

In our lattice gauge theory, the degrees of freedom on the vertices of the
space-time complex, is described by $g_i \in G$ where $i$ labels the vertices.
The gauge degrees of freedom are on the edges $ij$ which are described by
$h_{ij}\in GG$.

The action amplitude $\ee^{-S_\text{cell}}$ for a $d$-cell $(ij \cdots k)$ is
complex function of $g_i$ and $h_{ij}$: $V_{ij \cdots k}(\{h_{ij}\},\{g_i\})$.
The total action amplitude $\ee^{-S}$ for configuration (or a path) is given by
\begin{align}
\label{eS}
\ee^{-S}=
\prod_{(ij \cdots k)} [V_{ij \cdots k}(\{h_{ij}\},\{g_i\})]^{s_{ij \cdots k}}
\end{align}
where $\prod_{(ij \cdots k)}$ is the product over all the $d$-cells $(ij \cdots k)$.
Note that the contribution from a $d$-cell $(ij \cdots k)$ is
$V_{ij \cdots k}(\{h_{ij}\},\{g_i\})$ or $V^*_{ij \cdots k}(\{h_{ij}\},\{g_i\})$
depending on the orientation $s_{ij \cdots k}$ of the cell.
Our lattice theory is defined
by following imaginary-time path integral (or partition function)
\begin{align}
 Z_\text{gauge}=\sum_{ \{h_{ij}\},\{g_i\} }
\prod_{(ij \cdots k)} [V_{ij \cdots k}(\{h_{ij}\},\{g_i\})]^{s_{ij \cdots k}}
\end{align}
If the above action amplitude  $\prod_{(ij \cdots k)} [V_{ij \cdots k}(\{h_{ij}\},\{g_i\})]^{s_{ij \cdots k}}$ on closed  space-time complex ($\prt
M_\text{tri}=\emptyset $) is invariant under the gauge transformation
\begin{align}
\label{gaugeT}
 h_{ij} \to g'_{ij}=h_i h_{ij} h_j^{-1}, g_i \to g'_i =h_ig_i\ \ \ h_i \in GG
\end{align}
then the action amplitude $V_{ij \cdots k}(\{h_{ij}\},\{g_i\})$ defines a gauge
theory of gauge group $GG$.  If the  action amplitude is invariant under the
global transformation
\begin{align}
\label{symmT}
 h_{ij} \to h'_{ij}=g h_{ij} g^{-1}, g_i \to g'_i =gg_i\ \ \ g \in G ,
\end{align}
then the action amplitude $V_{ij \cdots k}(\{h_{ij}\},\{g_i\})$ defines a $GG$
lattice gauge theory with a global symmetry $SG=G/GG$.  (We need to mod out
$GG$ since when $h\in GG$, it is a part of  gauge transformation which does not
change the physical states, instead of a global symmetry transformation which
change a physical state to another one.)

However, in this paper, we are mainly considering a system with a global
symmetry $G$, where we gauged a subgroup $GG\subset G$.  The resulting gauge
connection $h_{ij}$ is treated as non-dynamical probe fields.
Such a system

Using a cocycle $\nu_d( g_0, g_1, \cdots ,g_d) \in \cH^d(G,\R/\Z),\ g_i\in G$
[where $\nu_d( g_0, g_1, \cdots ,g_d)$ is a real function over $G^{d+1}$], we
can construct an  action amplitude $V_{ij \cdots k}(\{h_{ij}\},\{g_i\})$ that
define a gauge theory with gauge group $SG$ and global symmetry $SG$.  The
gauge theory action amplitude is obtained from $\nu_d( g_0, g_1, \cdots ,g_d)$
as
\begin{align}
\label{Vnud}
& V_{01 \cdots d}(\{h_{ij}\},\{g_i\})=0,
\text{ if } h_{ij}h_{jk}\neq h_{ik}
\\
& V_{01 \cdots d}(\{h_{ij}\},\{g_i\})=
\ee^{2\pi \ii \nu_d( h_0 g_0, h_1 g_1, \cdots , h_d g_d)},
\nonumber\\ &
=
\ee^{2\pi \ii \om_d(g_0^{-1}h_{01} g_1, \cdots, g_{d-1}^{-1}h_{d-1,d}g_d )},
\text{ if } h_{ij}h_{jk}= h_{ik},
\nonumber
\end{align}
where $h_i$ are given by
\begin{align}
 h_0 &=1, & h_1 &=h_0 h_{01},
&
 h_2 &=h_1 h_{12}, & h_3 &=h_2 h_{23},  \cdots
\end{align}
and $\om_d$ is the nonhomogenous cocycle that corresponds to $\nu_d$
\begin{align}
 \om_d(h_{01},h_{12},\cdots,h_{d-1,d})=\nu_d(h_0,h_1,\cdots,h_d) .
\end{align}

To see the above action amplitude defines a $GG$ lattice gauge
theory with a global symmetry $SG$, we
note that the cocycle satisfies the cocycle condition
\begin{align}
&
\nu_d( g_0, g_1, \cdots ,g_d)=\nu_d( gg_0, gg_1, \cdots ,gg_d)\text{ mod } 1, \ \ \ g\in G
\nonumber\\
&
\sum_i \nu_d( g_0, \cdots , \hat g_i , \cdots ,g_{d+1})=0\text{ mod }1
\end{align}
where $g_0, \cdots , \hat g_i , \cdots ,g_{d+1}$ is the sequence $g_0, \cdots ,
g_i , \cdots ,g_{d+1}$ with $g_i$ removed.  Using such a property, one can
check that the above action amplitude $V_{01 \cdots d}(\{h_{ij}\},\{g_i\})$ is
invariant under  the global symmetry transformation \eq{symmT}.  We can also
rewrite the partition function as (see \eqn{Vnud})
\begin{align}
 Z=\sum_{ \{h_{ij}\},\{g_i\} }
\prod_{(ij \cdots k)} [V_{ij \cdots k}(\{g_i^{-1}h_{ij}g_j\},\{1\})]^{s_{ij \cdots k}}
\end{align}
which is explicitly gauge invariant.  Thus it defines a symmetric gauge theory
with a gauge group $GG$ and a global symmetry group $SG$.

We note that the action amplitude is non-zero only when $h_{ij}h_{jk}= h_{ik}$
or $h_{ij}h_{jk}h^{-1}_{ik}=1$.  The condition $h_{ij}h_{jk}h^{-1}_{ik}\equiv
\ee^{\ii \text{``gauge flux''}}=1$ is the zero-flux condition on the triangle
$(ijk)$ or the flat connection condition. The corresponding gauge theory is in
the weak-coupling limit (actually is at the zero-coupling).  This condition can
be implemented precisely only when $GG$ is finite.  With  the flat connection
condition $h_{ij}h_{jk}= h_{ik}$, $h_i$'s and the gauge equivalent sets of
$h_{ij}$ have an one-to-one correspondence.

Since the total action amplitude  $\prod_{(ij \cdots k)}
[V_{ij \cdots k}(\{h_{ij}\},\{g_i\})]^{s_{ij \cdots k}}$ on a sphere is always equal to 1
if the gauge flux vanishes, therefore $V_{ij \cdots k}(\{h_{ij}\},\{g_i\})$
describes a quantized topological term in weak-coupling gauge theory (or
zero-coupling gauge theory).  This way, we show that \emph{a quantized
topological term in a weak-coupling gauge theory with gauge group $GG$ and
symmetry group $SG$ can be constructed from each element of $\cH^d(G,\R/\Z)$}.

\begin{figure}[tb]
\begin{center}
\includegraphics[scale=0.5]{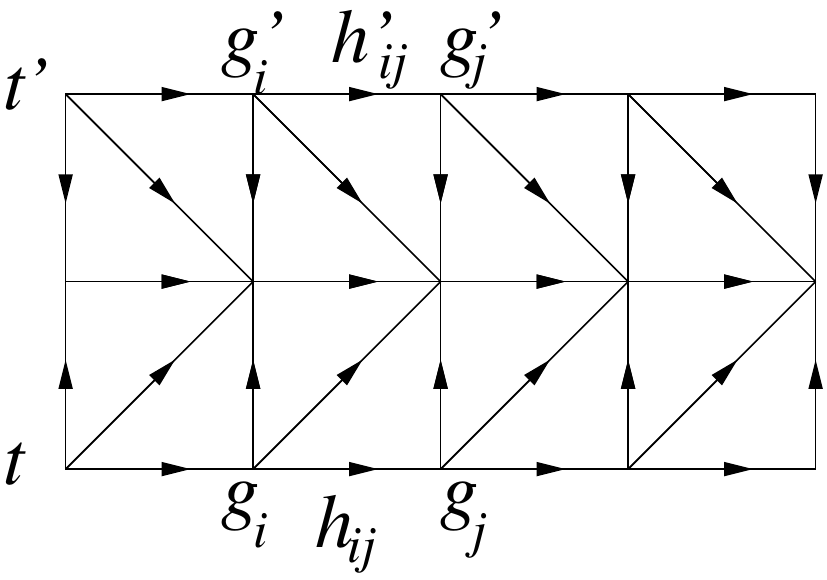} \end{center}
%Fig. 7
\caption{
Each time-step of evolution is given by the path integral on a particular form
of branched graph.  Here is an example in 1+1D. In SPT states, the gauge
connection $h_{ij}$ on the links is a non-dynamical probe field. In this case,
the gauge connection $h_{ij}=1$ on the time-links (\ie the vertical links).  In
SET states, the gauge connection $h_{ij}$ on the links is a dynamical field. In
this case, the gauge connection $h_{ij}$ can be non-trivial on any links.
}
\label{tStep}
\end{figure}

\begin{figure}[tb]
\begin{center}
\includegraphics[scale=0.45]{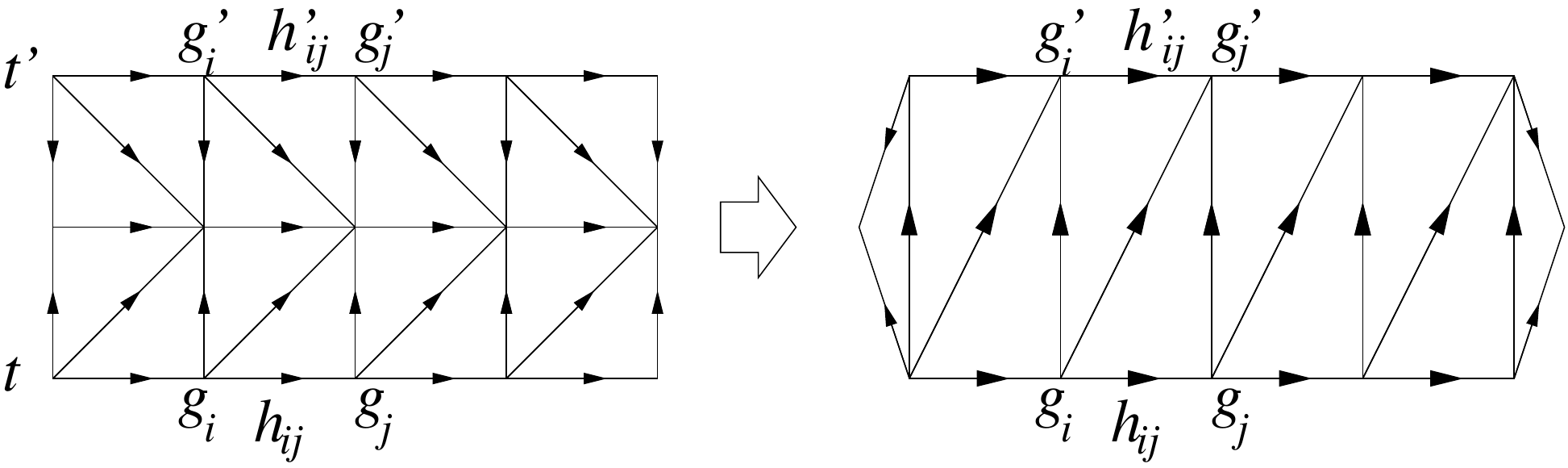} \end{center}
%Fig. 8
\caption{
The reduction of double-layer time-step to single-layer time-step on space with
boundary in an 1+1D example.
}
\label{stStep}
\end{figure}

\subsubsection{From path integral to Hamiltonian}
\label{pathham}

A path integral can give us an amplitude $Z[\{ g_i', h_{ij}' \}, \{ g_i,
h_{ij}\}]$ for a configuration $\{ g_i, h_{ij}\}$ at $t$ to another
configuration $\{ g_i', h_{ij}' \}$ at $t'$.  We like to interpret $Z[\{ g_i',
h_{ij}' \}, \{ g_i, h_{ij}\}]$ as the amplitude of an evolution in imaginary
time by a Hamiltonian:
\begin{align}
 Z[\{ g_i', h_{ij}' \}, \{ g_i, h_{ij}\}] =\< g_i', h_{ij}' | \ee^{-(t'-t)H} |
g_i, h_{ij} \> .
\end{align}
However, such an interpretation may not be valid since $ Z[\{ g_i', h_{ij}' \},
\{ g_i, h_{ij}\}]$ may not give raise to a Hermitian matrix.  It is a worrisome
realization that path integral and Hamiltonian evolution may not be directly
related.

Here we would like to use the fact that the path integral that we are
considering are defined on the branched graphs with a ``reflection'' property
(see \eq{eS}). We like to show that such path integral are better related
Hamiltonian evolution.  The key is to require that each time-step of evolution
is given by  branched graphs of the form in Fig. \ref{tStep}.  One can show
that $Z[\{ g_i', h_{ij}' \}, \{ g_i, h_{ij}\}]$ obtained by summing over all in
the internal indices in the  branched graphs Fig. \ref{tStep}
has a form
\begin{align}
&\ \ \ \
 Z[\{ g_i', h_{ij}' \}, \{ g_i, h_{ij}\}]
\\
&=\sum_{\{ g_i'', h_{ij}'' \}}
 U^*[\{ g_i'', h_{ij}'' \}, \{ g_i', h_{ij}'\}]
\
 U[\{ g_i'', h_{ij}'' \}, \{ g_i, h_{ij}\}]
\nonumber
\end{align}
and represents a positive-definite Hermitian matrix.  Thus the path integral of
the form \eq{eS} always correspond to a Hamiltonian evolution in imaginary time.
In fact, the above $Z[\{ g_i', h_{ij}' \}, \{ g_i, h_{ij}\}]$ can be viewed as
an imaginary-time evolution $T=\ee^{-\Del \tau H}$ for a single time step.

For most cases studied in this paper, $h_{ij}$ is a static probe field. In
those case, $h_{ij}$ are the same on all the time slices and $h_{ij}=1$ on the
vertical time links.  In this case, $Z[\{ g_i', h_{ij} \}, \{ g_i, h_{ij}\}]$
(with fixed $h_{ij}$) can still be viewed as an imaginary-time evolution
$T=\ee^{-\Del \tau H}$ for a single time step, where only $g_i$'s are dynamical.

For the ideal path integrals with the action-amplitudes described by the
cocycles, we can reduce the double-layer time-step to a single-layer time-step,
using the retriangulation invariance of the action-amplitudes if the space has
no boundary.  If the space does have boundary, we can still reduce the
double-layer time-step to a single-layer time-step, but with some extra terms
on the boundary (see Fig. \ref{stStep}).

\section{Physical properties of defects in 2+1D $Z_n$ SPT states}

If we view $h_{ij}$  in the last section as a static probe field, then the
formalism developed in the last section can be viewed as the path integral
description of SPT states with possible monodromy defects or other possible
twists described the ``gauge configuration'' $h_{ij}$ on the links.  In this
section, we are going to use such a formalism to study the physical properties
of defects in SPT states.

\subsection{Symmetry transformations and their non-factorization}

\begin{figure}[tb]
\begin{center}
\includegraphics[scale=0.35]{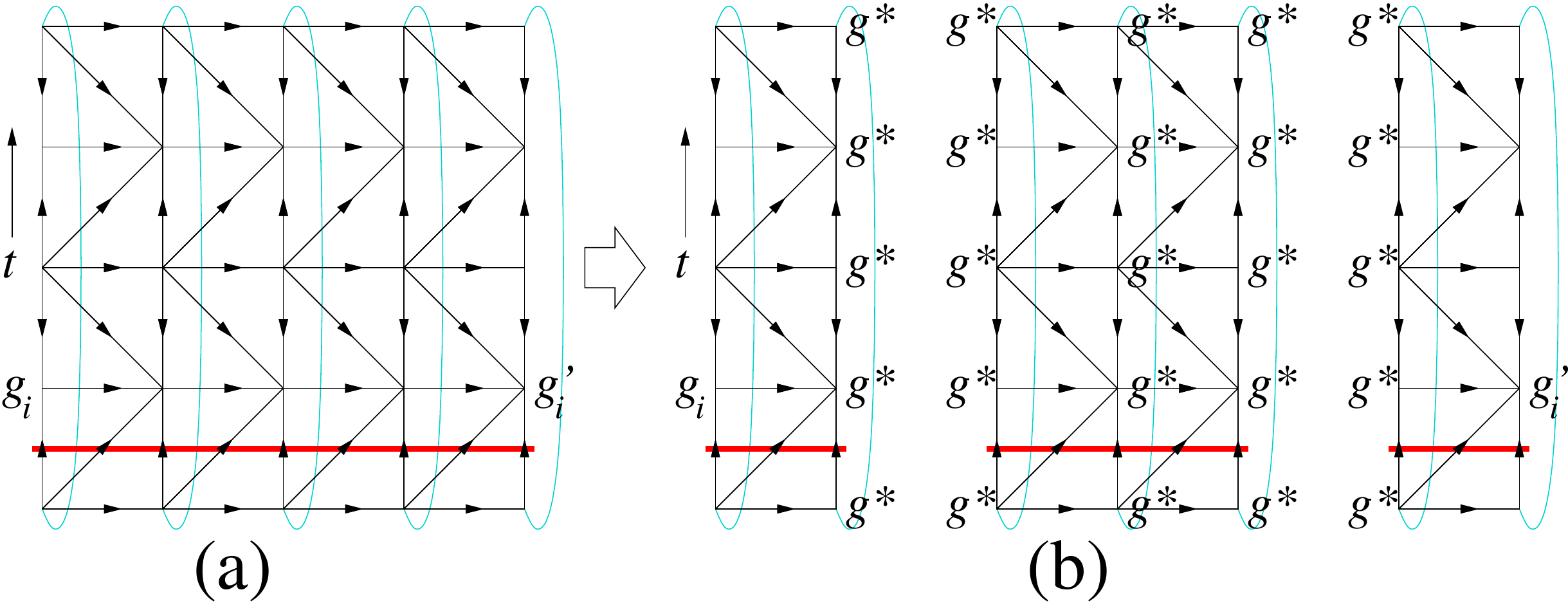} \end{center}
%Fig. 9
\caption{
(a) The trace $\Tr( T^NW_g)$ can be represented by a graph which is periodic in
time direction, with one layer of vertical time links with $h_{ij}=g$, in an
1+1D example.  Those vertical time links are marked by red-line crossing them.
(a $\to$ b) We can use the retriangulation invariance of the action-amplitudes
to set all the internal $g_i$ to a fixed $g^*$ without changing the
action-amplitude.  (b) For fixed $g^*$, we can rewrite one graph as three
graphs, where the middle graph just represents a phase factor.
}
\label{trace}
\end{figure}

First let us examine how symmetry transformations act on the defects.  Consider
a system with symmetry $G$.  The evolution operator $T=\ee^{-\tau H}$ satisfies
\begin{align}
 W_g TW_g^{-1} =T, \ \ \ \ g\in G
\end{align}
where $ W_g$ is a represent of the symmetry.  We like to examine
%use the decomposition  Fig. \ref{stStep} to study
the amplitude of the evolution from a configuration $\{ g_i, h_{ij}\}$ to its
symmetry $g$ transformed  configuration $\{ gg_i, h_{ij}\}$ (where we have
assumed that $gh_{ij}g^{-1}=h_{ij}$.) Or more precisely, we want to examine the
trace $\Tr( T^NW_g)$.  Such a trace can be expressed as a graph which is
periodic in time direction, with one layer
of vertical time links given by $h_{ij}=g$,
while other layers of vertical time links by $h_{ij}=1$ (see Fig. \ref{trace}).

For the ideal path integrals with the action-amplitudes described by the
cocycles, the action-amplitudes only depend on the $g_i$'s on the boundary.
(Here we assume that $h_{ij}$'s are fixed non-dynamical probe fields.  We can
use the retriangulation invariance of the action-amplitudes to set all the
internal $g_i$ to a fixed $g^*$ without changing the  action-amplitude.
(Usually, we may take $g^*=1$.) Thus the trace $\Tr( T^NW_g)$ can be
represented by the three graphs in Fig. \ref{trace}(b).  We see that the trace
$\Tr( T^NW_g)$ factorizes into independent boundary terms (one for each
boundary) and the non-dynamical bulk phase factor:
\begin{align}
\label{UnWg}
 &\Tr( T^NW_g)=
\\
&
 \Tr U^\text{bulk}_g
 \Tr( T_\text{bndry,1}^N W^\text{bndry,1}_g)
 \Tr( T_\text{bndry,2}^N W^\text{bndry,2}_g)
\nonumber
\end{align}
Note that $U^\text{bulk}_g$ is $1\times 1$ matrix described by the middle graph
in Fig. \ref{trace}(b).  In fact $U^\text{bulk}_g$ is a pure phase factor.
$T_\text{bndry,1}^N$ and $T_\text{bndry,2}^N$ describe the dynamic time
evolution on the two boundaries, which are independent of each other.  We see
that to total symmetry transformation $W_g$ has a form
\begin{align}
\label{Wgfact}
 W_g = U^\text{bulk}_g\  W^\text{bndry,1}_g \otimes W^\text{bndry,2}_g
\end{align}
If the phase factor $U^\text{bulk}_g=1$, then the total symmetry transformation
factorize on the two independent boundaries.  However, if $U^\text{bulk}_g$
form a non-trivial 1D representation of $G$, the total $G$ quantum number will
the sum of the  $G$ quantum numbers on each boundaries plus a shift generated
by $U^\text{bulk}_g$.  In this case, the total symmetry transformations do not
cleanly factorize into independent boundary terms.  It contains an extra phase
factor $U^\text{bulk}_g$.

\subsection{The low energy effective theory and low energy effective
symmetry at the monodromy defects}
\label{Zncharge}

%\begin{figure}[tb]
%\begin{center}
%\includegraphics[scale=0.8]{trans} \end{center}
%%Fig. 2
%\caption{ (Color online)
%The complex is formed by three tetrahedrons: $(0122')$, $(00'1'2')$, and
%$(011'2')$.
%The triangles
%$(012)$ and
%$(0'1'2')$ are on two time slices.
%}
%\label{trans}
%\end{figure}

\begin{figure}[tb]
\begin{center}
\includegraphics[scale=1.1]{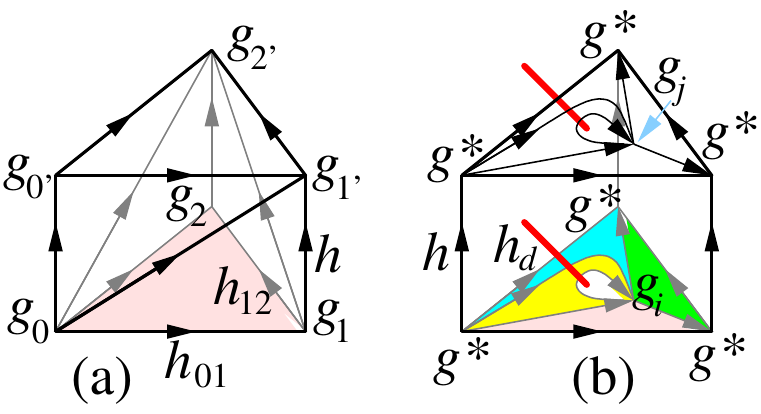} \end{center}
%Fig. 10
\caption{
(Color online) (a) The action amplitude on the complex in (a), for an ideal
fixed point action, is given by $C( g_0,g_1,g_2; g_{0'},g_{1'},g_{2'};
h_{01},h_{02},h)$.  The complex is formed by three tetrahedrons: $(0122')$,
$(00'1'2')$, and $(011'2')$.  The triangles $(012)$ and $(0'1'2')$ are on two
time slices.
(b) A defect is described by the dynamical variables
$g_i,g_j$.  When $h_{kl}$ on the links crossed by the red lines are non-trivial
($h_{kl}=h_d\neq 1$), the defect is a monodromy defect.  When $h=g$ on the
vertical links (the time links), it describes the insertion $W_g$ in the path
integral.  Note that $g_i,g_j$ are surrounded by $g^*$'s and the complex (b) is
formed by four complices of the type in (a), represented by the four colors of
the base triangles.
}
\label{defect}
\end{figure}

In this section, we are going to apply the formalism developed in the last
section to study the low energy effective theory and low energy effective
symmetry at the monodromy defects in 2+1D $Z_n$ SPT states.  The monodromy
defects are created by a $h_d$ twist ($h_d\in Z_n$).

A $Z_n$  monodromy defect is described by Fig. \ref{defect}(b).  The low energy
degrees of freedom in the defect are described by $g \in Z_n$.  Let us use
$g^{(k)} =\ee^{\frac{2\pi k \ii}{n}}, \ \ k=0,\cdots,n-1$, to describe the
$Z_n$ group elements.  The states on a  defect are described by $|g^{(k)}\>$.
To construct the path integral \eqn{UnWg} that describes
low energy dynamics of the defects,
let us first introduce
\begin{align}
&\ \ \ \
 C( g_0,g_1,g_2; g_{0'},g_{1'},g_{2'};
h_{01},h_{02},h)
\nonumber\\
&=
\ee^{2\pi\ii \om_3(g_0^{-1}h_{01}g_1,g_1^{-1}h_{12}g_{2},g_{2}^{-1}hg_{2'})}
\times
\nonumber\\&\ \ \ \
\ee^{-2\pi\ii \om_3(g_0^{-1}h_{01}g_1,g_1^{-1}hg_{1'},g_{1'}^{-1}h_{12}g_{2'})}
\times
\nonumber\\&\ \ \ \
\ee^{2\pi\ii \om_3(g_0^{-1}hg_{0'},g_{0'}^{-1}h_{01}g_{1'},g_{1'}^{-1}h_{12}g_{2'})}.
\end{align}
Physically, the above is the action amplitude for ideal fixed point system
described by \eq{Vnud}, on the complex in Fig. \ref{defect}(a).  Using $C(
g_0,g_1,g_2, g_{0'},g_{2'},g_{2'}, h_{01},h_{02},h)$, we can construct a
$|G|\times |G|$ matrix $U_\text{def}(g^*,h_d,h)$ whose matrix elements are
given by
\begin{align}
&
 [U_\text{def}(g^*,h_d,h)]_{g_j,g_i}
=\frac{ C( g_i,g^*,g^*; g_j,g^*,g^*; 1,1,h) }{
C( g^*,g_i,g^*; g^*,g_{j},g^*; 1,1,h) }\times
\nonumber\\
&
C( g^*,g_i,g_i; g^*,g_j,g_j; 1,h_d,h)
C( g^*,g_i,g^*; g^*,g_j,g^*; h_d,1,h).
\end{align}
Then the $|G|\times |G|$ matrix $T^{\Del\tau}_\text{def}(g^*,h_d)$
\begin{align}
 T^{\Del\tau}_\text{def}(g^*,h_d) = [U_\text{def}(g^*,h_d,1)]^\dag U_\text{def}(g^*,h_d,1)
\end{align}
will generate the imaginary-time evolution
for a single defect.
We have (for two defects)
\begin{align}
 \Tr (T^N)=
U_0^\text{bulk}
\Tr [ T^{\Del\tau}_\text{def}(g^*,h_d)]^N
\Tr [ T^{\Del\tau}_\text{def}(g^*,h_d)]^N,
\end{align}
where $T$ is the imaginary-time evolution operator $\ee^{-\Del\tau H}$ of the
whole system for a single time step, $T^{\Del\tau}_\text{def}$ is the
imaginary-time evolution operator for a single defect, and the bulk
contribution $U^\text{bulk}=1$.

Let us calculate $T^{\Del\tau}_\text{def}(g^*,h_d)$ for the monodromy defects
in the 2+1D $Z_n$ SPT state.  We will always choose $g^*=1$.  The cocycles in
$\cH^3(Z_n,\RZ)$ are labeled by $m=0,1,\cdots,n-1$, and are given by
\begin{align}
\label{om3Zn}
 \om_3(g^{(k_1)},g^{(k_2)},g^{(k_3)})
&=\ee^{m\frac{2\pi \ii}{n^2} k_1(k_2+k_3-[k_2+k_3]_n)},
\nonumber\\
g^{(k)} &=\ee^{\frac{2\pi k \ii}{n}},
\end{align}
where $[k]_n$ is a short-hand notation for
\begin{align}
 [k]_n \equiv \text{mod}(k,n).
\end{align}
In the following, we will only consider the $Z_n$ SPT phases described by $m=1$.

Let us first concentrate on 2+1D $Z_2$ SPT states.
Using the cocycles, we find that, for a 2+1D $Z_2$ SPT state,
\begin{align}
  U_\text{def}(g^*,h_d=1,h=1)
&=
\begin{pmatrix}
 1&1\\
 1&1\\
\end{pmatrix}
\nonumber\\
  T^{\Del\tau}_\text{def}(g^*,h_d=1)
&=
\begin{pmatrix}
 2&2\\
 2&2\\
\end{pmatrix}
.
\end{align}
We find that for a trivial monodromy defect,
the ground state on a defect is given by
$|g=1\>+|g=-1\>$, which is an expected result.
We also find that
\begin{align}
  U_\text{def}(g^*,h_d=-1,h=1)
&=
\begin{pmatrix}
 1&1\\
 -1&1\\
\end{pmatrix}
\nonumber\\
  T^{\Del\tau}_\text{def}(g^*,h_d=-1)
&=
\begin{pmatrix}
 2&0\\
 0&2\\
\end{pmatrix}
.
\end{align}
This mean that the a non-trivial  monodromy defect carry two degenerate states
$g=|\pm 1\>$.  However, the degeneracy can be lifted by perturbations that
respect the symmetry.

To study the $Z_2$ symmetry of the defects, let us consider
the path integral
\begin{align}
& \Tr (W_g T^N)= U_g^\text{bulk} U_0^\text{bulk}\times
\\
&\ \
\Tr (W^\text{def}_g [ T^{\Del\tau}_\text{def}(g^*,h_d)]^N)
\Tr (W^\text{def}_g [ T^{\Del\tau}_\text{def}(g^*,h_d)]^N)
,
\nonumber
\end{align}
where $ W_g,\ \ g\in Z_2$ is a representation of $Z_2$ acting on the total
system: $|\{g_k\}\> \to |\{gg_k\}\>$, and  $W^\text{def}_g$ describes how $Z_2$
symmetry transformation act on the low energy degrees of freedom on the defect.
We note that now the phase factor contribution from the bulk $ U_g^\text{bulk}
U_0^\text{bulk} $ has a $g$ dependence, and thus becomes non-trivial.

Let us first calculate  $W^\text{def}_g$.  Note that $\Tr
[T^{\Del\tau}_\text{def}(g^*,h_d)]^N$ is a trace of product of many $
U_\text{def}(g^*,h_d,h=1)$ operators.  To calculate $\Tr W^\text{def}_g [
T^{\Del\tau}_\text{def}(g^*,h_d)]^N $, we just need to replace one of the $
U_\text{def}(g^*,h_d,h=1)$'s by $ U_\text{def}(g^*,h_d,h=g)$.  Therefore, we
have
\begin{align}
\label{UWg}
&\ \ \ \
[U_\text{def}(g^*,h_d,1)]^\dag U_\text{def}(g^*,h_d,g)
\nonumber\\
&=
[U_\text{def}(g^*,h_d,1)]^\dag U_\text{def}(g^*,h_d,1)
W^\text{def}_g .
\end{align}
For $Z_2$ SPT state, we find
\begin{align}
  U_\text{def}(g^*,h_d=-1,h=1)
&=
\begin{pmatrix}
 1&1\\
 -1&1\\
\end{pmatrix}
,
\nonumber\\
  U_\text{def}(g^*,h_d=-1,h=-1)
&=
\begin{pmatrix}
 1&-1\\
 1&1\\
\end{pmatrix}
.
\end{align}
Eqn. \ref{UWg} becomes (for $h_d=-1$)
\begin{align}
 \begin{pmatrix}
 0&-2\\
 2&0\\
\end{pmatrix}
=
\begin{pmatrix}
 2&0\\
 0&2\\
\end{pmatrix}
W^\text{def}_{-1}
.
\end{align}
We find that
\begin{align}
 W^\text{def}_{-1}=\begin{pmatrix}
 0&-1\\
 1&0\\
\end{pmatrix}
=\ii \si^2 ,
\end{align}
for a non-trivial monodromy defect.

\begin{figure}[tb]
\begin{center}
\includegraphics[scale=0.6]{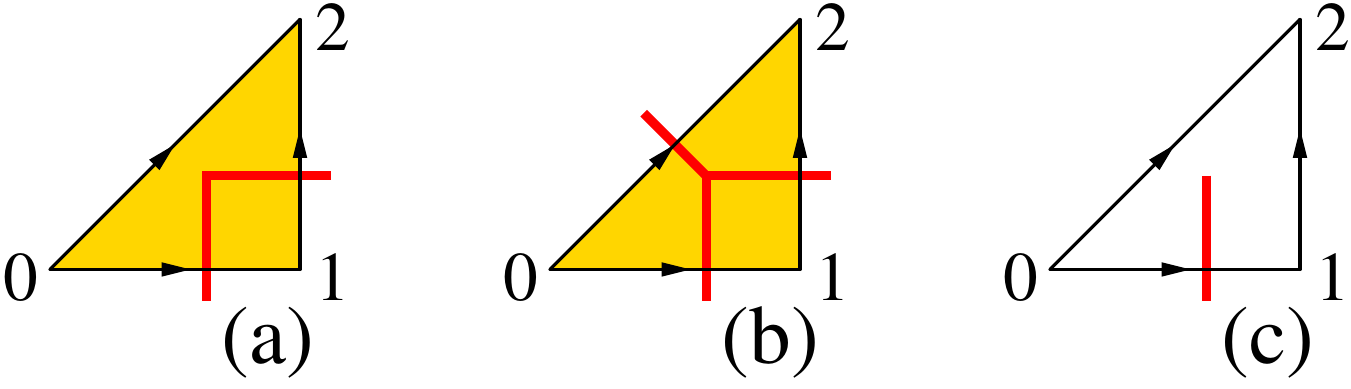} \end{center}
%Fig. 11
\caption{ (Color online)
A graphic representation of $U(-1, h_{01}, h_{12})$.  The edges crossed
by the red line have $h_{ij}=-1$.  The edges not crossed by the red line have
$h_{ij}=1$.  The gauge configurations in (a) and (b) have $U(-1, h_{01},
h_{12})=-1$.  The gauge configuration in (c)
and other configurations have $U(-1, h_{01}, h_{12})=1$.
}
\label{cocy}
\end{figure}

Next, let us calculate  the phase factor from the bulk, $U_g^\text{bulk}$.  For
this purpose, we introduce
\begin{align}
 U(g, h_{01}, h_{12})&=\frac{
\ee^{\ii 2\pi \om_3(h_{01},h_{12},h_{22'})}
\ee^{\ii 2\pi\om_3(h_{00'},h_{0'1'},h_{1'2'})}}{
\ee^{\ii 2\pi\om_3(h_{01},h_{11'},h_{1'2'})}}
\nonumber\\
&=C( g^*,g^*,g^*; g^*,g^*,g^*; h_{01},h_{12},g)
\end{align}
which is the action-amplitude on a single space-time complex in Fig.
\ref{defect}(a) with $g_i=g_{i'}=g^*=1$.  We find that (see Fig. \ref{cocy})
\begin{align}
\label{Ugggg1}
 U(-1, -1, -1)&=-1,
\nonumber\\
 U(g, h_{01}, h_{12})&=1\text{ otherwise}.
\end{align}
The total action-amplitude for the bulk is given by
\begin{align}
\label{UbU}
 U^\text{bulk}_g={\prod_{(ijk)}}'
U^{s_{ijk}}(g, h_{01}, h_{12})/
U^{s_{ijk}}(1, h_{01}, h_{12})
,
\end{align}
where $s_{ijk}$ describes the orientation of the triangle $(ijk)$, and
$\prod'_{(ijk)}$ is a product over all the triangles that are not monodromy
defects (\ie contain no $Z_2$-flux).  From Fig. \ref{z2gauge}, we see that
$U^\text{bulk}_{-1}=-1$ for two identical monodromy defects.  Therefore, the
low energy effective $Z_2$ symmetry transformation $W_g$ is given by
\begin{align}
 W_g=U^\text{bulk}_g  W^\text{def}_g\otimes  W^\text{def}_g
\end{align}
For $g=-1$, we have
\begin{align}
 W_{-1}=U^\text{bulk}_{-1}  W^\text{def}_{-1}\otimes  W^\text{def}_{-1}
=-\ii\si^2 \otimes \ii\si^2,
\end{align}
where the first $\ii\si^2$ acts on the states on the first monodromy defect and
the second $\ii\si^2$ on the second monodromy defect.

\begin{figure}[tb]
\begin{center}
\includegraphics[scale=0.4]{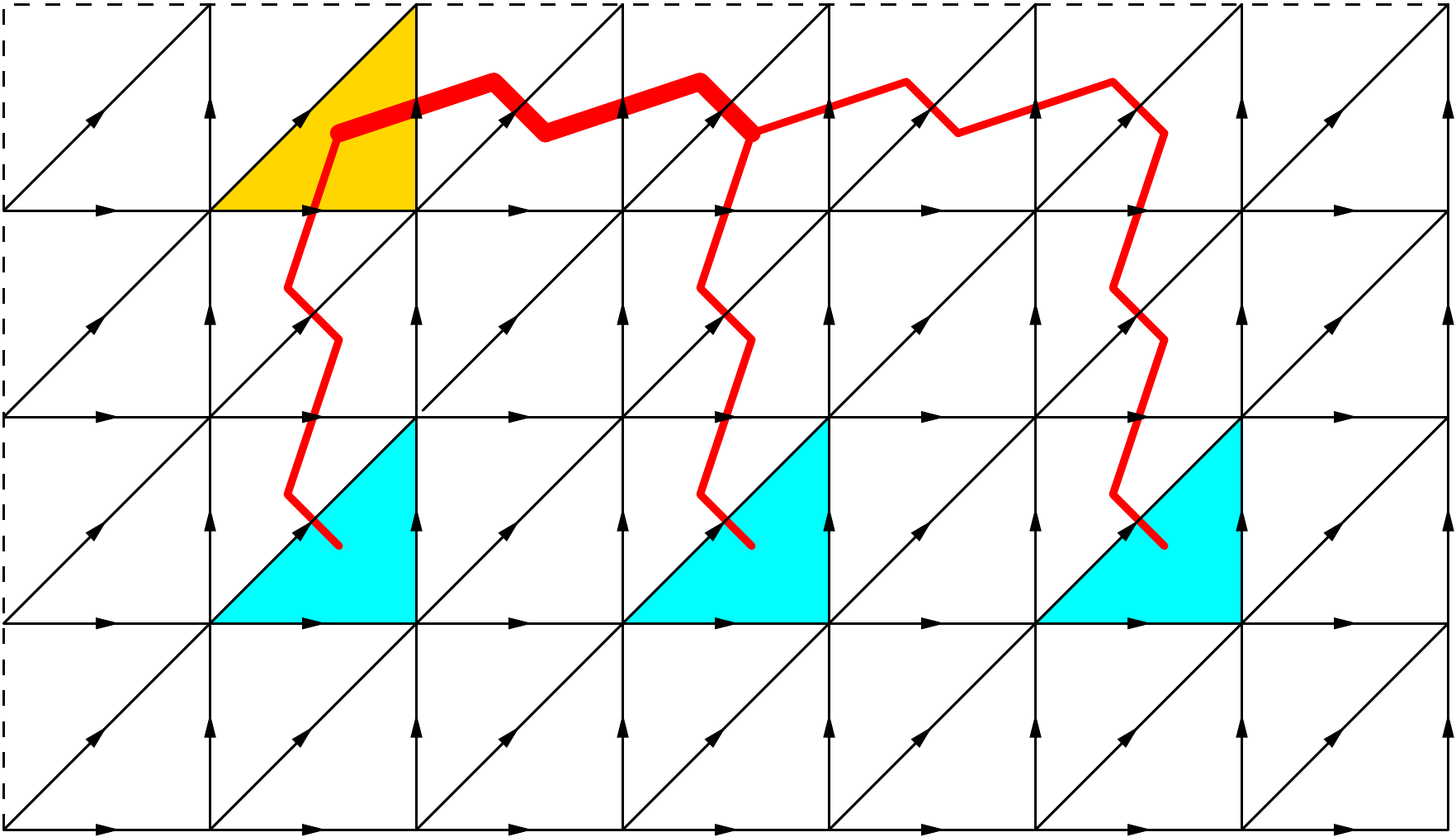} \end{center}
%Fig. 12
\caption{ (Color online)
A $Z_3$-gauge configuration with three \emph{identical} $Z_3$ monodromy defects
(blue triangles) on a torus.  The details of a  monodromy defect are given in
Fig. \ref{defect}(b).  The yellow triangle contributes a phase factor
$\ee^{2\pi \ii/3}$ to $ U^\text{bulk}_{g^{(1)}}$.
}
\label{ZnDefect}
\end{figure}

The above calculation can be generalized to $n$ \emph{identical}  monodromy
defects in a 2+1D $Z_n$ SPT state, described by the cocycle \eqn{om3Zn}.  We
find that the low energy effective $Z_n$ symmetry transformation $W_g$ is given
by
\begin{align}
 W_{g^{(1)}}&=U^\text{bulk}_{g^{(1)}}
\underbrace{
W^\text{def}_{g^{(1)}}\otimes  \cdots \otimes W^\text{def}_{g^{(1)}}}_{n\text{ terms}},
\nonumber\\
g^{(k)} & =\ee^{2\pi k \ii/n},
\end{align}
Here $ W^\text{def}_{g^{(1)}}$ is a $n\times n$ matrix acting on the states
on one $Z_n$ monodromy defect.
If we choose $|g^{(k)}\>$
to be the basis of the  states
on one $Z_n$ monodromy defect,
the action of $ W^\text{def}_{g^{(1)}}$  is given by
\begin{align}
 W^\text{def}_{g^{(1)}})|g^{(k)}\>
&= f_k |g^{(1)}g^{(k)}\>,\ \ \ k=0,1,\cdots,n-1,
\nonumber\\
 f_0&=\ee^{2\pi \ii /n}, \ \ \ \ \ f_{k>0} =1.
\end{align}
$U^\text{bulk}_{g}$ is a pure phase factor which is given by \eqn{UbU}.  For
the $Z_n$ SPT state described by the cocycle  \eqn{om3Zn}, we find that
\begin{align}
U(g^{(1)}, h_{01}^{(k)}, h_{12}^{(k')})
& =\ee^{2\pi \ii (k+k'-[k+k']_n)/n^2},
\nonumber\\
k,k' &=0,1,\cdots,n-1.
\end{align}
This gives us (see Fig. \ref{ZnDefect})
\begin{align}
U^\text{bulk}_{g^{(1)}}
=\prod_{k=0}^{n-1}
 \ee^{2\pi \ii (k+1-[k+1]_n)/n^2}
=
 \ee^{2\pi \ii /n}
\end{align}

We note that
%\begin{align}
%\label{Wgn}
$(W^\text{def}_{g^{(1)}})^n= \ee^{2\pi \ii /n}$.
%\end{align}
So we may say that each monodromy defect carries $\frac 1n+$integer $Z_n$
charges.  The fact that $U^\text{bulk}_{g^{(1)}} = \ee^{2\pi \ii /n} $ implies
that the bulk also carries an $Z_n$-charge 1.  So \topinv{\label{topZn}
$n$ \emph{identical}
elementary monodromy defects (\ie generated by the twist $h_d=g^{(1)}$) in 2+1D
$Z_n$ SPT states on a torus always carry a total $Z_n$-charge 2, if the $Z_n$
SPT states are described by the $m=1$ cocycle in $\cH^3(Z_n,\RZ)$ (see
\eqn{om3Zn}).}

Although we only present the derivation of the above result for a particular
choice of cocycles as in \eqn{om3Zn}, we have checked that the  result remains
to be valid for any choices of cocycles. In other words, the above result does
not change if we add a coboundary to the cocycle that describes the SPT state.

There is a simple way to understand the SPT invariant \ref{topZn}.  We
may view the $Z_n$ SPT state as a $U(1)$ SPT state with Hall conductance
$\si_{xy}=2\frac{1}{2\pi}$ in 2+1D.  An $Z_n$ monodromy defect corresponds to
$2\pi/n$ $U(1)$ flux which will nuclear $2\frac{1}{n}$ $U(1)$ charge.
$2\frac{1}{n}$ $U(1)$ charge correspond to $2\frac{1}{n}$ $Z_n$ charge.  Thus
$n$ identical $Z_n$ monodromy defects carry a total $Z_n$ charge $2$.

\bibliography{../../bib/wencross,../../bib/all,../../bib/publst,./tmp}

\end{document}